\documentclass[vecphys]{svmult}

\usepackage{makeidx}         
\usepackage{graphicx}        
\usepackage{psfig}           
\usepackage{multicol}        
\usepackage{url}             
\usepackage[bottom]{footmisc}
\usepackage{float, amsmath, natbib, epsfig}

\makeindex             

\begin{document}

\title*{Pulsars: Progress, Problems and Prospects}


\author{
Jonathan Arons
}


\institute{
     Departments of Astronomy and of Physics, and Theoretical Astrophysics Center \\ University of California, Berkeley \\
     and \\
     Kavli Institute of Particle Astrophysics and Cosmology \\
     Stanford University \\
     \texttt{arons@astro.berkeley.edu}
}

\maketitle

\begin{abstract}

Pulsars are the quintessential dogs that don't  bark in the night - their observed loss of rotational energy mostly disappears into the surrounding world while leaving few traces
of that energy loss in observable photon emission.  They are the prime example of compact objects which clearly lose their energy through a large scale Poynting flux. 

I survey recent successes in the application of relativistic MHD and force-free electrodynamics to the modeling of the pulsars' rotational energy loss mechanism as well as to the structure and emission characteristics of Pulsar Wind Nebulae.  I suggest that unsteady reconnection in the current sheet separating the closed from the open zones of 
the magnetosphere is responsible for the torque fluctuations observed in some pulsars, as well as for departures of the braking index from the canonical value of 3. I also discuss the theory of high energy pulsed emission from these neutron stars, emphasizing the significance of the boundary layer between the closed and open zones as the active site in the outer magnetopshere. I elaborate on the conflict between the models currently in use to interpret the gamma ray and X-ray pulses from these systems with the electric current flows found in the spin down models. Because the polar cap ``gap'' is {\it the} essential component in the supply of plasma to pulsar magnetospheres and to pulsar wind nebulae, I emphasize the importance of high sensitivity gamma ray observations of pulsars with core components of radio emission and high magnetospheric voltage, since observations of these stars will look directly into the polar plasma production region and will probe the basic plasma parameters of these systems. I also discuss the current state  of understanding and problems in the shock conversion of flow energy into the
spectra of the synchrotron emitting particles in the Nebulae, emphasizing the possible role of heavy ions in these processes.   I comment on the prospects for future developments and improvements in all these areas.

\end{abstract}

\section{Introduction} \label{sec:intro}
This paper was originally entitled ``Pulsar Emission: Where to Go'' in the theory of  pulsar radio and high energy emission.  ``Where to Go'' on these topics depends greatly on where one thinks the energy and the particles are. Therefore, I mostly discuss questions of dynamics, and return to the emission physics through the lens of the successes and problems of dynamical models of rotation powered pulsars' magnetospheres. Also, I emphasize results and problems common to all pulsars. I will give an impressionistic rather than a comprehensive review, more in the spirit of setting goals as I see them rather than providing a scholarly survey\footnote{Much of what I have to say derives from collaborations, most recently with  Elena Amato, Phil Chang, Niccolo Bucciantini, Eliot Quataert, Todd Thompson and especially Anatoly Spitkovsky; in earlier years, with Ted Scharleman, Bill Fawley, Colin Norman, David Alsop,  Don Backer,  Brian Gaensler, Yves Gallant, Vicky Kaspi, Bruce Langdon, Claire Max and Marco Tavani.  However, I am solely responsible for the views expressed in the subsequent pages.}. I have focused on issues that can be addressed by timing and by high energy photon observations. For a survey of the current status of models of pulsar radio emission, see the article by Usov in these proceedings.

\section{Pulsar Electrodynamics: Follow the Energy} \label{sec:electrodynamics}

\begin{figure}[H]
\begin{center}
\hspace{10\unitlength}
\begin{picture}(300,100)(0,15)
\put(0,-100){\makebox(300,200)[tl]{\includegraphics[width=4in]{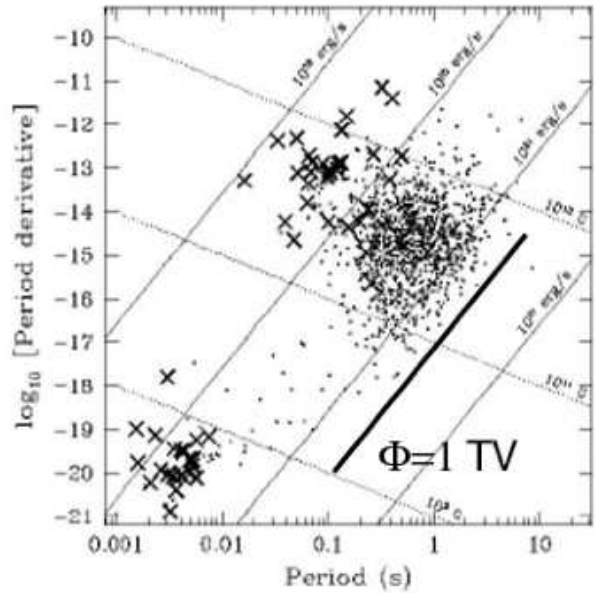}}}
\end{picture}
\end{center}
\vspace{7.5cm}
\caption{Observed RPP periods and period derivatives, from \cite{kaspi06}.  ``X'' marks a pulsar with $P, \dot{P}$ measured from X-rays  as well as radio observatons. The line
$\Phi = 10^{12}$ Volts = 1 TV is the locus in the $P \dot{P}$ diagram where the rotation induced voltage drops to $10^{12}$ V, clearly marking a boundary beyond which pulsar emission is unlikley }
 \label{fig:PPdot}
\end{figure}

Astrophysical understanding comes from using observation and theory to find and follow the flow of energy, mass and momentum in the macroscopic systems of interest.  For pulsars, this has been a challenge.  They are quintessential ``dogs that don't bark in the night'', with only a small fraction of the energy they broadcast into the Universe appearing in directly observable forms. The interpretation of the regular pulse periods from sources distant enough to require stellar and sub-stellar luminosity in the radio (and in the infrared, optical, X-ray and gamma ray) discovered by the radio astronomers led immediately to the understanding that the observed periods are the consequence of rotation of massive stellar flywheels (neutron stars). The steady lengthening of the pulse period, shown in Figure \ref{fig:PPdot}, led immediately to a permanently
successful model of that spindown, the electromagnetic torques exerted if the stars are sufficiently well magnetized. 

One can readily estimate the magnitude of such torques from the observation that rotation
of a stellar magnetic field $B$ induces a poloidal  electric field of magnitude 
$E \sim (\Omega r /c)  B_p$, with $\Omega = 2\pi /P$ and $B_p$ the poloidal magnetic field - from the point of view of the torque, that field is well approximated by a dipole with
dipole moment $\mu = R_*^3 B_p$. The winding up of the magnetic field as the conducting star rotates requires the existence of a toroidal magnetic field of magnitude
$B_\phi \sim (\Omega r /c) B_p$ .  This $E$ field corresponds to energy loss in a Poynting flux 
$c {\boldsymbol E} \times {\boldsymbol B} /4\pi $.  If the electromagnetic energy density exceeds all the material energy densities, one obtains the total energy loss
${\dot E}_R$ and therefore the torque $\dot{J} = {\dot E}_R / \Omega$ by summing the
Poynting flux over a sphere of radius $R_A$, expected to be comparable to the light
cylinder distance $R_L = c/\Omega$, where the electromagnetic inertia $B^2 /4 \pi c^2$
causes the poloidal field to depart from the imposed stellar (dipole) field by an amount
on the order of $B_p$ itself. Then $R_A$ is the smallest radius where $B_\phi$ becomes comparable to $B_p$, and $B_p \approx \mu/R_A^3$. Therefore 
\begin{equation}
\dot{E}_R \sim 4\pi R_A^2 c \frac{E(R_A) B_\phi (R_A)}{4\pi} 
 \approx \left (\frac{\Omega^4 \mu^2 }{c^3}  \right) \left( \frac{R_L}{ R_A} \right)^2.
\label{eq:approx-spin}
\end{equation}
For radio emitting Rotation Powered Pulsars (RPPs), stars are known with 
$\dot{E}_R =I \Omega {\dot \Omega}$ from as small as $10^{30}$ ergs/s to as large as $10^{39}$ erg/s - $I$ is the stellar moment of inertia, $I \approx 10^{45}$ cgs for
currently acceptable equations of state for neutron stars.
\vspace*{-1cm}

Modeling of RPPs has one great advantage over modeling of other compact objects - observations of $P, \dot{P}$ determine the energy supply, to within
the uncertainties in the moment of inertia.  In contrast, modeling of accreting
black holes always suffers from major uncertainty as to whether the systems
are, or are not, accreting at a well determined rate, e.g., the Eddington limit.
This fact runs through much of what I discuss below - with the total energy
budget known, the effort turns to aspects of the machine's physics at a level
of sophistication not sustainable in many other aspects of compact object
physics.

\subsection{Force-Free Model: Heuristics}

Expression (\ref{eq:approx-spin}) makes no reference at all to charged particles, and
indeed the first theories of RPP spindown (some invented before RPPs were discovered) 
invoked the electrodynamics of a vacuum rotator
as an explanation of the observed $\dot{P}$ (\citealt{deutsch55, pacini67, ostriker69}).
Except for geometric factors, vacuum theories yield expression (\ref{eq:approx-spin}), but with the special addition that as the angle $i \equiv \angle ({\boldsymbol \mu}, \; {\boldsymbol \Omega})$ becomes small, so does the torque, in proportion to $\sin^2 i$.
Application of this model to the spindown data  for normal RPPs yields dipole moments
on the order of $10^{30}$ cgs, corresponding to surface dipole fields 
$B_* \equiv \mu R_*^3 \sim 10^{12}$ Gauss  
for ``normal'' neutron stars\footnote{``Normal'' means neutron stars discovered via their ``normal'' radio emission,  an obvious selection effect.  More recent 
discoveries, of millisecond radio pulsars and of X-ray selected objects, have revealed neutron stars 
with magnetic moments from $\sim 10^{33}$ cgs down to ``zero'' in the X-ray burst
sources, which effectively means $\mu < 10^{26}$ cgs (\citealt{kaspi06}). In particular, X-ray cyclotron lines confirm the existence of $10^{12}$ Gauss surface fields.} 

Vacuum models have large electric fields parallel to the magnetic field at the stellar surface, a fact which led Deutsch (in the context of the oblique rotator $i \neq 0$) to suggest that a vacuum rotator has to form a charged
magnetosphere, as charged particles move from the surface to short out ${\boldsymbol E} \cdot {\boldsymbol B}$. Simultaneously with the appearance of the vacuum torque models after RPP discovery,
\cite{gold69} independently made the same observation in the context of the aligned rotator. They went further to suggest that a charge separated outflow forms, creating a conduction current ${\boldsymbol J} = \eta_R {\boldsymbol v}$, where $\eta_R$ is the charge density required to force $E_\parallel = {\boldsymbol E} \cdot {\boldsymbol B}/B$ to zero, $\eta_R = -{\boldsymbol \Omega} \cdot {\boldsymbol B}/2\pi c$ + relativistic corrections. They also introduced the idea that the magnetosphere is ``force-free'', that
is, the electromagnetic energy density is so large that all inertial, pressure and dissipative forces can be neglected, a concept consistent with the fact that RPPs are non-barking dogs - the large energy loss manifested in spindown does not appear
in any radiative emission associated with the magnetosphere (here defined as the region
interior to $R_A$, probably $\approx R_L$.)   Conceived of as a system which is strictly 
steady in the corotating frame  - after all, pulsars form superb clocks, therefore the
rotating lighthouse picture should apply, which it does, at least to averages of many 
pulses - the flow of the charges decompose into any velocity parallel to $\boldsymbol B$
plus rotation ${\boldsymbol \Omega} \times {\boldsymbol r}$. The same charge density
and velocity decomposition apply to the magnetohydrodynamic (MHD) model introduced
by \cite{michel69} in the same year, with the difference that the MHD model assumes a density large compared to $\eta_R /q$.  Charge separated/MHD outflow and magnetospheric dynamics  occurs for densities equal to or less than/greater than 
\begin{eqnarray}
\frac{|\eta_R|}{e} & \approx & \frac{\mu}{R_L^3 P c e} \left( \frac{R_L}{R_A} \right)^3
     \frac{B}{B(R_A)}  \nonumber \\
     & \approx & 6 \times 10^3 \frac{\mu_{30}}{P_{100}^3}   \left( \frac{R_L}{R_A} \right)^3
        \left[ \left(\frac{R_A}{r} \right)^3 + \left(\frac{R_A}{r} \right)^2 \right] \; {\rm cm}^{-3},
\label{eq:GJdensity}
\end{eqnarray}
if the particles have Larmor radii small compared to $r$. Here $P_{100} = P/100$ msec
and $\mu_{30} = \mu /10^{30}$ cgs.
Interpreted as a particle outflow,
this density corresponds to a particle loss rate 
\begin{equation}
\dot{N}_R = 4\pi R_A^2 c \frac{|\eta_R (R_L)|}{e} = \frac{I_R}{e} = \frac{c \Phi}{e} 
   = 2.7 \times 10^{32} \frac{\mu_{30}}{P_{100}^2} \; {\rm s}^{-1} .
\label{eq:GJflow}
\end{equation} 
The poloidal electric current $I_R$ is the current expected such that the induced 
magnetic field becomes comparable to the dipole field at the light cylinder.  In Goldreich
and Julian's charge separated picture of the aligned rotator, the charges in $I_R$,
composed of the charges in the fully charge separated plasma flowing parallel to the poloidal magnetic field, provides the support for $B_\phi$ and the Poynting flux.  
$\Phi = \sqrt{\dot{E}_R /c} = 1.3 \times 10^{15} \mu_{30}/P_{100}^2$ Volts is the
magnetospheric electric potential.

Observations of the synchrotron emission from young pulsar
wind nebulae (PWNe) ({\it e.g.} \citealt{rees74, ken84, gallant94, gaens02}) reveal particle injection rates $\dot{N}$ (in the form of electrons
and - probably - positrons) corresponding to densities in a wind outflow 
$n = \dot{N}/4\pi r^2 c$  (at distances {\it much} larger than $R_L$) a factor of $10^3$ 
and more larger than
the density of the charge separated flow predicted by the charge separated wind idea.
Thus the nebular observations suggest MHD models with a  quasi-neutral plasma (which can only be electron-positron pairs, see \S \ref{sec:pairs}), appear to be a good starting place for understanding these systems.

\subsection{Force-Free Model: Results}

Thus the simplest idea is that a dense plasma exists everywhere in the magnetosphere
and beyond, with the plasma energy density much lower than $B^2 /8\pi$ - for the young, high voltage pulsars, plasmas with energy density remotely comparable to that of the EM fields and still under the rotational control of the stars would lead to pulsed photon emission orders of magnitude greater than what is observed. The force free idea was
elegantly formulated in the early 70s in the ``pulsar equation'' for the aligned rotator
(\citealt{michel73a, scharl73}), a variation
of the Grad-Shafronov equation familiar from the theory of magnetic confinement
(\citealt{bateman78}):
\begin{equation}
\left(1-\frac{\varpi^2}{R_L^2} \right) \left( \frac{\partial^2 \psi}{\partial \varpi^2 }
    + \frac{\partial^2 \psi}{\partial z^2} \right) - 
      \left(1 + \frac{\varpi^2}{R_L^2}\right)  \frac{1}{\varpi} \frac{\partial \psi}{\partial \varpi} +
       I(\psi) \frac{\partial I}{\partial \psi} = 0.
\label{eq:psr_eq}
\end{equation}
Here $\psi$ is the magnetic flux, with the poloidal magnetic field related to $\psi$ by
${\boldsymbol B}_p = - \varpi^{-1} \hat{\boldsymbol \phi} \times {\boldsymbol \nabla} \psi$, $\varpi$ is the cylindrical distance from the rotation ($z$) axis, while the toroidal magnetic field is $B_\phi = I(\psi) /\varpi$, with $I$ the current enclosed within a circle 
around the $z$ axis of radius $\varpi$.  These fields are supported by charge and current densities all derivable from $\psi$, once a solution of (\ref{eq:psr_eq}) is
determined. The electromagnetic structure of this non-pulsing model was (and is) thought to capture the essence of what is needed to make a full, oblique rotator model. 
The model says nothing about particle energetics - thus it, and its oblique rotator descendants, provides a geometric platform and an accounting of the dominant electromagnetic energy flow tapped by the subdominant dynamical processes that
lead to the observable emissions.  In particular, it does provide a basic model for the
invisible processes that lead to pulsar spindown.

Solution of (\ref{eq:psr_eq}) in the simplest relevant case (a star centered dipole 
with rotation axis parallel to the magnetic moment) has
taken a remarkably long time.  Solutions appeared immediately for a) a strictly co-rotating magnetosphere $I(\psi) = 0$, which is not relevant since it does not spin down (this dog really doesn't bark!) and implies particle motions faster than the speed of light at $\varpi > R_L$ and b) a star centered monopole, with an elegant result obtained by \cite{michel73} whose most important element is the poloidal current function 
\begin{equation}
I(\psi) = c \Phi \frac{\psi}{\psi_0}  \left( \frac{2 \psi_0 -  \psi }{\psi_0} \right),
\label{eq:mon-current}
\end{equation} 
where  $ \psi_0 \approx \pi \varpi_{cap}^2 (2 \mu /R_*^3) \approx (\mu /R_L) (R_L/R_Y)$ is the open magnetic flux in one hemisphere of the monopole  - $R_Y$ is the equatorial radius of the
$Y$ point in the magnetic field which marks the largest extent of the closed
magnetic field lines in the rotational equator, $\varpi_{cap} \approx (R_* /R_Y)^{1/2}$ is
the cylindrical radius of the magnetic polar cap, and 
$\psi_0 = R_Y \Phi = \mu /R_Y$.  
Finding these solutions 
required inspired guessing of $I(\psi)$. In the years between
1973 and 1999, many attempts were made to solve (\ref{eq:psr_eq}) by 
guessing various forms for $I(\psi)$ and applying ever more clever analytic
techniques to this fundamentally non-linear model.  None yielded anything 
credible - see \cite{mestel99} for a summary of much of this work.

\begin{figure}[H]
\begin{center}
\hspace{10\unitlength}
\begin{picture}(300,100)(0,15)
\put(-40,-100){\makebox(200,200)[tl]{\includegraphics[width=2.25in]{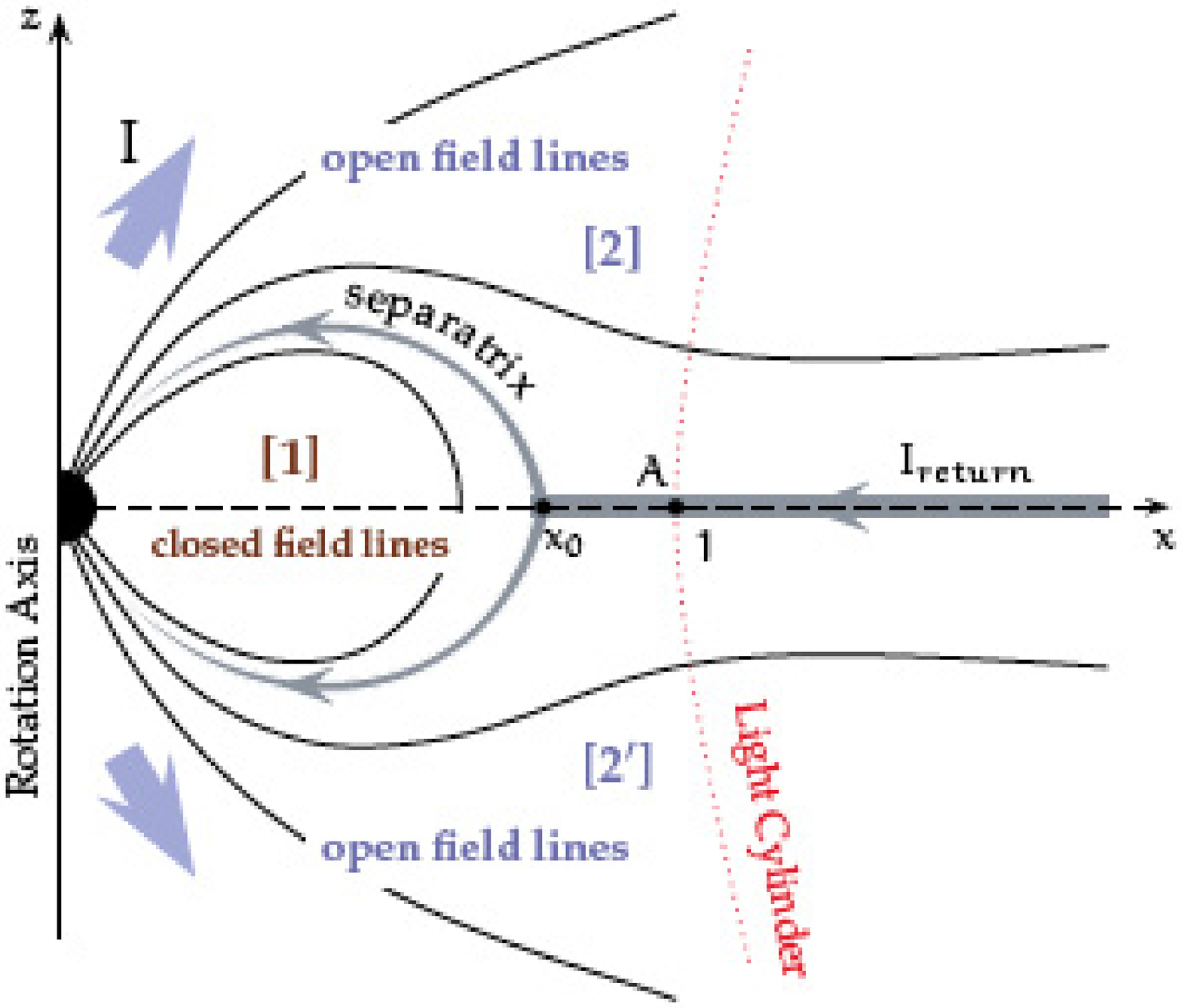}}}
\end{picture}
\hspace{3\unitlength}
\begin{picture}(300,100)(0,15)
\put(130,-25){\makebox(200,200.00)[tl]{\includegraphics[width=2.5in]{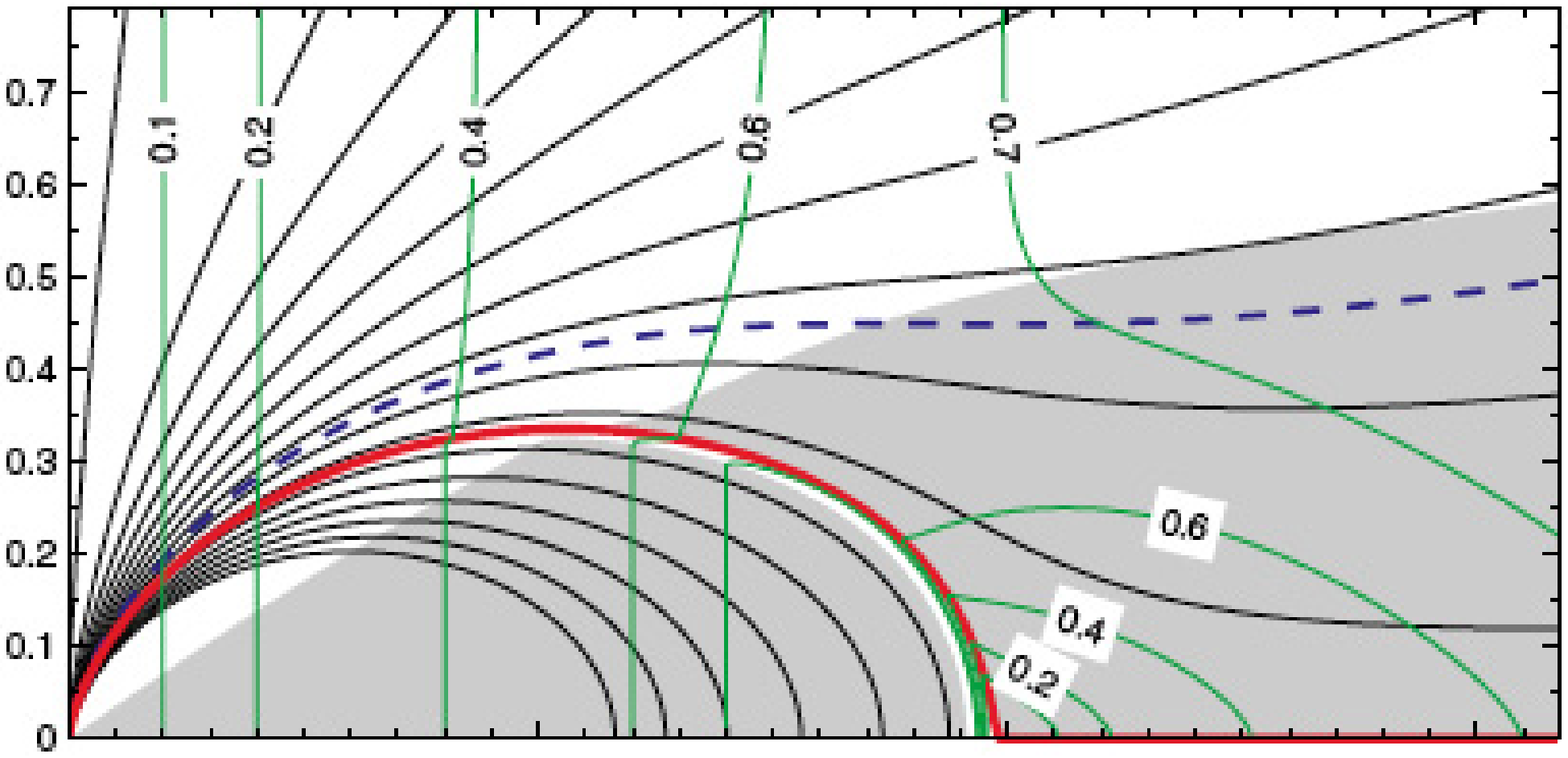}}}
\end{picture}
\hspace{1\unitlength}
\end{center}
\vspace{-2.3cm}
\caption{{\bf  Left}:  Cartoon of the aligned rotator's magnetosphere, showing the
primary polar current and the return current flowing along the separatrix a the equatorial
current sheet. The anti-aligned case, with ${\boldsymbol \mu}$ anti-parallel to 
${\boldsymbol \Omega}$ is shown. The aligned case has the same topology, with the
sign of the current flows reversed.
{\bf Right}: Field lines (magnetic flux surfaces) of the aligned rotator solution, for the case $R_Y = 0.992 R_L$}
 \label{fig:aligned}
\end{figure}

The situation changed when \cite{contop99} took seriously the nonlinear
eigenvalue and eigenfunction character of (\ref{eq:psr_eq}) and its associated boundary and regularity conditions and successfully applied an iterative numerical technique to find ${\boldsymbol B}$ and $I(\psi)$ to produce a result with ${\boldsymbol E} \cdot {\boldsymbol B} = 0$ and $E^2 - B^2 < 0$ everywhere - the latter condition is required if the model is to be taken seriously as a representation of a physical magnetosphere, since the ${\boldsymbol E} \times {\boldsymbol B} $ velocity must be subluminal, for a physical model\footnote{$E^2 - B^2 > 0$ is possible, in principle. However, in the absence of losses particles then accelerate to
energy $\sim q\Phi$ in distances not greater than $R_L$ 
In the younger pulsars, the acceleration becomes radiation reaction limited,
implying radiation emission from RPPs far in excess of what is observed.}. This solution, in which the last closed field line is assumed to  
have equatorial radius equal to $R_L$,  has been reproduced with increasing
numerical accuracy by \cite{gruzinov05}, \cite{mckinn06}, \cite{timokhin06} and \cite{spit06}. It exhibits a number of long expected features (\citealt{michel75}). In particular, the last closed field line has a Y-type neutral point on the equator, with return current flowing (mostly) in an unresolved current sheet along the boundary of the closed zone, then extending as an equatorial current sheet to radii $\varpi > R_Y$.  Figure  
\ref{fig:aligned}, taken from \cite{timokhin06}, shows the poloidal magnetic geometry of the aligned rotator. The solution illustrated has $\psi_0 = 1.23 \mu /R_L$, in the case
$R_Y/R_L = 0.992$, in excellent agreement with \cite{gruzinov05} and with
\cite{contop99}, who assumed $R_Y = R_L$ exactly.  All authors agree on the
spindown energy losses of the aligned rotator, 
$ \dot{E}_R = k \Omega^4 \mu^2 /c^3, \; k = 1 \pm 0.1$.

As predicted by \cite{michel74}, the asymptotic structure 
($r = \sqrt{\varpi^2 + z^2} \gg R_L$) approaches that of
the (split) monopole, as appears most clearly in McKinney's and Timokhin's results. Thus the poloidal current flow is almost that of the monopole, a point discussed further below. Also, as shown by \cite{goodwin04}, \cite{contop05} and \cite{timokhin06}, the steady state force free magnetopshere has a whole range  of possible solutions, parameterized by $R_Y /R_L \leq 1$. 

\vspace*{-1.3cm}

\begin{figure}[H]
\begin{center}
\hspace{10\unitlength}
\begin{picture}(300,100)(0,15)
\put(-20,-100){\makebox(300,200)[tl]{\includegraphics[width=2in]{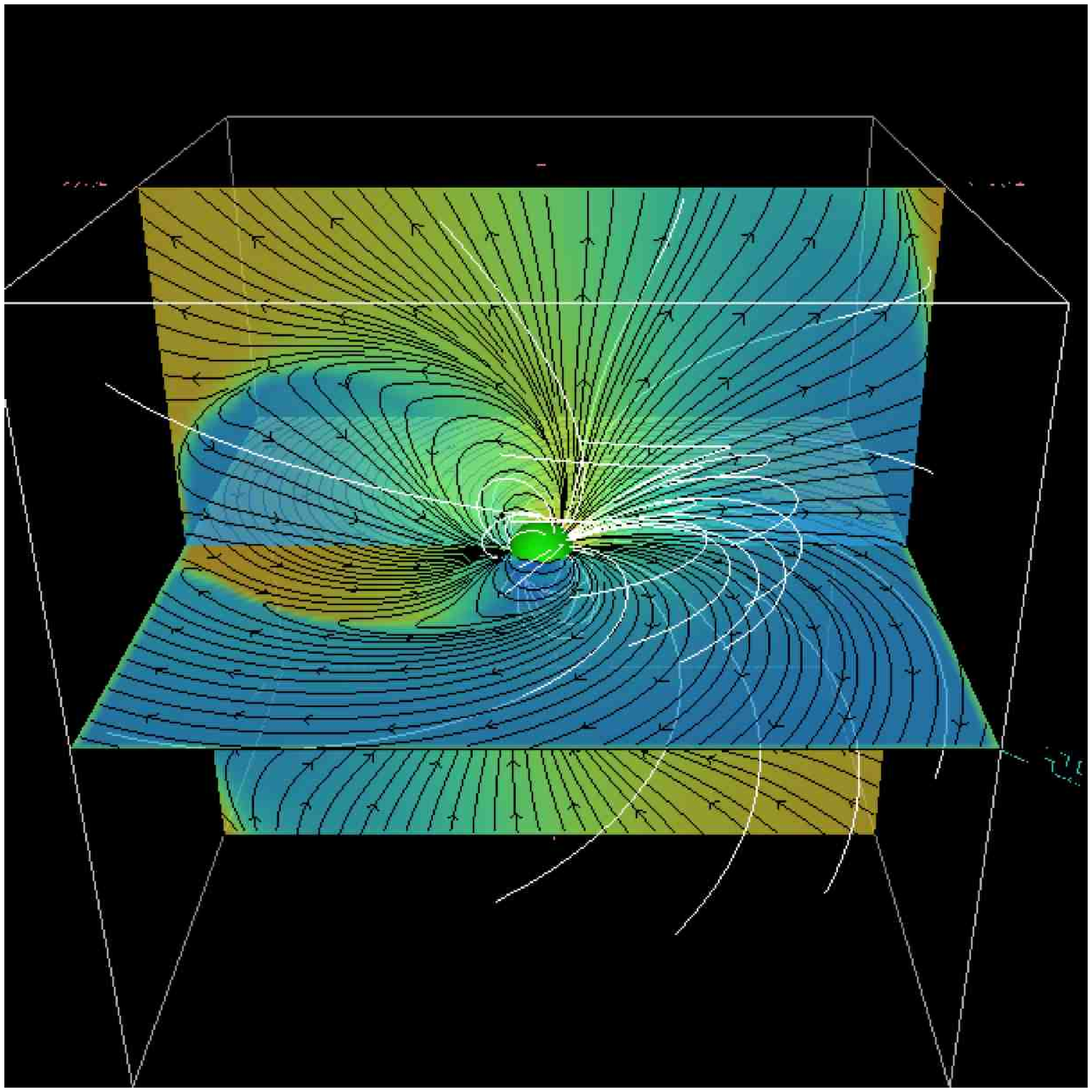}}}
\end{picture}
\hspace{3\unitlength}
\begin{picture}(300,100)(0,15)
\put(160,5){\makebox(300,200)[tl]{\includegraphics[width=2.25in]{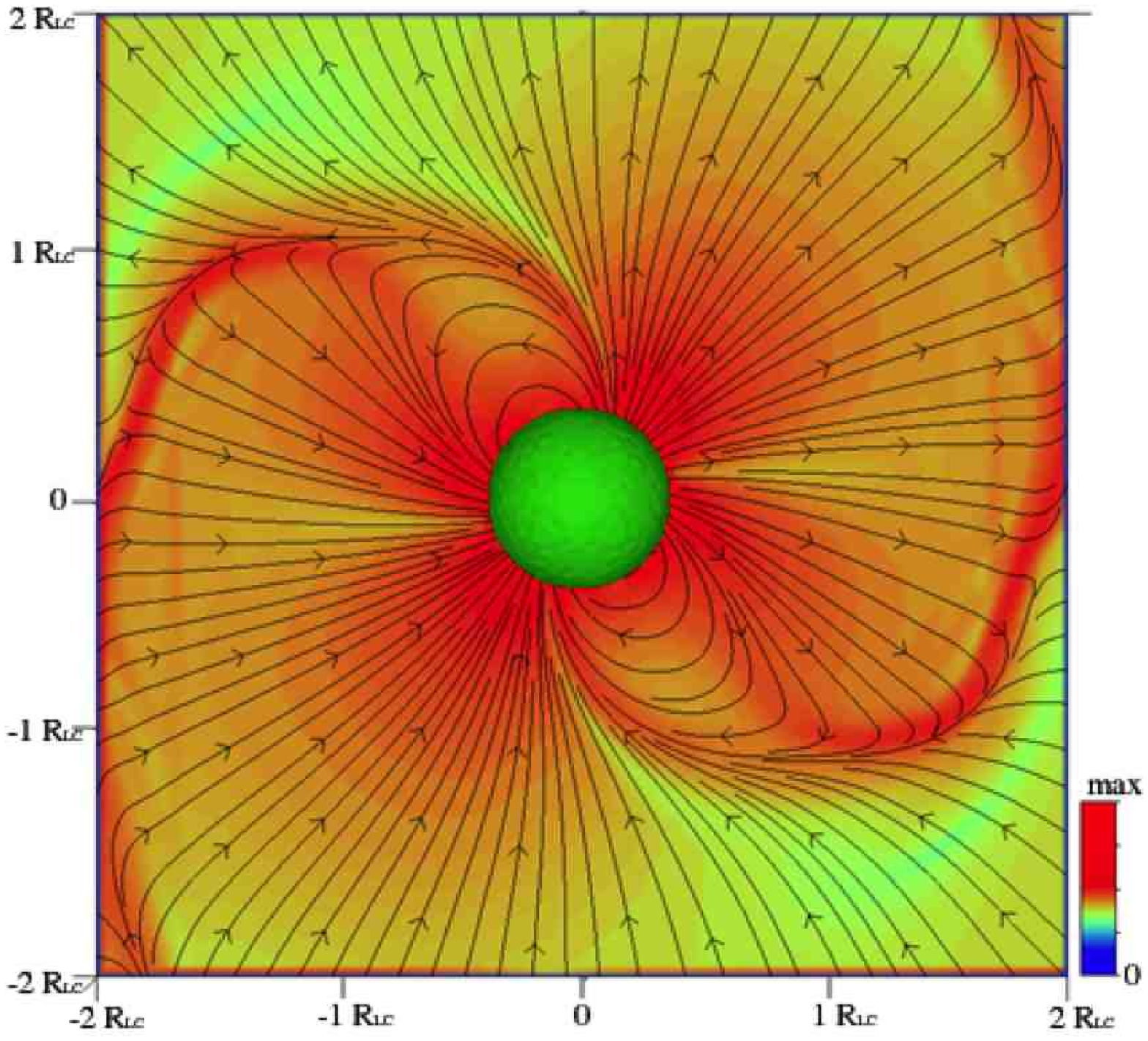}}}
\end{picture}
\hspace{1\unitlength}
\end{center}
\vspace{-1.3cm}
\caption{{\bf  Left}:  A snapshot of a force free simulation of a RPPs magnetosphere, for $r < 2 R_L$ (from \citealt{spit06}).
{\bf Right}: Total current $(c/4\pi){\boldsymbol \nabla} \times {\boldsymbol B}$, the sum of conduction
and displacement currents.}
 \label{fig:oblique}
\end{figure}

By solving
the time dependent force free equations, \cite{spit06} showed that the force-free
magnetosphere evolves with $R_Y \rightarrow R_L$, starting from a static 
vacuum magnetic dipole on a star instantaneously set into rotation with angular 
velocity $\Omega$; at $t=0$, the electric field on the stellar surface was set 
equal to 
$-({\boldsymbol \Omega} \times {\boldsymbol r} ) \times {\boldsymbol B}$. 
The
rate of approach of $R_Y$ to $R_L$ depends on the artificial resistivity used
to control singular behavior at current sheets. \cite{komiss06} and \cite{bucc06} 
found  similar results using a relativistic MHD model ({\it i.e.} inertial forces included). Spitkovsky's method allowed the current sheet to have an arbitrary shape. Thus, he also succeeded in finding the force free solution for arbitary $i$; the resulting 3D model of the magnetic field appears in Figure \ref{fig:oblique}. Within the assumption of a magnetosphere everywhere filled 
with plasma of density sufficient to short out parallel electric fields (and
no physics that might support such electric fields in a plasma of density greater
than $|\eta_R| /e$), a full solution for the electromagnetic structure of the RPP's magnetosphere (both aligned and oblique rotators) is now available, after 38 years of discussion. Spitkovsky's result,
\begin{equation}
\dot{E}_R = k\frac{\mu^2 \Omega^4}{c^3}(1 + \sin^2 i), \; k = 1 \pm 0.1,
\label{eq:spindown}
\end{equation}
contains all the aligned rotator studies as a special case. As is clear from Figure
\ref{fig:oblique}, the magnetic topology of the oblique rotator (closed field lines
terminating at a Y line, current sheet extending from the Y line separating
regions of oppositely directed field in the wind) is a rotationally distorted version of the
simpler aligned rotator geometry. 
\vspace*{-0.5cm}
\noindent 

\subsection{Beyond the Force-Free Model: Plasma Sighs and Whispers}

Until recently, the magnetosphere was assumed to have $R_Y = R_L$, an assumption consistent with the observation that radio emission from low altitude 
appears to occupy a polar flux tube which, if modeled as being in a static vacuum dipole's geometry, is bounded by a closed zone which appears to extend to the light cylinder (\citealt{rankin90, kramer98}), {\it i.e.}, have a polar cap opening angle 
$\theta_c = (R_* /R_Y)^{1/2}$ with $R_Y = R_L$; of course, the observations and the simplified model do not come close to {\it proving} that 
$R_Y = R_L$, or even that static dipole model for the $B$ field is accurate all the way to
$r = R_Y$. But if $R_Y /R_L \leq 1$ is some constant, then since $R_L$ increases as the pulsar spins down, there must be net conversion of open field lines to closed field lines (transfer of open magnetic flux to closed flux)  on the spindown time scale. This topological change requires reconnection and a violation of ideal MHD, at least in local regions - the likely culprit is the Y-line and the current sheet, as has been observed by  \cite{contop05} and \cite{contspit06}. Having made this observation, \cite{contop05} suggests that for pulsars near the death line in Figure \ref{fig:PPdot}, reconnection proceeds sporadically - the magnetosphere ``coughs'' -
because of failure in the supply of plasma from the polar cap accelerator and pair creation region - see \S\ref{sec:pairs}.   He applies this idea to a scenario for the major outburst observed
in the magnetar SGR1806-20 on December 27, 2004 (\citealt{hurley05, palmer05}).  

In fact reconnection is likely to have an unsteady, ``bursty'' character for all pulsars\footnote
{Reconnection measured in the laboratory and in space plasmas, and observed in solar plasmas, does occur with bursty, often explosive, behavior.} - the magnetosphere should be noisy at some level all the time. Figure \ref{fig:magnetosphere} shows a snapshot of the magnetosphere of a rotating neutron star with large plasma supply, taken from \cite{bucc06}.  This relativistic MHD (not force-free) model was designed to represent the wind from a young, rapidly rotating magnetar, with the wind driven by the enormous thermal pressure at the neutron star's surface.  The wind, formed by plasma flowing out on open field lines, converges on the equatorial current sheet. That
convergence causes driven reconnection, operating in a bursting mode - the sheet
forms ``plasmoids'', islands  of reconnected poloidal field with closed, O-point magnetic topology\footnote{In these axisymmetric models, the islands are magnetic torii.} which flow away at the local Alfven speed, $\sim c$. Numerical resistivity in the code provided the dissipation required to allow the transformations of field topology shown - the GEM study of non-relativistic reconnection (\citealt{gem01}) demonstrated that any non-ideal effect allows rapid driven reconnection with inflow velocity into the separatrix (the current sheet separating the closed field from the open field regions in Figures \ref{fig:aligned} 
and  \ref{fig:oblique}) being $(0.1-0.2) v_A$, outside the restrictive  bounds of incompressible MHD with uniform resistivity.  Note that the reconnection sporadically transforms the Y-line into a X-line, with the current sheet then containing a series of {\it dynamical} X-lines, all leaving the star - \cite{lyub90} objection to the formation of a {\it stationary} X-line (the field lines inside the separatrix on the open side of the X-line are not anchored to the magnetosphere) is answered simply by the fact that the plasmoids indeed are not anchored to the magnetosphere and fly away, but constantly reform. The Poynting flux was found to be time dependent, fluctuating around the mean by 
$\sim$ 30\%.

The Bucciantini {\it et al.} model was not designed specifically for classical RPPs, or for magnetars in their currently observed state - future work on reconnection in configurations with electron-positron plasma, where the Hall effect, so important in
the nonrelativistic studies, is absent (and is replaced either by pressure anisotropy  as the facilitator of rapid reconnection, as pointed out by \citealt{bessh05}, or by particle inertia), in a state suggested
by the models of plasma supply in RPPs' magnetospheres, are required to quantify this ``noisy magnetosphere'' picture. In particular, extending such
modeling to the full 3D rotator in strongly magnetized MHD has not yet been done.  Also, {\it it remains to be demonstrated that noise in the current sheet at 
\vspace*{1.15in}
\begin{figure}[H]
\begin{center}
\unitlength = 0.0011\textwidth
\hspace{1\unitlength}
\begin{picture}(300,100)(0,15)
\put(-50,70){\makebox(350,200){\includegraphics[width=1.75in]{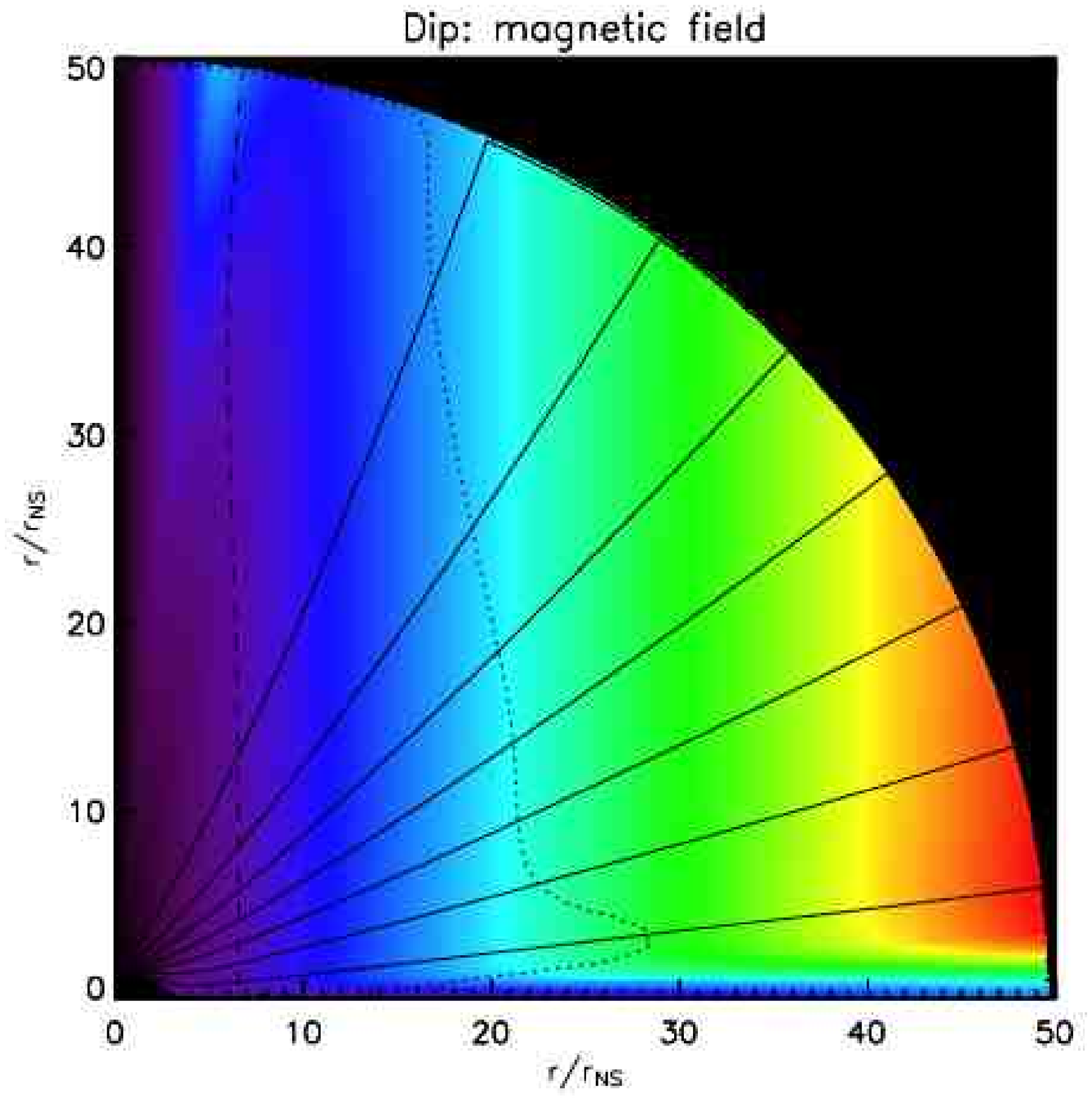}}}
\end{picture}
\hspace{5\unitlength}
\begin{picture}(275,200)(0,15)
\put(-20,70){\makebox(275,190){\includegraphics[width=1.75in]{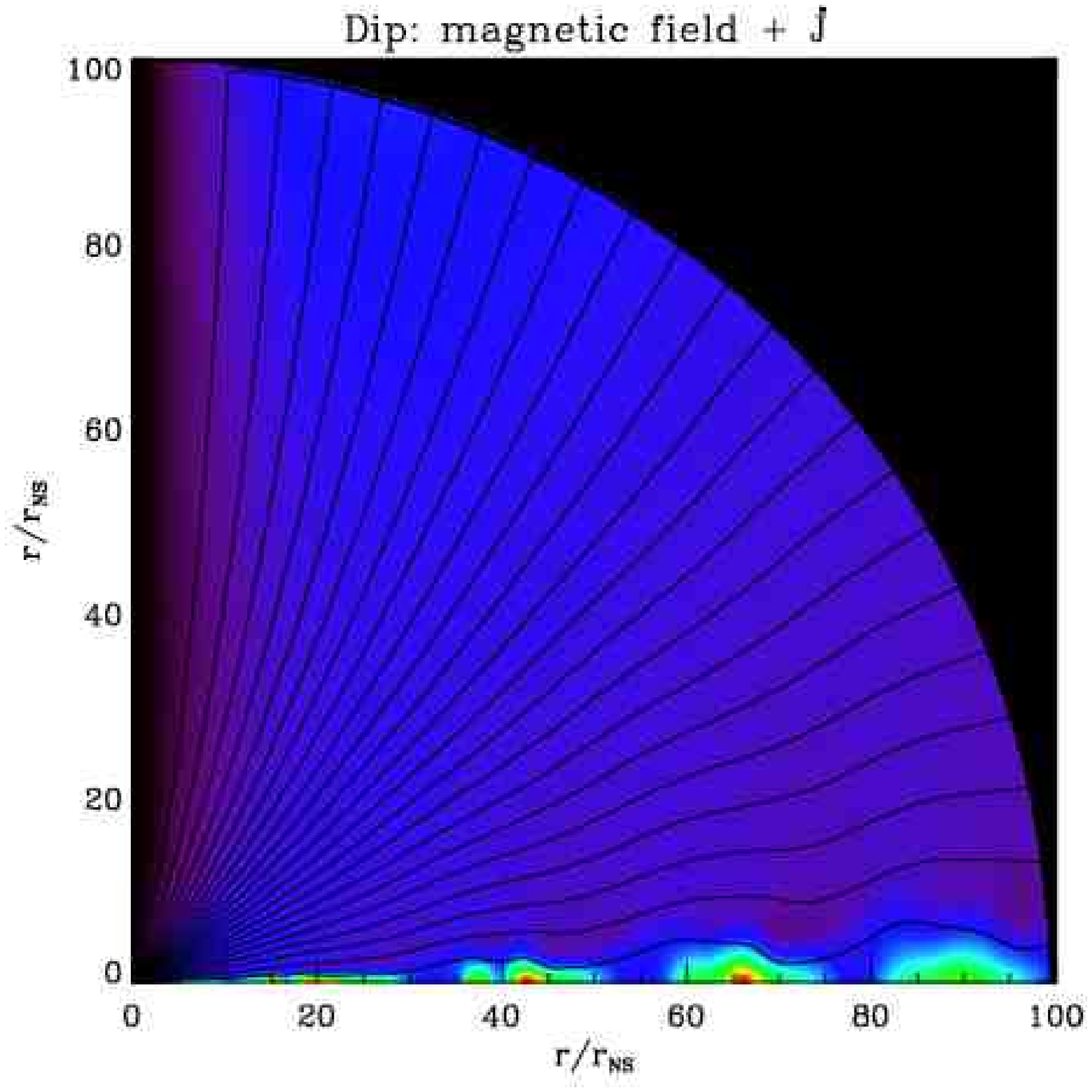}}}
\end{picture}
\hspace{3\unitlength}
\begin{picture}(275,200)(0,15)
\put(10,70){\makebox(275,190){\includegraphics[width=1.675in]{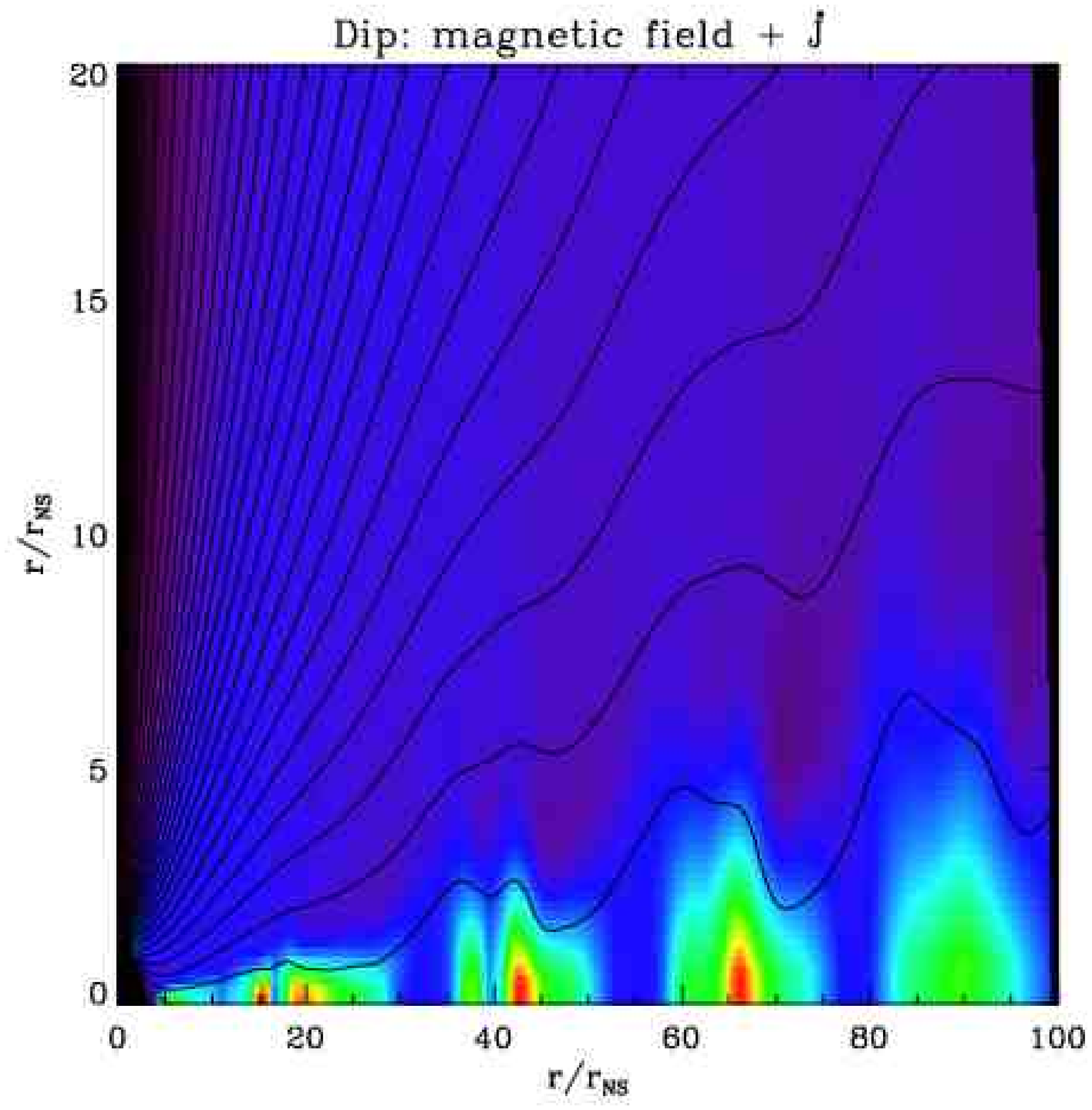}}}
\end{picture}
\hspace{1\unitlength}
\end{center}
\vspace{0.9cm}
\caption{{\bf Left}: Magnetic structure of a relativistic magnetosphere with mass outflow, in a case when the magnetic pressure at the light cylinder exceeds the relativistic plasma inertia by a factor approaching 20, a record high for MHD calculations. The contours represent poloidal magnetic field lines, while colors represent the ratio $B_{\phi}/B_r$.  Reconnection have been artificially suppressed. {\bf Middle}: Snapshot of the magnetic structure when the reconnection components of the electric field are not suppressed. Reconnection occurs because of numerical resistivity introduced by the finite difference scheme.  Plasmoids emerge along the equatorial current sheet, grow and flow out at the local Alfven speed ($\sim c$).  {\bf Right}: Blow up of the plasmoid structure.  \label{fig:magnetosphere}}
\end{figure}
\noindent and beyond $R_Y$ communicates back to the inner magnetopshere and the star}, through (kinetic) Alfven waves traveling back along field lines at and near the separatrix (see \S \ref{sec:pairs}).

But if this picture does apply
to RPPs, it has a number of consequences for observables and outstanding questions, some of which I touch on further in \S \ref{sec:pairs}.  From the point of view of the basic energetics embodied in spindown, the
fluctuating Poynting flux may imply a fluctuating torque.  Noise in pulsar spindown has been known since the early days - it limits the ability to time pulsars coherently.  If
the magnetic field interior to but near $R_Y$ fluctuates by tens of per cent on time scales comparable to the  rotation period, and these fluctuations represent variations in the poloidal current that communicates stress to the star, then the torque is noisy with magnitude the same as is inferred from representing the observed random walks in the rotation frequency 
(\citealt{helfand80, cordes80a, cordes80b}) as being the consequence of white noise in the electromagnetic torque (\citealt{aro81b} -
reconnection may provide the mechanism for magnetopsheric instability and torque fluctuations that was not specified in this early attempt at scenario building). 

Magnetospheric noise opens the possibility that $R_Y /R_L$ evolves.  One can readily show (\citealt{bucc06}) that the braking index, defined as
$n = \bar{ \Omega } \ddot{\bar{ \Omega}}  / 
\dot{ \bar{ \Omega} }^2$ (with averages indicating the usual average over subpulses taken
in measuring pulsars' periods, here taken to be the same as an average over 
plasmoid emission and torque fluctuations), in a magnetosphere with evolving $R_Y /R_L$ but fixed $\mu$ and $i$, is
\begin{equation}
n = \frac{ \bar{ \Omega} \ddot{\bar{\Omega } } }
 {\dot{ \bar {\Omega} }^2 }
   = 3 + 2 \frac{\partial \ln \left(1 + \frac{R_Y}{R_L} \right)}
       {\partial \ln  \bar{\Omega}}.
\end{equation}
Braking indices less than 3 (\citealt{liv05} and references therein) thus may indicate a progressive lag of the closed zone's expansion, measured by the 
radius of the $Y$ line, behind the expansion of the light cylinder as a pulsar spins 
down. 

This is hardly the only thinkable explanation of $n < 3$.  Magnetic
moment evolution has long been advocated as the origin of small braking indices,
going back to the crustal field growth model of \cite{blandford83} - which doesn't
actually work in those authors' formulation, the threshold for growth set by crustal resistivity is too high - to Ruderman's model for growth of 
$\mu_\perp = \mu \sin i \propto \sqrt{P}$
due to interaction of interior magnetic flux tubes with the quantized vorticity of the
superfluid interior (e.g., \citealt{ruderman06}). When combined with expression (\ref{eq:spindown}),
this model yields
\begin{equation}
n = \frac{3  + 4 \frac{\Omega_0}{\Omega} \tan^2 i_0}
      {1 + 2 \frac{\Omega_0}{ \Omega} \tan^2 i_0},
\label{eq:Bbrake}
\end{equation}
where $\Omega_0, \; i_0$ are the angular velocity and obliquity at the time when the
arrays of quantized magnetic flux tubes and vortex tubes have both formed, thought
to be perhaps $10^3$ years after the neutron star's birth.  This model can produce
any braking index between 2 and 3.

\begin{figure}
\begin{center}
\hspace{10\unitlength}
\begin{picture}(300,200)(0,15)
\put(-35,-15){\makebox(200,200)[tl]{\includegraphics[width=4.5in]{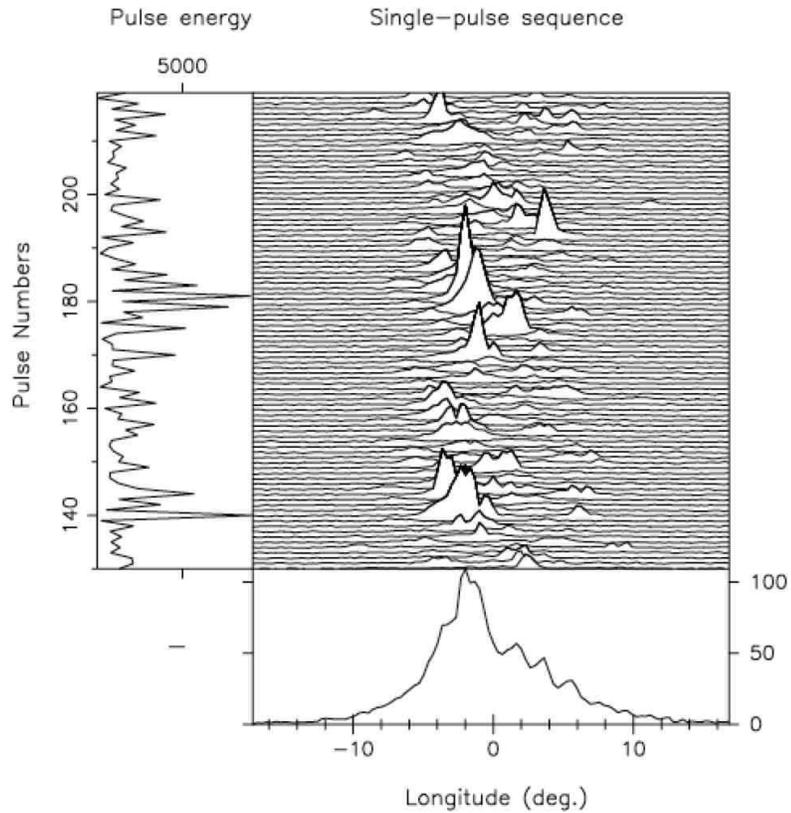}}}
\end{picture}
\end{center}
\vspace{6cm}
\caption{A series of individual pulses from PSR B0943+10 (center). The average pulse is
shown in the bottom panel, as a function of pulse number and longitude, with
$360^\circ$ or longitude corresponsing to one rotation of the star. This star shows
organized drifting of the pulses through the pulse window. From \cite{desh99}. \label{fig:subpulses}}
\end{figure} 
Fluctuations and oscillations  in the corotating frame of the currents at the light cylinder also offer a possible explanation for the long known fact that pulsars are {\it flickering}  lighthouses. The well-known pulse stability that allows exquisitely precise timing applies to average pulses, formed by summing hundreds to thousands of individual pulses. However, individual pulses arrive with at varying times within the pulse
window, usually at random in those pulsars with ``core'' emission characteristics, and either at random or with an organized drift of the arrival times through the pulse window,
in stars with ``conal'' emission characteristics (see \citealt{rankin83} for pulsar beam
classification).  Figure \ref{fig:subpulses} shows an example of pulse to pulse variability. Typically one or at most two pulse components are within the pulse window at any one time, suggesting the individual pulse variability time is on the order of the rotation period. That time scale is consonant with the Alfven wave transit time from the low altitude emission region to the radius of the Y-line, where current fluctuations are formed, a coincidence suggesting the subpulse variability (both random and drifting) is a consequence of current variations created by activity in the outer magnetosphere
(\citealt{aro81b}). In this picture, drifting and chaotic subpulses are both the consequence of the same dissipative dynamics of the currents coupling the magnetosphere to the wind, with drifting subpulses reflecting limit cycle behavior of magnetospheric reconnection while chaotic subpulses represent a more random, bursty behavior of the field lines topological changes. Objects such as PSR B0943+10, in which
transitions from organized drifting to chaotic single pulse behavior and back are observed (\citealt{rankin06}), are particularly telling laboratories for investigation of the connection between current flow and pulsar emission, and thus offer insight into magnetospheric dynamics.

\subsection{Electrospheres?}

There is, however, a ghost hiding inside the force-free/MHD magnetospheric machine. These theories assume that a plasma dense enough to enforce $E_\parallel = 0$
is present everywhere in
the magnetosphere, an assumption which relies upon the  pair creation physics 
summarized in \S \ref{sec:pairs}. Pair creation assumes relativistic beams contribute
a substantial fraction - possibly all - of the electric currents embodied in the force-free and MHD models, since only these beams can (plausibly) emit the gamma rays that
convert to $e^\pm$. One can think of pair creation as an instability of the current flow originally hypothesized in the fully charge separated scenario of \cite{gold69}. But, as 
was recognized not long after the charge separated outflow scenario was suggested,  the charge separated wind model must fail, so long as charged particle flow across  field lines is forbidden  - many field lines of a dipole (not a monopole) must
pass through a surface where ${\boldsymbol \Omega} \cdot {\boldsymbol B} = 0$. The charge (and plasma) density of the charge separated medium on the exterior (larger) radius side of this ``null surface'' has sign opposite to that of the plasma that can be supplied from the stellar surface by particle motion parallel to ${\boldsymbol B}$. The plasma in this exterior region
has no source, if the only allowable charged particle motions are sliding along the magnetic field plus bulk flow ${\boldsymbol E} \times {\boldsymbol B}$ drift (\citealt{holo73, holo75}). Thus, one expects such a  magnetosphere to open large gaps, and probably have
no charged particle wind - certainly no wind with particle flux greatly in excess of
$\dot{N}_R = c \Phi /e$. 

\begin{figure}[H]
\begin{center}
\unitlength = 0.0011\textwidth
\hspace{10\unitlength}
\begin{picture}(300,200)(0,15)
\hspace*{1\unitlength} \put(-365,-45){\makebox(200,200)[tl]{\includegraphics[width=2.4in]{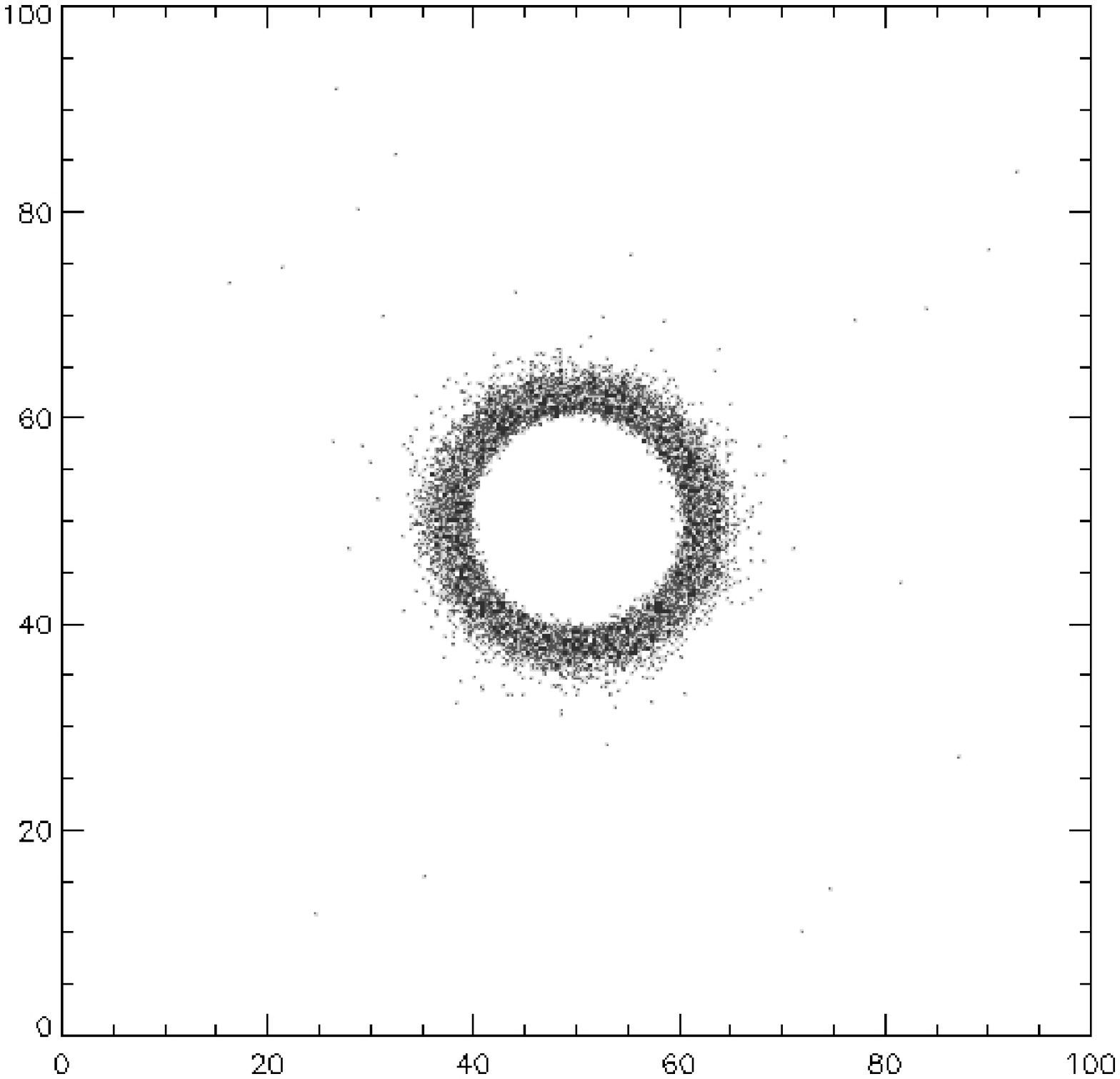}}}
\hspace{10\unitlength}
 \put(135,-45){\makebox(200,200)[tl]{\includegraphics[width=2.4in]{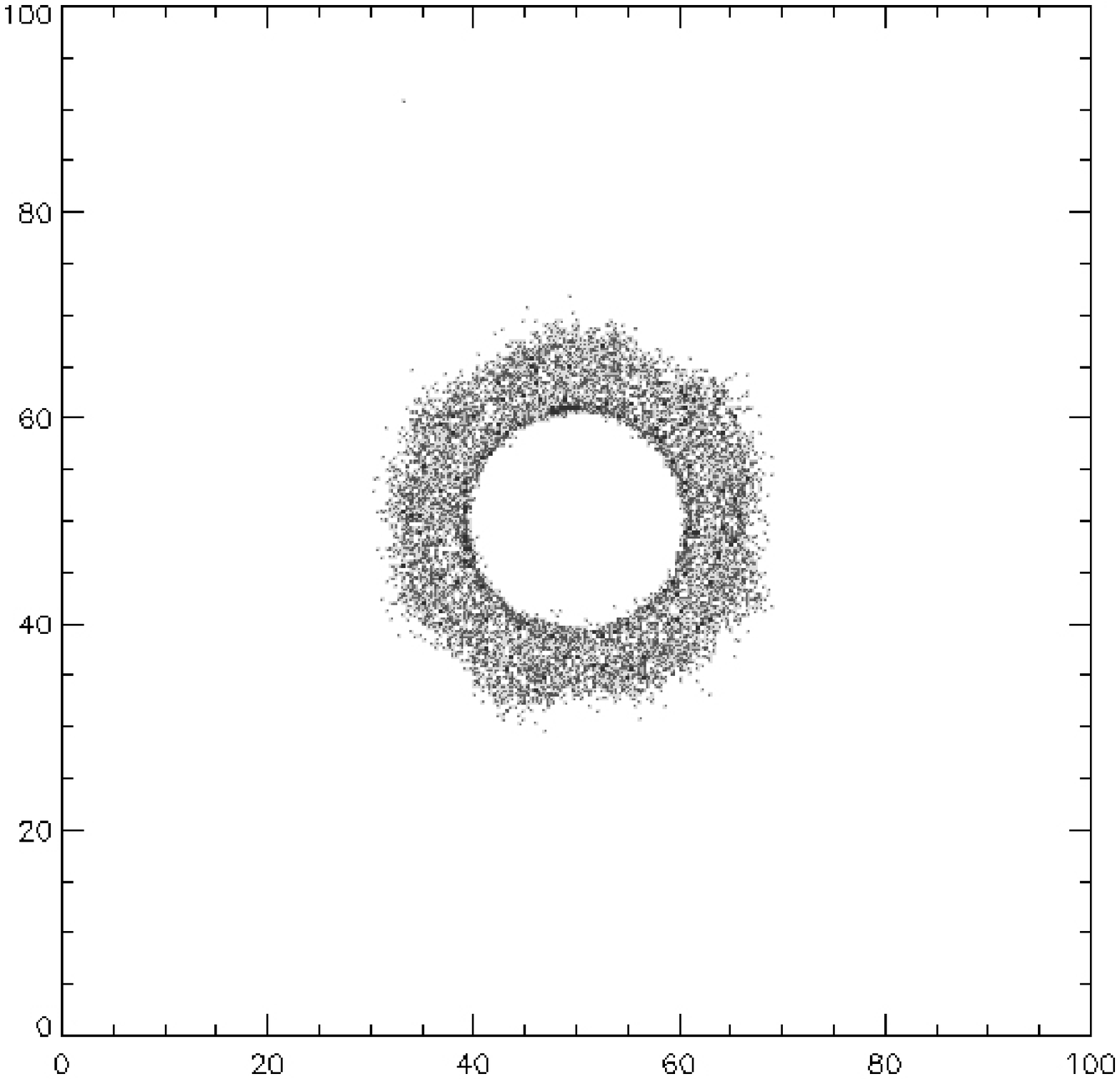}}}
\vspace*{0.2cm}
\hspace*{1\unitlength} \put(-365,-555){\makebox(200,200)[tl]{\includegraphics[width=2.4in]{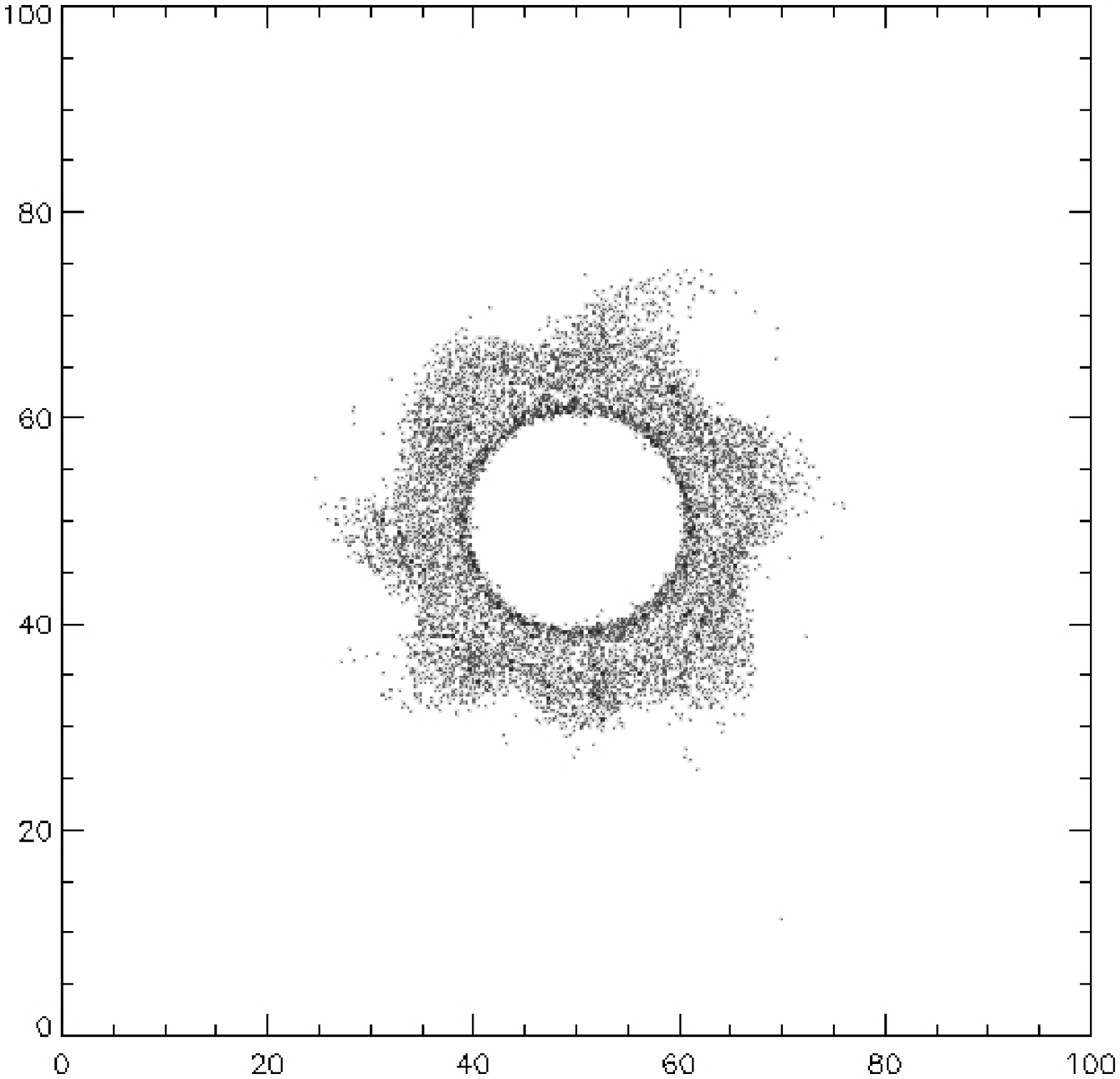}}}
\hspace{10\unitlength}
 \put(125,-555){\makebox(200,200)[tl]{\includegraphics[width=2.4in]{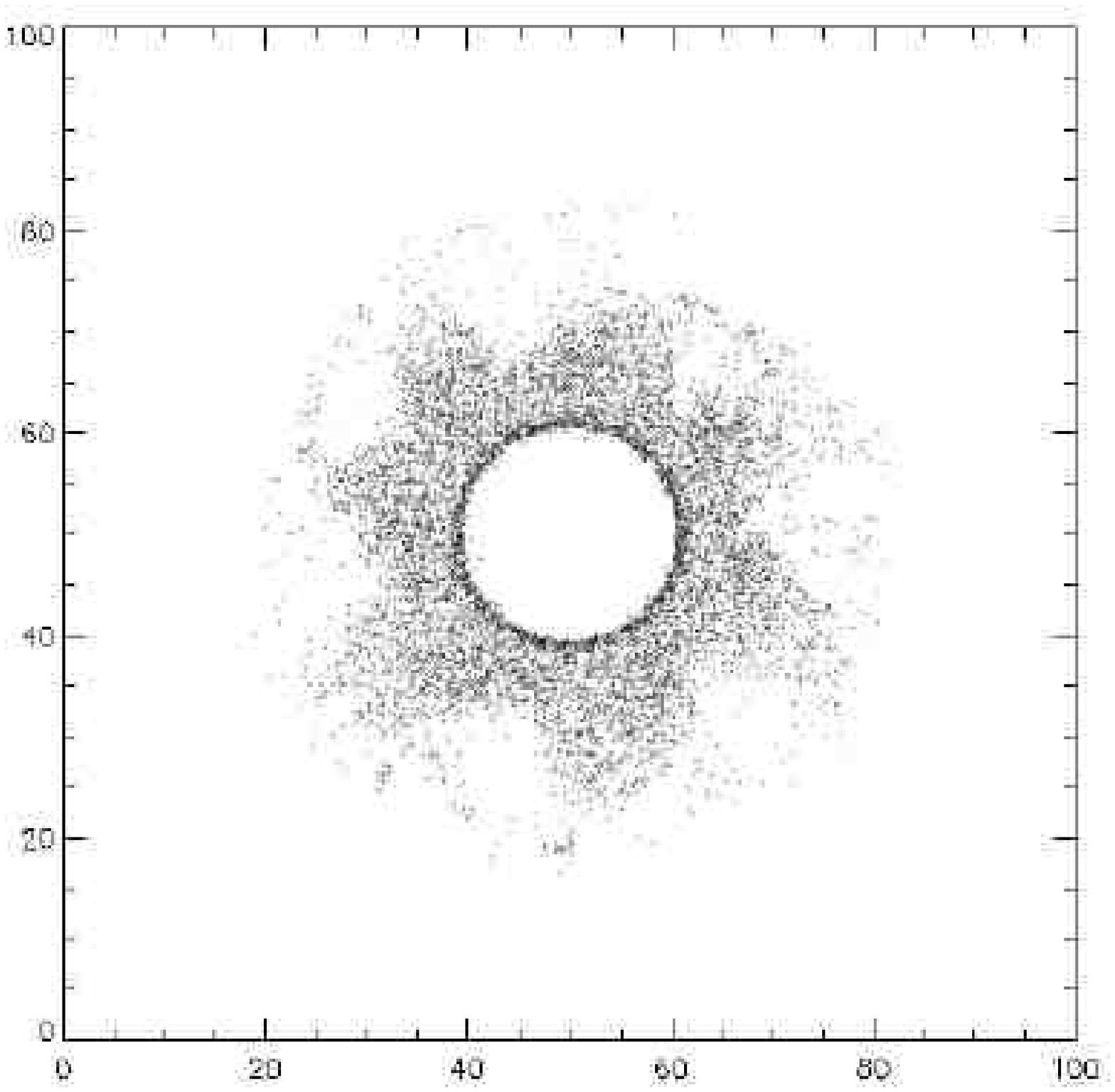}}}
\end{picture}
\end{center}
\vspace{10.5cm}
\caption{Spreading of the equatorial charged disk in the electrosphere of an aligned
rotator under the influence of the diocotron instability, from a PIC simulation of the flow,
from \cite{spit02}.  The figure shows a series of snapshots of the disk's density in the
rotational equator of the neutron star, which fills the central circle - the spatial scale is in units of computational cells, with 10 cells equaling one neutron star radius.  The time sequence goes from left to right, with the lower row following the upper row.
The simulation
begins with the small disk of the equilibrium electrosphere. At later times the disk spreads and develops nonaxisymmetric rolls and fingers, characteristic of Kelvin-Helmholtz instabilities, to which diocotron instability is closely related. \label{fig:diocotron}}
\end{figure}

Such ``electrospheres''
(\citealt{krause85}) do {\it not} appear to be unstable to pair creation (\citealt{ petri02a}), thus do not collapse to the plasma
filled state hypothesized in the force free models, by pair creation alone. However,
they are unstable to
cross field transport even without pair creation.  A large gap
separates the equatorial regions of the electrosphere from the stellar surface, leading to differential rotation of the equatorial plasma.  This differential rotation is
subject to the diocotron instability - a variation of a Kelvin-Helmholtz instability
(\citealt{spit02, petri02b}). Simulations (\citealt{spit02, petri03, petri06}) suggest the resulting time dependent (in the corotating frame) 
${\boldsymbol E} \times {\boldsymbol B}$ drifts create cross-field ``diffusion'' which
may relax the charge separated magnetosphere to something approximating the
state envisaged by Goldreich and Julian. Expansion of the equatorial plasma is illustrated in Figure \ref{fig:diocotron}.

Bottoms-up models based on these results have not been investigated.
The fact that young pulsars supply their nebulae with particle
fluxes greatly in excess of $\dot{N}_R$ lends support to the perhaps more practical
view that the filled magnetosphere model has consequences in reasonable accord with observations of
high energy pulsed emission and of pulsar wind nebulae, thus deserves the main
focus of scientific attention.  Such models may have charge separated current flows possibly
unstable to pair creation. 

The force free solutions and their possible extensions have a number of implications for 
 emission models, and for the
pair creation models that underlie the emission physics.

\subsection{Magnetic Geometry of Radiating Layers}
\hspace*{0in}
The oblique rotator solution determines a polar cap/polar flux tube size and shape.
These are noncircular, and have centroid displaced from the magnetic axis (Spitkovsky,
personal communication). Such changes in magnetic geometry from the conventional assumptions need to be folded into radio beaming and polarization, and polar cap X-ray emission models, which often
invoke {\it ad hoc} changes in polar cap size and shape, perhaps created by surface 
anomalies in the surface magnetic field ({\it e.g.}, \citealt{melik06}), in order to explain departures from the simplest, static dipole geometry. The need for such extra parameters in the models needs to be evaluated in the context of realistic magnetic
geometry of the rotating dipole.

\hspace*{0in}
The surface of last closed field lines (the separatrix) and of 
the return current flowing along that separatrix has been determined within the
force-free approximation.  Particle acceleration in gaps (regions of low density where a parallel electric field 
$E_\parallel = {\boldsymbol E} \cdot {\boldsymbol B} / B$
forms because of charge starvation below the
Goldreich-Julian density) on the open field lines close to this surface has been advanced as the origin of the pulsed gamma rays observed from a small number of pulsars (\citealt{mus04, hiro06} and references therein) by the EGRET experiment (\citealt{thomp00}) and by other high energy detectors, with a substantial increase in the population observed expected with the launch of the GLAST telescope (\citealt{mclaugh00}).
\vspace*{-1.4cm}

\vspace*{-1.3cm}
\begin{figure}[H]
\hspace{10\unitlength}
\begin{picture}(100,100)(0,15)
\put(-10,-15){\makebox(100,100)[tl]{\includegraphics[width=2in]{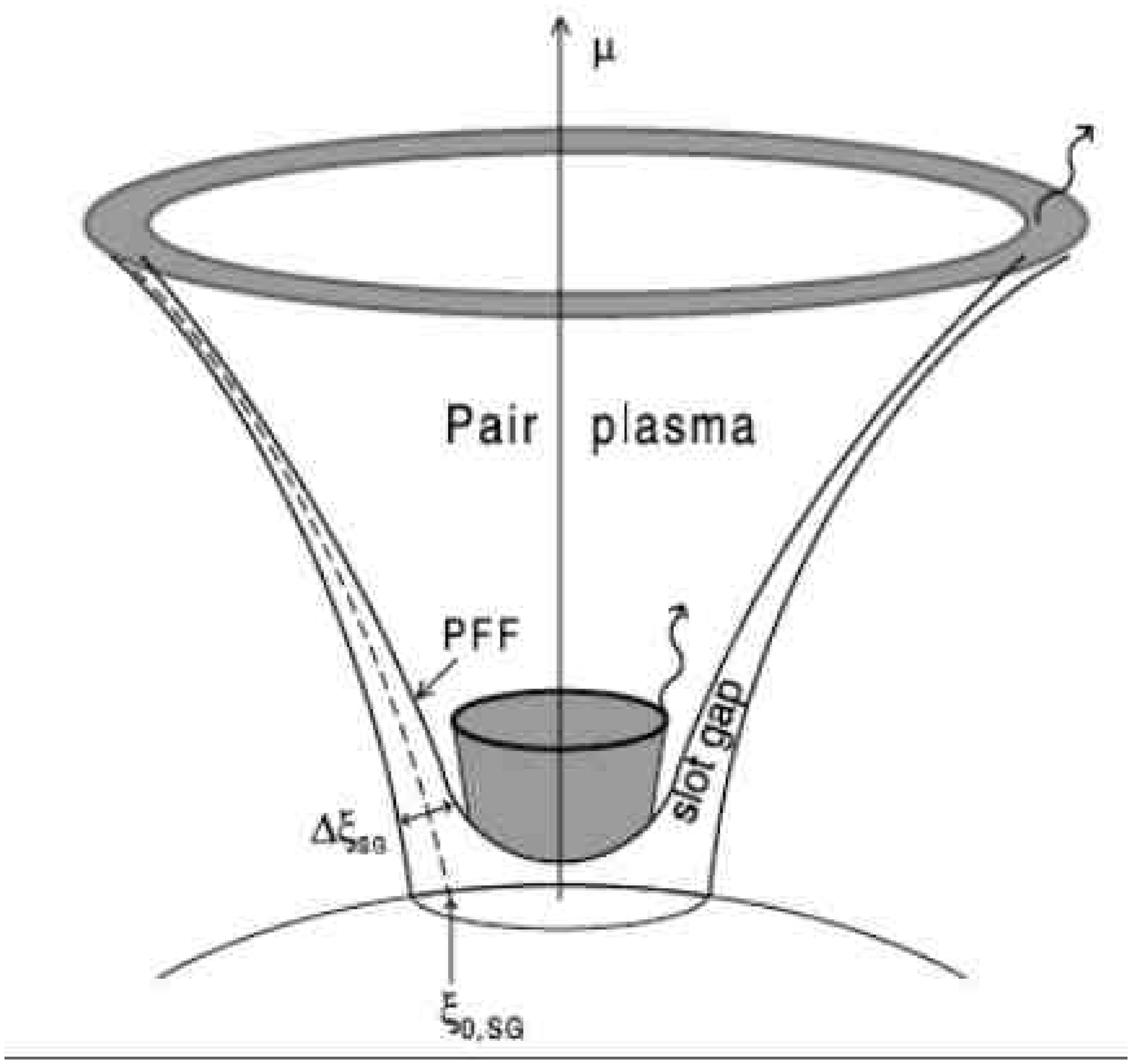}}}
\end{picture}
\hspace{10\unitlength}
\begin{picture}(100,100)(0,15)
\put(35,-65){\makebox(135,135)[tl]{\includegraphics[width=2.5in]{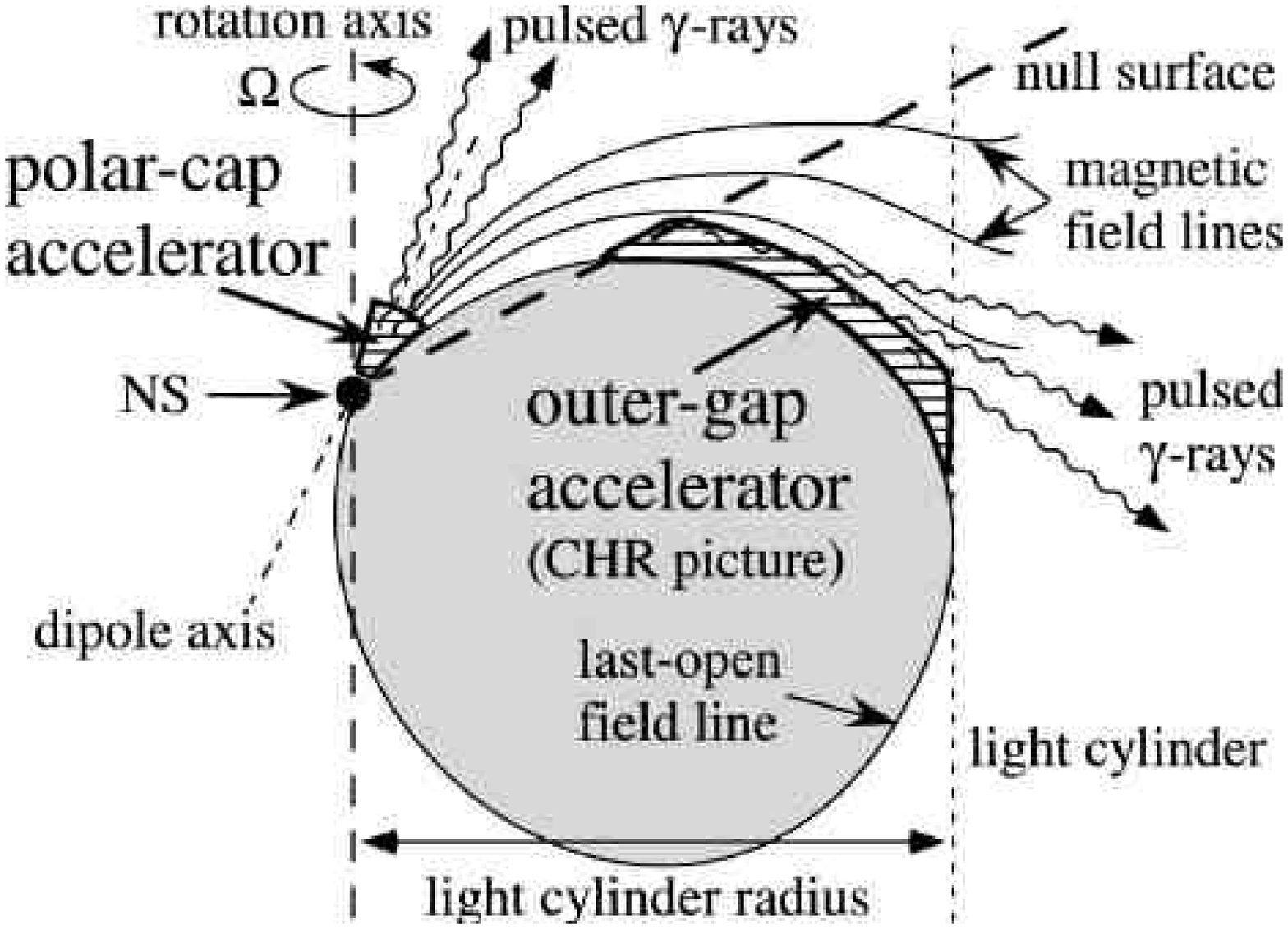}}}
\end{picture}
\vspace{3.3cm}
\caption{Location of proposed charge starvation gaps in the magnetopshere, employed
in models for pulsed high energy emission. Left:  slot gap model of \cite{aro79, aro83b, mus03, mus04}; figure from \cite{mus03}. Right:
classical outer gap geometry of \cite{cheng86, chiang94, romani95, romani96}; figure from \cite{hiro06} \label{fig:gapgeom}}
\end{figure} 
These models use
magnetic geometry borrowed from the vacuum oblique rotator, with plausible
but {\it ad hoc} prescriptions for the shape of the separatrix and for the
choice of field lines assumed to participate in the gap.  Strictly steady flow (and therefore electrostatic accelerating electric fields in the corotating frame) are assumed, with a variety of higher order drift effects included on the high energy particles' orbits, in order to capture the rather highly refined beaming dynamics at large radius, including  pulse structure controlled by caustic formation through delicate cancellations of aberration and light travel time with field line sweep back along a last closed flux surface of assumed form (\citealt{dyks03}). Manipulation of a
variety of free parameters (especially the thickness of the assumed accelerating
layer) allows fits of the resulting radiation spectra and pulse
profiles to observations  with greater or lesser success.  Since such geometric constructions are sensitive to the exact form of the geometry guessed, a fruitful exercise would be to place such models, or alternates such as the gapless model involving the return current itself outlined below, in the context of the force-free solutions, where the location and shape of the separatrix is not a free function, or is a function only of $R_Y / R_L$. At the very least,
such model construction would be a step toward probing the basic structure
of the magnetosphere, a task made possible since all the phenomenological gap models contain parallel voltage drops 
$\Delta \Phi_\parallel = \int E_\parallel ds$ 
small compared to $\Phi$, therefore allowing the force-free theory to be a good
zeroth order platform for parallel accelerator and radiation model construction.

\subsection{Current Flow Profile and Gap Electrodynamics}

The force free electrodynamic solutions also exhibit an important result which affects {\it all} the gap models constructed for the last 30 years, starting with the vacuum surface and outer gaps of \cite{rud75, cheng86} and their many 
successors, as well as the space charge limited flow beam models, both with and without slot gaps, of \cite{aro79, aro81a, aro83b} and the many successors of this modeling idea. 

All these schemes embody the idea that 
$E_\parallel$ appears as a result of the magnetosphere's attempt to restore charge neutrality in the co-rotating frame by accelerating non-neutral beams of particles with density comparable to the Goldreich-Julian density $\eta_R/q$, with that adjustment to perfect charge neutrality in the corotating frame being incomplete - an $E_\parallel$
due to charge starvation.  In
such configuations, the field aligned current 
$J_\parallel = {\boldsymbol J} \cdot {\boldsymbol B}/B $ adjusts to a value controlled by the local electrostatics of the hypothesized region of charge starvation, not to a value
determined by the energetically dominant magnetospheric dynamics.  

For currents emerging from the star's atmosphere,  the accelerating $E_\parallel$ appears from the gravitational depletion of density below the Goldreich-Julian value, due to the low temperature and low radiation pressure which prevents the filling of the magnetosphere and formation of a wind with a charge neutral plasma pushed up by pressure and centrifugal forces from the stellar surface. The current supplies the charge needed to support a polarization electric field which (almost) cancels the vacuum field. The residual (``starvation'') electric field still supports enough of a voltage drop to allow the accelerated particles to emit gamma rays that can convert to electrons and positrons. The resulting gap structure, elaborated assuming strictly steady flow in the co-rotating frame, thus enforces a current density almost uniform with distance from the magnetic axis, with value close to the canonical value $c\eta_R$.  Since this current fills (almost) the whole polar cap,  the total current from a polar cap is 
$I \approx c\eta_R \pi r_{cap}^2 \approx c  (\Omega B_{cap} \cos i /2\pi c) \pi R_*^3 /R_Y
 = c \Phi$, which suggested that such a gap {\it might} be an element of the magnetospheric circuit, although with the peculiar property that the charge density (and
 therefore the current density) is an eigenfunction of the local electrostatics. The return 
 current is not included in these local acceleration models, being explicitly or implicitly 
 assigned to the current sheet.
 
 Such local 
 determination of $J_\parallel$ is not
 what one expects on energetic grounds, since the current density reflects the induced
 magnetic field, through which all the spindown energy loss flows. That energy flow is 
 much larger than the energy flowing through 
 the proposed electrostatic accelerator, which is thought of as a small perturbation of the 
 force-free structure. In the absence of further information
 from full magnetospheric solutions, or specific features coming from phenomenological 
 models of radio or high frequency pulse obsevations which characterize the
 current flow in more detail, the hope expresssed by the approximate correspondence
 between the total gap current and the current of the magnetsopheric circuit has stood
 unchallenged\footnote{That one might be able to use observations to probe the
 current flow structure has been an almost untouched subject.  One of the few counterexamples was provided by \cite{hibsch01a}, who pointed out that thin return current layers might create observable signatures in the radio polarization data. Their  predicted signature of the return current layer may have been seen in PSR J1022+1001 
 (\citealt{ramach03}).}. 

\begin{figure}[H]
\hspace{10\unitlength}
\begin{picture}(100,100)(0,15)
\put(-10,-15){\makebox(100,100)[tl]{\includegraphics[width=2in]{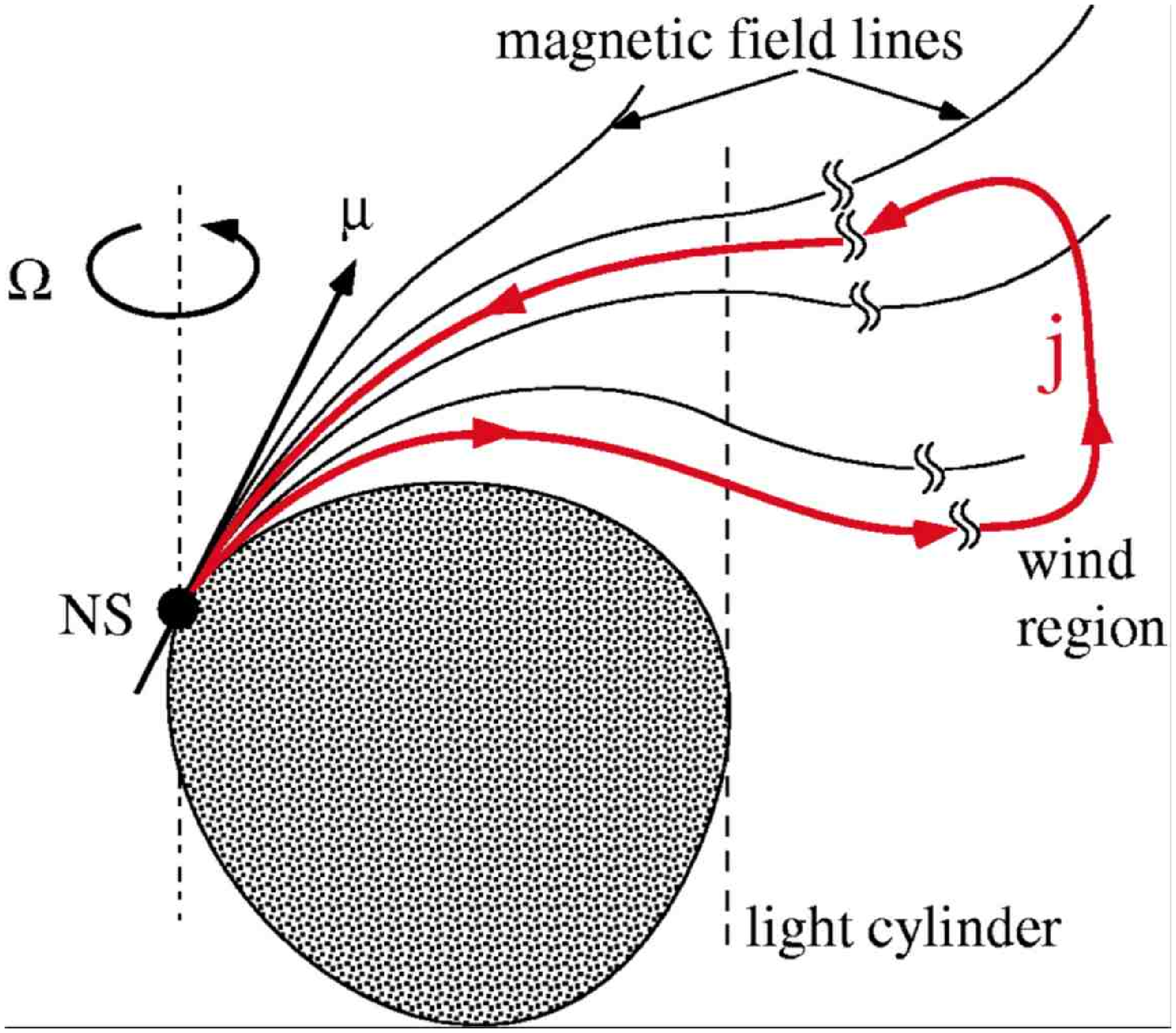}}}
\end{picture}
\hspace{10\unitlength}
\begin{picture}(100,100)(0,15)
\put(55,-5){\makebox(100,100)[tl]{\includegraphics[width=1.8in]{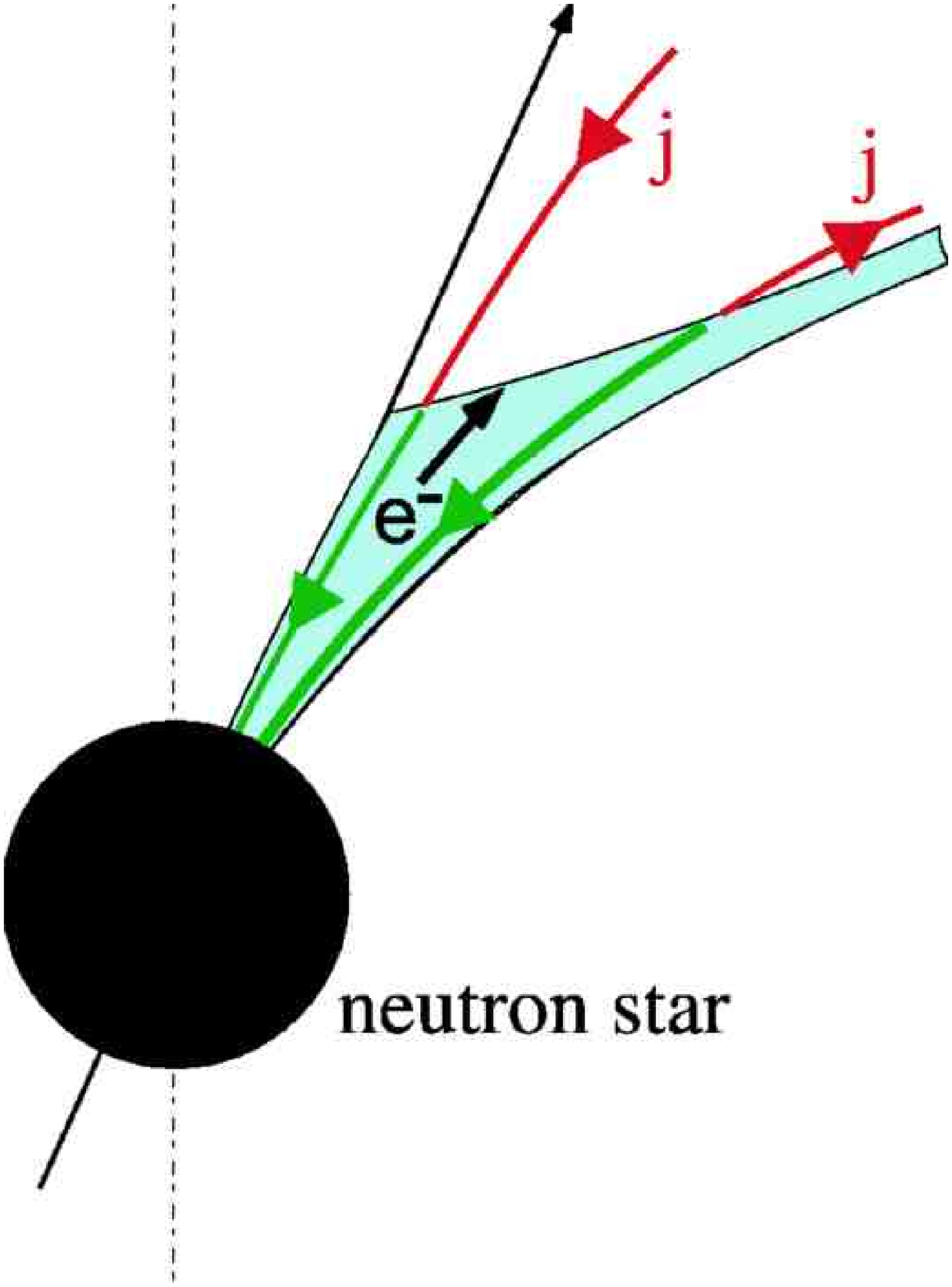}}}
\end{picture}
\vspace{3.3cm}
\caption{Current flow patterns hypothesized for modern outer gap models 
(\citealt{hiro06}), which include possible outward ion flow on field lines near the
separtrix (in $i < 90^\circ$ geometry). 
Left: global pattern assuming a closed circuit, with currents closing in the
wind. Right: hypothetical current flow from the star through the null surface. No attempt is made to account for the return current in the current sheet in these models, nor has there been any correspondence made between these hypothetical current flows and the currents  in the force free solutions. \label{fig:ogcurrent}}
\end{figure}

For outer gaps, whose morphology appears in Figure \ref{fig:gapgeom},  the current density is also established by the effect of pair plasma shorting out the starvation $E_\parallel$, since within the gap counterstreaming electron and positron beams coexist, with acceleration ceasing at the end points where the pair density rises to be approximately equal to the Goldreich-Julian density - or not at all, along field lines closest to the (assumed) conducting boundary formed by the last closed field line, where pair creation is weak, for reasons traceable to the assumed geometry.  Figure \ref{fig:ogcurrent} shows the place outer gaps might have in a hypothetical picture of the global circuit.
The starvation electric field can be sustained only if the $e^\pm$ beams have density not exceeding $c \eta_R /e$, therefore the current density necessarily approximates the Goldreich-Julian current {\it density}.  Since the gap must be thin in the poloidal direction across $B$ (otherwise the photon emission from the gap would not make a narrow pulse, the express purpose for which the gap model was designed), the total gap current has to be small compared to the magnetospheric current - outer gaps cannot close the whole magnetopsheric circuit, if they are to have any pretensions as a successful model for pulsed gamma rays. Their biggest success, which they share with the slot gap model, is the assignment of the radiating geometry for gamma ray pulsars
to a thin sheet which closely follows the last closed flux surface of the oblique rotator.

Outer gaps, as regions of field aligned acceleration relying upon starvation electric fields, can occur only on field lines not supplied with a dense plasma either from the polar caps, as comes from the polar and slot gap space charge limited acceleration region, or from the recirculation of polar outflow by reconnection flows in the vicinity of the Y-line.  Also,
the outer gap, if it exists, sends almost all of its pair plasma back to the stellar surface,
rather than supplying the wind (\citealt{hiro06}).  Thus the outer gap has a hard time being a major supplier of the known large ($\dot{N}_\pm \gg \dot{N}_R $) particle
fluxes known to be injected into the young Pulsar Wind Nebulae, as discussed further
in \S \ref{sec:pairs}.  Outer gaps also run the risk of supplying too much energy in precipitating particles to the stellar surface, thus powering too much thermal emission.
Outer gap modelers have mostly swept these issues under the rug.

\vspace*{-1.3cm}

\subsection{Gap Subversion: Non-Uniform Current Profiles 
\label{sec:gapsubversion}}
All gap models (vacuum polar cap gaps, space charge limited flow polar cap and slot gaps,
outer gaps) function as suppliers of plasma which come as close as possible to restoring the
charge density to the Goldreich-Julian density $\eta_R$.  They rely on starvation
electric fields, since the gaps' charge densities fall below $\eta_R$ (by a little or a lot, depending on the model). Since the voltage drops developed in the gaps are highly relativistic, such plasma takes the form of relativistic particle beams, with the resulting current density parallel to $B$ being 
$J_\parallel = c \eta_R (1-h)$, where $h \ll 1$ for polar cap and slot gaps (the
stellar surface supplies a beam of charge density almost equal to $\eta_R$, thus
trapped particle backflow formed at the PFF is small), while in traditional outer gaps, $h \approx -1$,  (the pair formation front at each end has to trap
plasma with the full Goldreich-Julian density, forming two counterstreaming beams of
approximately equal density). This current is almost constant, as a function of distance across the magnetic flux surfaces (formally, $J_\parallel (\psi) \approx$ constant).  
All the models assume strictly steady current and plasma flow in the co-rotating frame (electrostatic approximation), even though the models' authors frequently indulge in discussion of time dependence that they think should be part of their
proposals. Steady current flow {\it has} been found in evolutionary force-free and relativistic MHD models (\citealt{spit06, komiss06}) - these show no signs of variability in the corotating frame (they also cannot 

 \vspace*{-1.3cm}
\begin{figure}[H]
\hspace{10\unitlength}
\begin{picture}(100,100)(0,15)
\put(-40,-35){\makebox(100,110)[tl]{\includegraphics[width=2.15in]{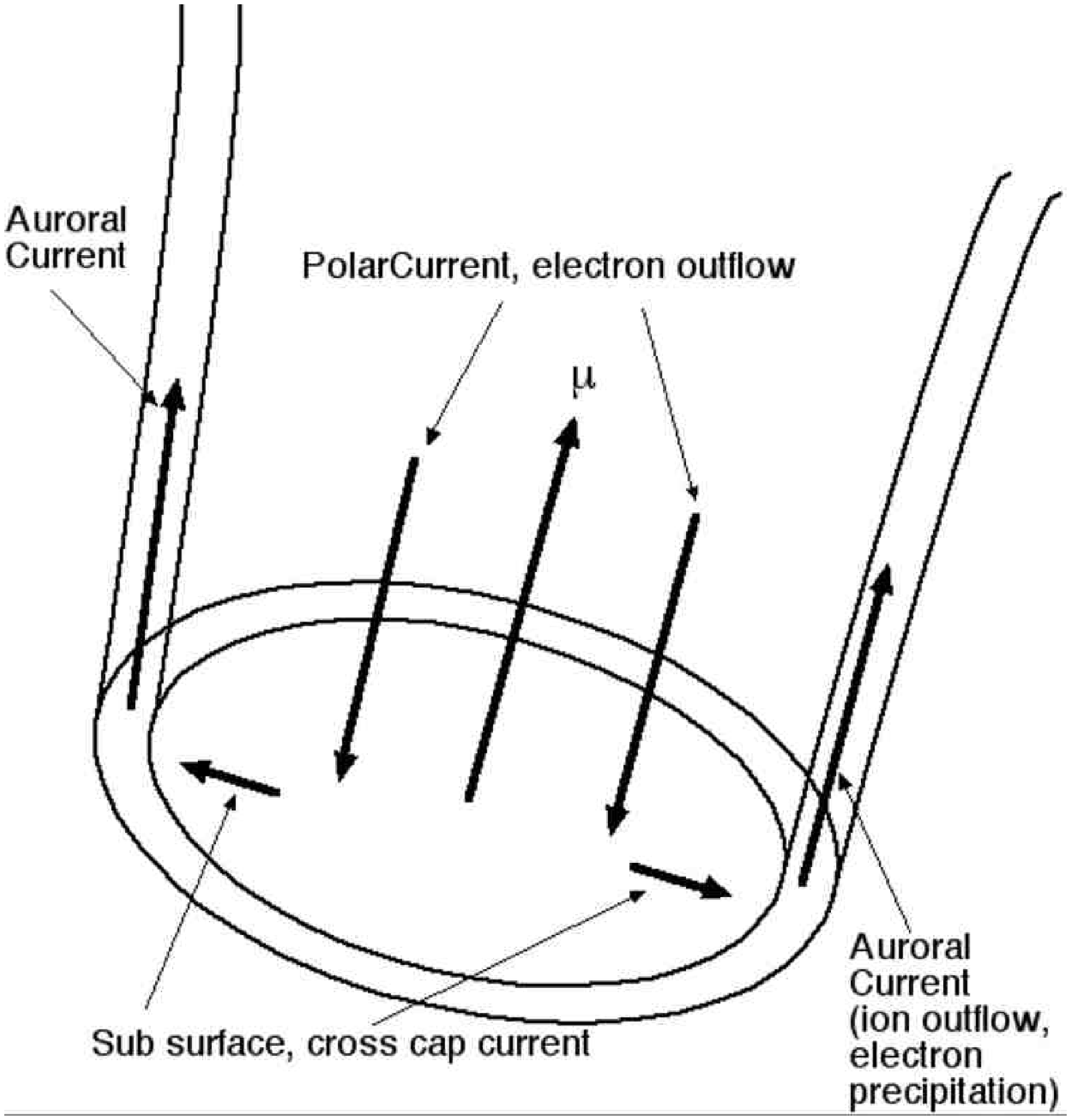}}}
\end{picture}
\hspace{10\unitlength}
\begin{picture}(100,100)(0,15)
\put(-80,-15){\makebox(110,100)[tl]{\includegraphics[width=2.65in]{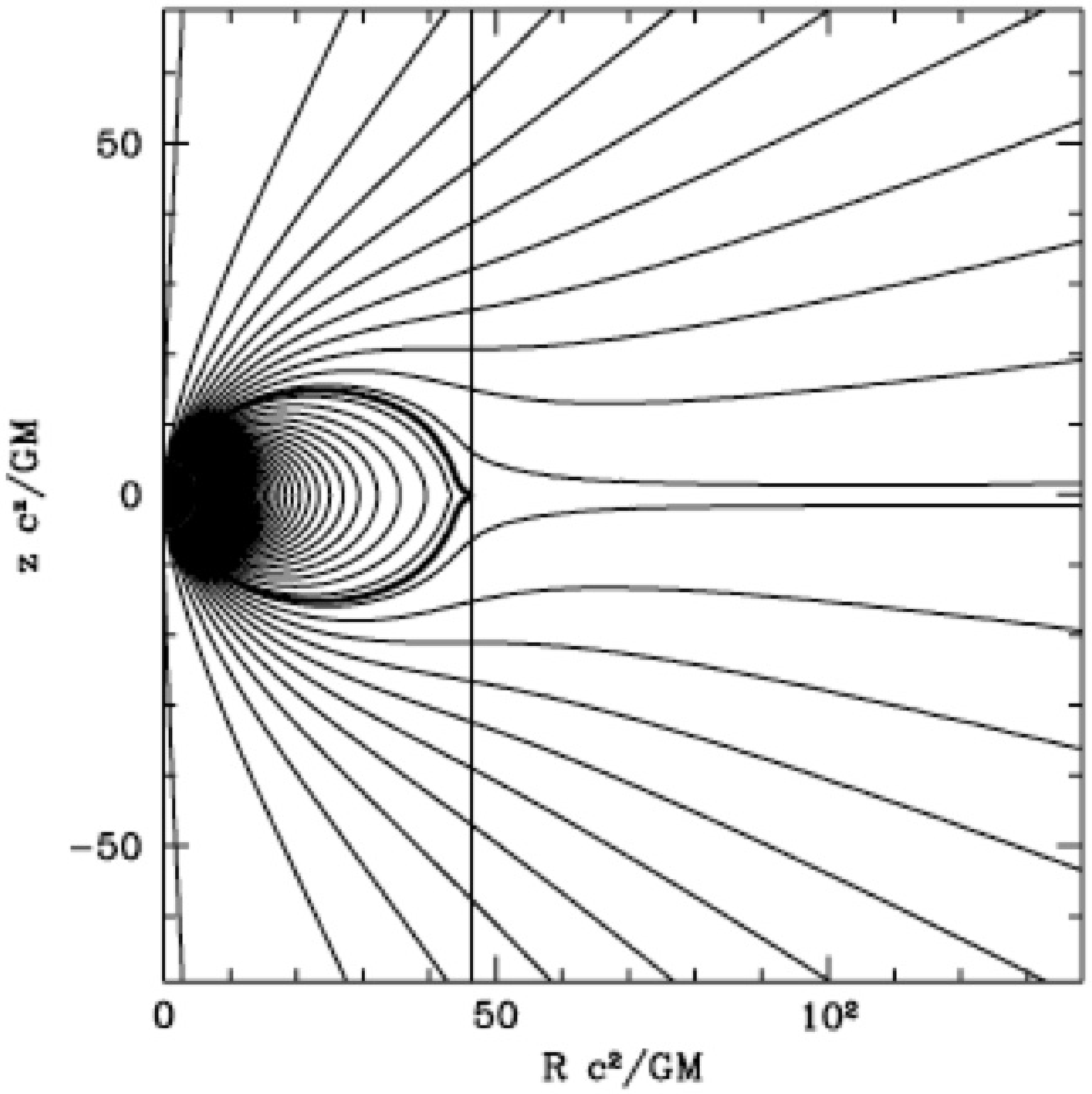}}}
\end{picture}
\hspace{10\unitlength}
\begin{picture}(100,100)(0,15)
\put(20,-110){\makebox(100,100)[tl]{\includegraphics[width=3.9in]{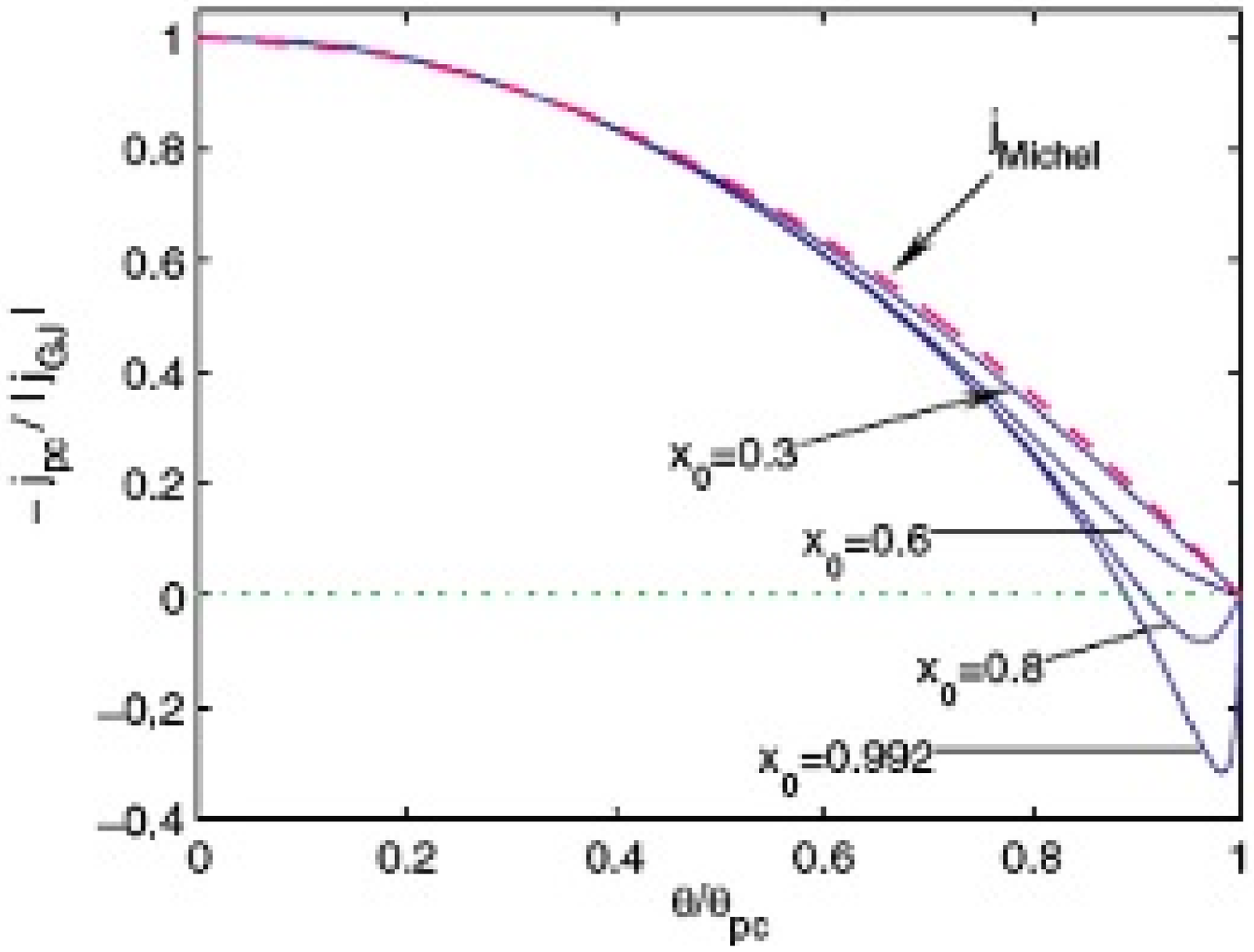}}}
\end{picture}
\vspace{8cm}
\caption{Polar current flow patterns hypothesized and found from force free models. Top Left: Polar current flow with all return current in the current sheet, and illustrating
balancing the charge loss with an ion current extracted from the surface, in response to electrons and plasma precipitating from the Y-line (\citealt{aro81b, aro83a}) Top Right: Magnetic field lines of the aligned rotator when $R_Y/R_L = 1.0$, from \cite{mckinn06}, showing the asymptotically monopolar character of the poloidal field structure. Bottom: Current flow in a force free solution of the aligned rotator, for various values of
$x_0 = R_Y/R_L$, from \cite{timokhin06}. $\theta$ is the magnetic colatitude of a field
line's footpoint, and $\theta_{pc}$ is the magnetic colatitude of the polar cap's edge,
defined as the intersection of the separatrix between closed and open field lines with the stellar surface. \label{fig:polarcurrent}}
\end{figure} 

\noindent capture reconnection physics, which probably requires higher resolution simulations than have been employed so far, and probably also
incorporating more physical models of dissipation and inertia. But, even though 
global and local theories both have stationary lighthouse behavior built in, the consequences of gap electrostatics for the current flow distribution are entirely {\it inconsistent} with the results of force free modeling, because of a serious mismatch between the current profiles found in the gap and in the global models.

Figure \ref{fig:polarcurrent} shows the poloidal current density, as a function of magnetic flux, exhibiting the fact that a fraction ($\sim 20$\%) of the return current flows on open field lines just within the boundary of the closed zone if $R_Y = R_L$; the rest of the return current lies within the unresolved current sheet separating the closed from the open field lines. Thus, as far as current flow is concerned, 30 years of research on the force free rotator can be summarized by saying that to within 20\% accuracy, the current flow distribution
of the force free dipole is that of the monopole, whose total current appears in
(\ref{eq:mon-current}) - as far as the open field lines are concerned, the dipolar 
magnetosphere is the monopole mapped onto a polar cap, in each hemisphere. The current density as a function of distance from the magnetic axis then is (ignoring the 
small piece of the return current required on open field lines)
\begin{equation}
J_\parallel = \frac{dI}{d\psi} = j_{GJ} \left(1 - \frac{\psi}{\psi_{cap}} \right)
        = j_{GJ} \left(1 - \frac{\varpi_*^2}{\varpi_{pc}^2} \right),
\label{eq:polcapJ}
\end{equation}
where $j_{GJ} =  c\eta_R (\varpi_* = 0 ) = \Omega B_{cap} \cos i /2\pi$ 
and $\varpi_*$ is the cylindrical distance of a  field line's footpoint from the dipole axis.
If the polar current is a charge separated, steadily\footnote{Steady on times long compared to the polar cap transit timescale $\varpi_{cap}/c \sim 10 \mu$ sec} flowing beam extracted from the star's atmosphere
by $E_\parallel$ with charge density $\eta = J_\parallel /c \beta$, with $\beta \approx 1$
except in a thin region at the surface of thickness  10 or so times the atmospheric scale height, then the difference of this beam charge density from the Goldreich-Julian 
density is of order $\eta_R$ itself over most of the polar flux tube:
\begin{equation}
\frac{J_\parallel}{c} - \eta_R = \frac{j_{GJ}}{c} \left(1 - \frac{\psi}{\psi_{cap}} \right) -
   \frac{j_{GJ}}{c}  = + \frac{| j_{GJ }| }{c} \frac{\varpi^2}{\varpi_{cap}^2}
    \rightarrow \frac{| j_{GJ} |}{c} , \varpi \rightarrow \varpi_{cap}.
\label{eq:currentgap}
\end{equation}
Expression (\ref{eq:currentgap}) means that the parallel electric field is not almost shorted out by the space charge density of the particle beam from the surface,
thus returning the space charge limited flow to an environment with an electric field
akin to what was envisaged for vacuum gaps (if the plasma forming the return current on 
the polar flux tube's boundary behaves as a perfect conductor, as has been assumed in all models to date), but now with the requirement (since the stars have dense, thermal
X-ray emitting atmospheres) that $E_\parallel (r= R_*) = 0$. 

The resulting huge acceleration inevitably leads to massive pair creation, in the manner
of the \cite{sturrock71} picture of a high energy beam coexisting with 
massive pair creation
and acceleration with almost all of $\Phi$ being dropped by the electric field within
a height about equal to the polar cap width.  Such a situation has both interesting possibilities and large problems, both theoretical and observational.

Pairs forming in an approximately vacuum electric field will short out
$E_\parallel$ at a height such that the voltage drop $\Delta \Phi_\parallel $ is sufficient to allow the accelerating particles of the beam from the surface to emit magnetically convertible gamma rays. Typically, $\Delta \Phi_\parallel  \sim 10^{12} \; {\rm V} \ll \Phi$.
Because the difference charge density (\ref{eq:currentgap}) is a large fraction of
$-\eta_R$ over most of the open flux tube, the pairs have to supply most of the charge density needed to shut down $E_\parallel$ at the pair formation front (PFF).  Because 
the pair formation front is now formed at lower altitude than is the case in space charge limited flow models with the current density determined by the local electrostatics, pair 
cascades may be better able to supply the total plasma flux inferred from pulsar wind
nebulae, as discussed in \S \ref{sec:pairs}.  However, the pairs generally are born
with energies small compared to $e \Delta \Phi$. Therefore, particles with charges having the same sign as the particles of the beam extracted from the stellar atmosphere are added
to the outbound beam, while particles with the opposite sign of charge are trapped
electrically and go backwards - backflow from the PFF enhances the current.  If the potential is monotonic, the particle backflow collides with the stellar surface with number flux  $\approx (c |\eta_R|/e) (\varpi /\varpi_{cap})^3$ and energy/particle $\sim e \Delta \Phi$. 
The particles in the backflow lose their energy after penetrating 
several hundred gm/cm$^2$ into the crust, heating the atmosphere from below. Then 
each pole would have thermal X-ray emission with luminosity/pole 
$L_{Xpole} \approx 0.5 \dot{E}_R (\Delta \Phi_\parallel /\Phi)$.  With $\Delta \Phi_\parallel \sim 10^{12}$ V (the voltage defining the theoretical death line in Figure \ref{fig:PPdot}),
these heated polar caps would emit substantially more thermal X-rays than are observed
in many RPPs.  Similar backflow in ``spark'' models runs into the same difficulty.

The traditional space charge limited polar caps, in which the current carrying beam extracted already has density close to $|\eta_R|/e$, greatly reduce
this emission - in modern models, in which dragging of inertial frames controls the
difference between the beam's density and $|\eta_R /e|$, the reduction is by a factor 
$\sim 0.4 GM/R_* c^2 \approx 0.06$. This reduction is enough to give polar cap X-ray emission either in accord with observations of some pulsars, or small enough to be hidden by magnetospheric non-thermal emission.  But the price paid is an electric current
(formed by response to backflow from the PFF as well as by emission from the atmosphere) over the whole polar cap (stationary or non-stationary) which is large compared to ($\ref{eq:polcapJ}$) except near the magnetic axis. Furthermore, just as in the early spark gap models of \cite{rud75},
the difference charge density in (\ref{eq:currentgap}) implies a nonzero $E_\parallel$
incident on the star's surface - really, the top layers of the atmosphere 

A number of ideas have been advanced to resolve this conundrum. 
\newcounter{gaposc}
\begin{list}
{$\bullet$}{\usecounter{gaposc}}
\item The
PFF has a different structure than has been found in studies starting with
\cite{aro79} all the way through recent work on the full slot gap (\citealt{mus03, mus04}).
If somehow electrons could be made to precipitate from pair plasma above the PFF and enter the current flow region, the current would be reduced (\citealt{timokhin06}).  However,
this is quite hard to achieve, the electric field below the PFF acts to expel such particles
- they could not enter unless they already had energy 
$\sim e\Delta \Phi_\parallel$. For curvature gamma ray emission generated pair cascades, the pairs are born with energies far below $e \Delta \Phi_\parallel$,
which make formation of a PFF that reduces $J_\parallel$ impossible. 

\item The PFF and the acceleration region below it is fully time dependent, with
the current flickering on the polar cap transit time $\sim \varpi_{cap}/c \sim 10 \mu$sec, in the manner of the \cite{sturrock71} and \cite{rud75} diode instability and spark scenarios, respectively.   Given  stars with atmospheres which make charges freely available to the magnetospheres above, the most significant gap question is, how does the
star adjust to provide the magnetospherically rerquired current density, while at the same time adjusting the charge density to reduce $E_\parallel$ below its vacuum value in a manner that does not do violence to thermal X-ray emission (for example).

\item All the models of relativistic field aligned acceleration considered to date have
assumed unidirectional flow, or at most counterstreaming beams with no trapped particles in the acceleration zone - that is, monotonic accelerating potentials and 
electric fields. The possible presence of trapped particles, implicit in the non-monotonic, non-relativistic acceleration model of a singly charged fluid of \cite{mestel85}, can break the straight jacket of current being proportional to charge density, even if the plasma is completely non-neutral. Furthermore, the outer magnetosphere can modify the local polar cap (and outer gap, if it exists) electric field through return current control of the currents and charge densities in the polar flux tube boundary layer separating the open from the closed field lines, which opens the possibility of relaxing the assumption of perfect conductivity on the boundaries of the acceleration region used in all models to date.    All
models to date have assumed the open field lines are bounded by a surface whose
behavior mimics a perfect conductor, which makes a very specific statement
about the surface charge density along the separatrix.  As with the total current and current density, there is no reason for the surface charge to adopt this locally determined
value - it depends not only on the charges contained within the open flux tube, but also
on the dynamics controlling the formation of the return current, both at the stellar surface and at the magnetopsheric Y line.  I discuss these speculative points further below. 

\end{list}

The magnetosphere is a high inductance system - the magnetic
fields induced by particle and displacement current flow  can change on times no shorter than the Alfven transit time between surface and Y line and back, $T_A \approx 2R_Y/v_A \approx (2/\Omega) (R_Y/R_L)$, very much greater
than the polar cap transit time, or the plasma period based on the Goldreich-Julian density 
$\omega_{pGJ} = \sqrt{4\pi e^2 |\eta_R|/m_\pm e} = \sqrt{2 \Omega \Omega_{c\pm}}
= \Omega \sqrt{2 \Omega_{c\pm}/\Omega} \gg \Omega$ [here $\Omega_{c\pm} = eB/m_\pm c$ is the nonrelativistic cyclotron frequency $\sim 10^{19} (B/10^{12} {\rm G})$ 
s$^{-1}$ of an electron or positron].  Thus one possibility is that the atmosphere
supplies the current demanded through pair creation discharges, which launch charge bunches accompanied by pulses of much denser pair plasma, a popular idea frequently mentioned in the cartoon approximation but rarely studied.

Homogeneous and spatially 1D models of time dependent pair creation have been
studied by \cite{alber75}, \cite{fawley78} and \cite{levinson05} - the first considers only
time dependent oscillations in a uniform medium (a ``0D'' model), the second, using a 1D Particle-in-Cell
method, considers the propagating transitions between vacuum $E_\parallel$ and
an $E_\parallel \approx 0$ region created by a burst of pair creation (motivated by the
\citealt{rud75} spark scenario), while the third
considers nonlinear uniform oscillations similar to those studied by \cite{alber75} and
also spatially inhomogeneous nonlinear limit cycle oscillations. \cite{levinson05} incorporated
the existence of the magnetospheric current as a fixed constant in the model, as is
appropriate since the oscillations occur on the time scale $\omega_{pGJ}^{-1}$, much shorter than all possible time scales of magnetospheric variability. They point to the
interesting possibility that the charge and current oscillations might become chaotic,
a topic of substantial interest to possible radio emission mechanisms, but present
no specific calculations that exhibit such behavior. 

None of these explicitly time dependent models showed approach to the steady flow
in the corotating frame assumed at the start in the models of \cite{aro79} and \cite{mus04}, in which the current density is fixed by the local electrostatics. Significantly, none of these local, 0D or 1D time dependent models included the effects of pair outflow from the system, nor included the Poynting fluxes (in effect, the collective radiation losses) from the regions of pair oscillation.  Considered as an instability, the oscillations studied are probably sensitive to the loss of plasma and Poynting flux from the system,
since the pair creation (in 1D) has the character of a spatial amplifier.  Thus the question of whether pair oscillations approach a steady state with the local value of the magnetospheric current density emerging as a steady flow or as an average over the oscillations remains open, requiring calculations which are able to give a more complete account of the coupling to the stellar surface and the losses from the region of oscillatory dynamics. Giving an account of the effects of plasma flow onto the star and the consequent heating and X-ray emission is of substantial importance, since the Levinson {\it et al.} model suggests a local current density $c\eta_R$ accelerated through $\sim 10^{13}$ V colliding with the surface, which leads to observable  thermal X-ray emission which may, or may not, be in excess of what is observed from many stars.

On the opposite side, \cite{aro83b} and 
\cite{mus97} point to how their steady flow, spatially inhomogenous models (``gap-PFF'') might become unstable, due to
inhomogeneity of the pair creation, in older stars where the pair creation gain lengths are large.  Such instability would be of the spatial, traveling wave amplifier variety, a possibility also of interest to the outer gap models, whose local time dependence has also begun to be studied. \cite{hsu06} have opened the first door to time dependent (in the co-rotating
frame) outer gap models, showing that their models, which do incorporate particle outflow from the accelerator region - mostly toward the star - converge to a steady state flow. Strictly speaking, however, these models are inconsistent, allowing for full time dependence
of the current and plasma densities but treating the electric field as electrostatic, which is
quantitatively incorrect in a relativistic, multi-dimensional (2D, in their model) system.
The approach to a steady state is attributed to the screening of $E_\parallel$ if too many
pairs materialize, while underscreening results in an increrase of $E_\parallel$. This mechanism is the same as underlies the nonlinear limit cycle oscillations appearing
in the polar cap pair oscillation model of Levinson {\it et al.}. Thus the approach to a steady state in the time dependent outer gap model more likely owes its origin to the
spatial loss of plasma from the acceleration region, an effect broadly akin to transit time 
damping of electromagnetic oscillations in a plasma. 

Given the large, almost vacuum conditions above much of the polar cap implied by 
(\ref{eq:currentgap}), coupling to the stellar atmosphere almost certainly requires consideration  of trapped particles' contribution to the charge density, in either steady flow or time dependent current flow conditions.

A full theoretical resolution of pair 
creation driven oscillations in {\it any} ``gap'' geometry awaits more definitive study, along 
with any sort of serious attempt to relate such oscillations to observable phenomena - 
current and torque fluctuations, radio microstructure, variability in thermal X-rays created by surface bombardment, etc.  All such modeling needs to be set into the
context of the global force-free models - so long as the potential drop in a local accelerator $\Delta \Phi_\parallel$ is small compared to the magnetospheric potential
$\Phi$, field aligned accelerators of any sort (employing starvation $E_\parallel$ in
all the schemes available in the literature) appear as small departures from force-free
conditions, allowing the use of force-free models as the basic zeroth order description
of the current flow and magnetic geometry.

\subsection{Gamma Ray Tests of Existing Gap Models \label{sec:gaptests}}

The modern force-free magnetsophere models open the possibility of using the upcoming gamma ray observatory GLAST (\citealt{gehrels99, carson06}) 
to test and improve our understanding of 
pulsars' magnetospheres, along the lines suggested above or in other directions.
Gamma rays afford the possibility of probing the magnetosphere using well understood
radiation processes, leaving the modeling and the synergy between models and
observations living in the domain of the geometry and the acceleration physics - 
``gapology'', in the existing theoretical frameworks.
In  particular, the advent of the force free-models should allow the outer gap and slot gap modelers to significantly reduce the geometric uncertainties in their constructions
of the beaming profiles and energy dependent light curves,
thus allowing much more stringent empirical tests of the hypothesized geometric
scenarios - if the necessary extensions of the models to 3D, and perhaps to time dependence, are incorporated.  Even more important, the improved sensitivity
of GLAST over past gamma ray telescopes will allow, for the first time, a direct test of whether polar caps and polar cap pair creation occur in a significant population of pulsars. Given that no proposed outer gap
(or slot gap, for that matter) model makes a significant contribution to pair creation and 
gamma ray emission for periods much in excess of $\sim 200$ msec ($\Phi < 10^{15}$
Volts), the much heralded association of pair creation with pulsar photon emission
and, more significantly, with relativistic wind formation must come from activity in the
polar cap region just above the surface. Indirect evidence for such pair creation
comes from the simple $\Phi = 10^{12}$ V radio pulsar death line shown in Figure
\ref{fig:PPdot}, which corresponds roughly to where polar cap/slot gap acceleration models predict pair creation to cease (\citealt{sturrock71, rud75, aro79}).  Previous gamma ray telescopes lacked the
sensitivity to probe the predicted gamma ray emission, which, in the models, is absorbed
at energies above 1 GeV (in simple, star centered static dipole geometry) through 
gamma ray conversion to $e^\pm$ pairs.  

\vspace*{-0.65cm}
\begin{figure}[H]
\begin{center}
\unitlength = 0.0011\textwidth
\hspace{10\unitlength}
\begin{picture}(200,200)(0,15)
\put(-200,95){\makebox(100,100)[tl]{\includegraphics[width=3in]{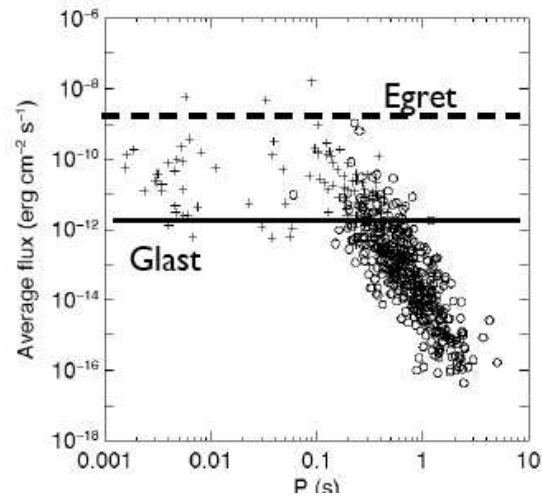}}}
\end{picture}
\end{center}
\vspace{4.3cm}
\caption{Sensitivity of the EGRET and GLAST telescopes to low altitude (below and just above the PFF)
gamma ray emission at energies $\varepsilon > 100$ MeV, in a simple dipole model for the low altitude geometry (which affects the magnetic curvature and therefore the maximum photon energy that can escape) and unidirectional space charge limited flow
{\it with current density $J_\parallel \simeq c\eta_R$}, not the force-free current given
(approximately) by (\ref{eq:polcapJ}). The different symbols refer to the different major contributors to gamma ray emission and absorption - see \cite{hibsch02} for the details. \label{fig:glast}}
\end{figure}     

Figure \ref{fig:glast}, taken from \cite{hibsch02}, gives a simple version of this opportunity, in good accord with more recent
evaluations of pulsar gamma ray emission from the inner magnetosphere. 
Testing the existing polar gap acceleration and gamma ray emission  models, or better, improved models that take proper account of the magnetospheric current system,
can be done best by studying pulsars that show {\it core} component radio emission,
since one looks down the ``barrel of the gun'' into the low altitude magnetic field,
where the core emission arises (see \citealt{kramer98} for the evidence that the core emission arises from a few kilometers above the surface in a substantially dipolar $B$
field).  

I can safely predict that GLAST observations of pulsars deeper in the
$P-\dot{P}$ plane than could be detected by previous gamma ray telescopes,
which should begin to become available in 2008-09, will 
stimulate a small host of improved gap models which take advantage of the geometric and current flow constraints coming from the force-free models.  They might stimulate investigations into origins of $E_\parallel$ based on physics differing from the starvation models that have been explored for the last 30 years - for example, invoking the 
$E_\parallel$ accompanying the kinetic Alfven waves that couple the outer magnetosphere to the star,
a conceivable acceleration mechanism that might augment or even replace outer gap
and slot gap accelerators, especially if plasma precipitating from reconnection flows at
the Y-line floods the regions envisaged for these gaps and poison their starvation electric fields. 

\begin{figure}[H]
\begin{center}
\unitlength = 0.0011\textwidth
\hspace{10\unitlength}
\begin{picture}(170,170)(0,15)
\put(-200,60){\makebox(100,100)[tl]{\includegraphics[width=2.75in]{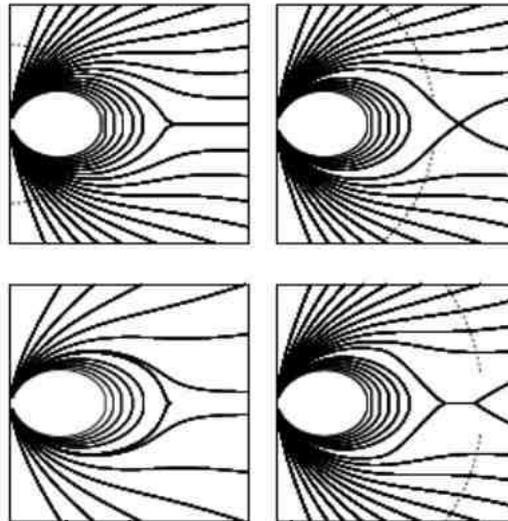}}}
\end{picture}
\end{center}
\vspace{4.5cm}
\caption{A cartoon of a pulsar magnetosphere undergoing sporadic reconnection,
from \cite{contop05}.\label{fig:rec}}
\end{figure}

To amplify this issue, which is a prospect for future research, consider the hypothetical global reconnection flow illustrated in Figure \ref{fig:rec}, shown here for an aligned rotator but just as applicable to the Y-line of the oblique rotator illustrated in Figure \ref{fig:oblique}.
\vspace*{-1cm}
As appears in the \cite{bucc06} model, pairs supplied from the polar gap supplying the
wind should allow reconnection of the current sheet to occur all the time. It has not yet been shown that the reconnection propagates back into the magnetosphere in the
manner envisaged in Figure \ref{fig:rec}. In the somewhat analogous problem of
reconnection at the Y line in the rapidly rotating Jovian mangnetosphere,
\cite{yin00} showed, using a PIC simulation of the electron-ion plasma at the
Y-line, that reconnection qualitatively akin to that illustrated in Figure 
\ref{fig:magnetosphere} occurs, with finite but not large reaction along the separatrix
interior to the sporadically forming X-line. That back reaction includes generating
precipitating $J_\parallel$  {\it and} plasma with density well in excess $|J_\parallel|/ec$ 
on and around the separatrix.  The field aligned currents are part of the kinetic Alfven
waves that couple the time variable Y-line to the inner magnetosphere and the star,
thus generating a time variable torque. The space charge in these boundary layer flows
can alter, in a major way, the electric field within the polar cap accelerator (as well as
poison both outer and slot gaps), while the parallel electric fields in the kinetic Alfven waves offer a new mechanism for field aligned acceleration in the boundary layer geometry already known, from the outer and slot gap models, to be extremely useful in understanding gamma ray pulsars' beaming morphology. If these waves have a chaotic time series, the chaos in the resulting polar $J_\parallel$ offers a good candidate for
understanding the random arrival times of radio subpulses within a pulse window; if the
reconnection induced waves have a limit cycle time series, the phenomenon of subpulse
drifting can be reproduced, assuming the radio emission intensity and beaming is a
direct product of the field aligned current density.  The electric fields in these
Alfven waves have a central role in the formation and extraction of the return current
required to maintain the average charge balance of the star. Finally, the shear between
the plasma flow in the polar flux tube and the neighboring boundary layer offers a
promising candidate for the collective radiation mechanism in ``conal'' radio emission.

An important constraint which must be met by any model of outer magnetospheric variability is that magnetic fluctuations must not broaden the beaming of photons
emitted with momenta parallel to the instantaneous magnetic field beyond the
characteristic pulse widths observed in the gamma-ray and associated optical and X-ray emission (and radio emission, in the Crab), thought to come from $ r > 0.5 R_L$ 
(\citealt{romani96}) - although this number, derived from a geometry based on a vacuum dipole with a phenomenological prescription for the location of the last closed field lines, 
will change when the magnetic field of the oblique force free rotator is put to use as the geometric platform for the beaming.  If reconnection (or any other mechanism) causes the magnetic field
at the light cylinder to fluctuate by an amount $\delta B $ at $r \sim R_L$, then the Alfven 
waves traveling along the boundary layer have, from conservation of energy flux,
amplitude $\delta B = B(R_L) (\delta B /B)_{r=R_L} (r/R_L)^{3/2}$ - for convenience,
I have here assumed $R_Y = R_L$. These are shear waves,  with 
$\delta B \perp B$, thus causing the local magnetic direction to vary about the mean
by an angle $\delta \theta \approx \delta B / B$.  The observed sharpness of the 
gamma-ray pulsars' light curves then suggests 
$\delta \theta < 0.2 (B/\delta B)_{r=R_L}^{2/3}$, while order of magnitude application of this idea to the observed torque fluctuations in the Crab pulsar and others suggests
$(B/\delta B)_{r=R_L} < 2$ (\citealt{aro81b}).  The correspondence of the limits on
the emitting radius from geometric fluctuation pulse broadening, from measurements
of the torques and from
modeling the beaming geometry provides an example of how  gamma ray observations of pulse profiles and radio observations of torque variability
over a substantial range of the $P-\dot{P}$ diagram can be used to 
seriously constrain both the electrodynamics and acceleration physics of these magnetospheres.

I have left out all discussion of the hoary problem of pulsar radio emission and transfer -
that would require an additional paper - other than the few comments above concerning
radiation beaming and single pulse fluctuations, which appear to me to provide probes
into the magnetospheric dynamics. In connection with the dynamical importance of the boundary layer bedween closed and open field lines - the location of the return current -
it is perhaps worth emphasizing that this region is likely to be the dynamical realization
of the site of ``conal'' emission, with velocity shear between the boundary layer plasma 
and the plasma filling the open flux tube (and that filling the closed zone) as a prime 
candidate for the free energy driving the collective radio emission process(es).

Perhaps a few bold souls will explore these issues more quantitatively {\it before} 
the GLAST observations 
become available, thus offering up their predictions to the sharp knives of experimental
tests. There is an urgent need for {\it physical} models of the boundary layer between
the open and closed regions, either with or without gaps, which account for the coupling with the stellar surface as well as the transition from the magnetosphere to the wind. This
is a collection of non-trivial problems - predictions of future progress are uncertain.

\section{Follow the Mass \label{sec:pairs}}

While there has been lots of attention to pair creation and particle acceleration {\it
within} pulsars' magnetopsheres from the community interested in observing and modeling these stars' lightcurves and SEDs, the most obvious evidence for particle 
particle acceleration and pair creation comes from observations and models of
Pulsar Wind Nebulae (PWNe).  These have recently been well reviewed from the observational standpoint by \cite{slane05} and \cite{gaens06}. The young PWNe and their pulsars - those still not crushed by the reverse shock - provide calorimetric information on both the energy and mass loss budgets of the underlying pulsars. Indeed,
since the earliest days, the energy budget has been used to constrain the moment of inertia of the neutron stars, thus the equation of state of nuclear matter. The mass loss budget provides a powerful constraint on all models
of plasma behavior within the magnetosphere, whether designed to explain specific observations or constructed to investigate basic theoretical issues,

\subsection{Observations and Consequences \label{sec:conseq}}

\begin{figure}[H]
\begin{center}
\hspace{10\unitlength}
\begin{picture}(100,100)(0,15)
\put(-45,-15){\makebox(100,100)[tl]{\includegraphics[width=2in]{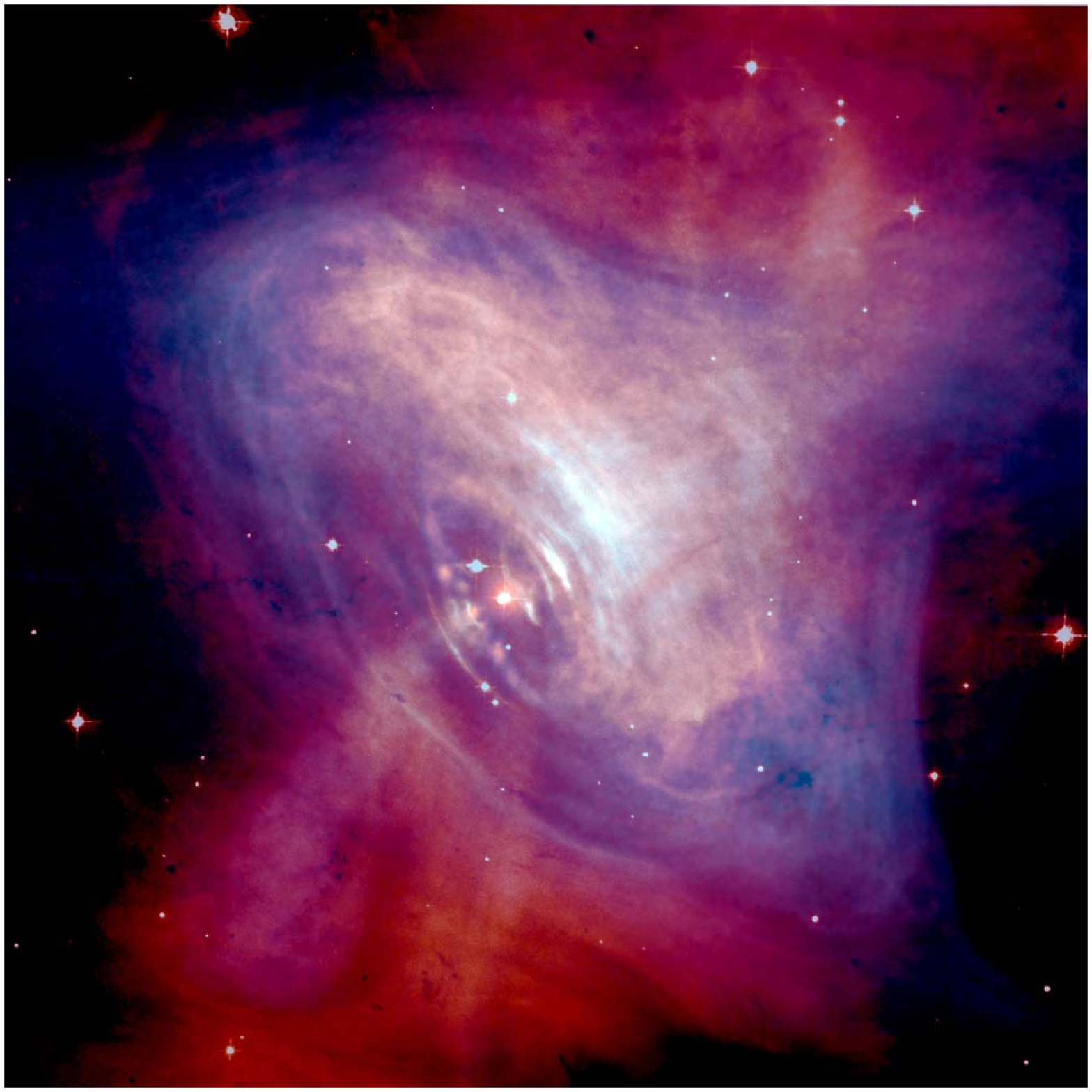}}}
\end{picture}
\hspace{10\unitlength}
\begin{picture}(100,100)(0,15)
\put(10,-65){\makebox(150,150)[tl]{\includegraphics[width=1.75in]{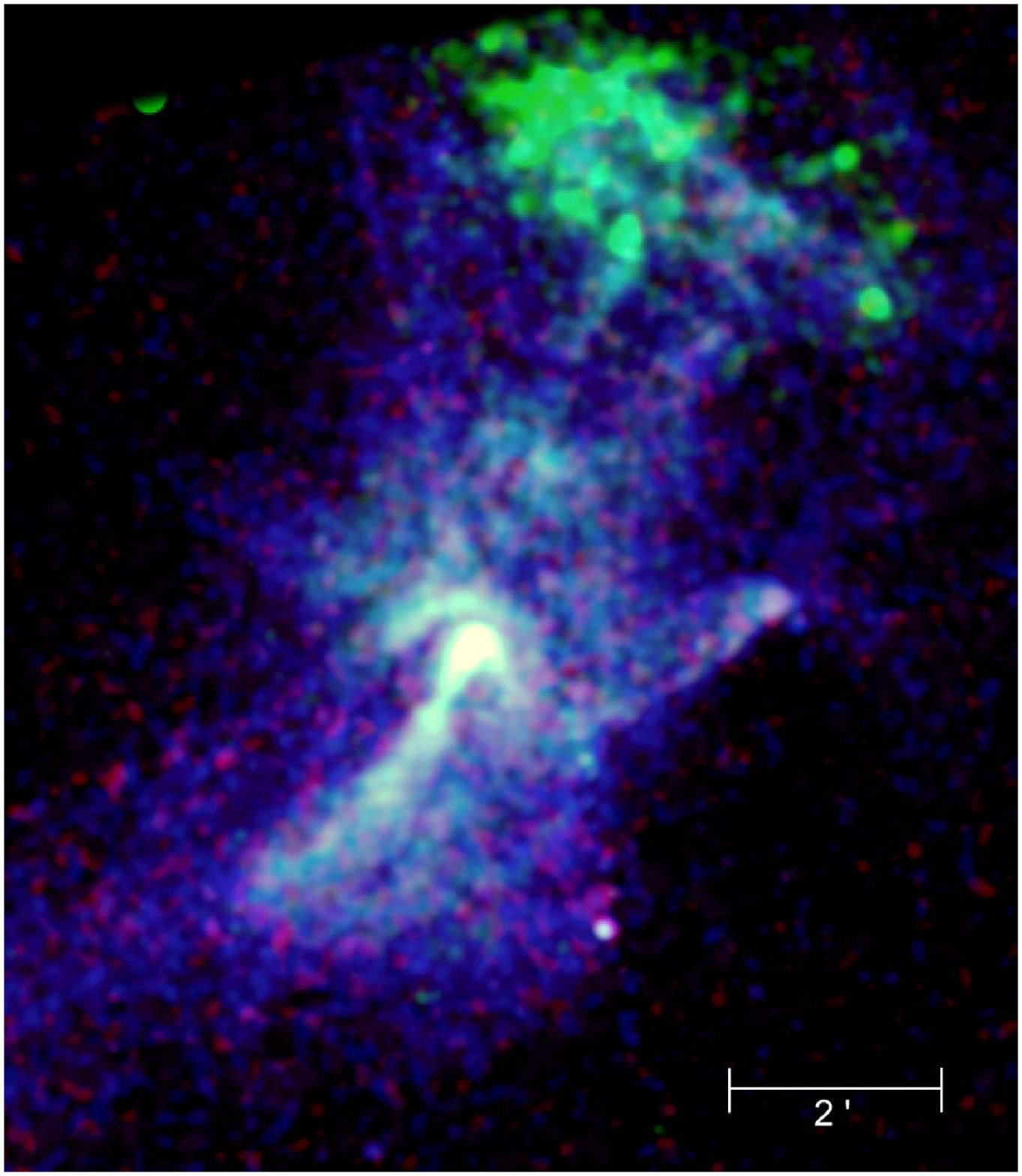}}}
\end{picture}
\end{center}
\vspace{3cm}
\caption{Left: composite of Hubble Space Telescope and Chandra images of the inner
$1^\prime$ of the Crab Nebula, showing the torus-plume structure around (torus) and along (plumes) the pulsar's rotation axis.  Right: Chandra image of the supernova remnant MSH 15-52 (G320.4+1.2), showing its torus-plume structure. \label{fig:hfpwne}}
\end{figure}

There has been major observational progress on these systems, coming most of all from
high resolution optical and X-ray imaging, as shown in Figure \ref{fig:hfpwne}, and from related studies of temporal variability.

Observations and models of the PWNe tell us about the particle loss rates $\dot{N}_\pm, \; \dot{N}_{beam}$ from the neutron stars.  In the case of the Crab, with its strong magnetic field and rapid synchrotron cooling of the particles radiating photons at energies above infrared frequencies, the now well resolved optical, X-ray and the unresolved gamma ray sources require a particle input of around $10^{38.5}$ s$^{-1}$, about $10^4$ times the basic electric current flow $c\Phi /e \sim 10^{34.5}$ s$^{-1}$ for this pulsar (e.g. \citealt{spit04}).  Similar conclusions have been reached for 
other PWNe as has been done, for example, in G320 around PSR B1509-58 (\citealt{gaens02, delaney06}), even though in this case the radiative
losses are not as rapid and therefore inferring $\dot{N}$ is not as straightforward. See \cite{dejager07} for pair injection rate inferences for
several other pulsars/PWNe. These
inferred rates come from examining the
\vspace*{-1.3cm}
\noindent brightness of the X-ray and (when they can be seen) optical nebulae, augmented by hard X-ray and gamma ray (GeV to TeV) observations of the young nebulae, when the hard photon telescopes can detect
anything - for a recent summary of the rapidly developing TeV observations of TeV
PWNe, see \cite{gallant06} and \cite{funk07}. The TeV observations are sensitive to
particles of energy comparable to those that give rise to synchrotron X-ray emission.
10 MeV to 10 GeV observations constrain the highest energy particles, which produce
synchrotron X- and $\gamma$-ray emission. A survey of PWN emission in this energy range awaits GLAST.

While the pair production rates found from polar cap models based on starvation electric fields do seem adequate for the understanding of the high energy photon emission in
PWNe (\citealt{hibsch01c} - these results remain the only attempt to survey pair creation across 
all observed  $P-\dot{P}$), other models have been developed specifically for the purpose of explaining gamma ray pulsar SEDs and light curves), there are substantial indications
that something of qualitative significance is missing.  It has long been known 
({\it e.g.} \citealt{shklovsky68}) that
the total radiating particle content of the Crab Nebula (mostly in the form of radio emitting electrons or pairs, which lose energy only because of adiabatic expansion) requires an injection rate averaged over the 1000 year history of the system on the order of $10^{40} - 10^{41}$ s$^{-1}$  electrons plus positrons, in order to understand the total radio 
emission from the Nebula. Recently \cite{dejager07} has revisited
this same question in the light of the TeV observations of VelaX, G320 and the newly discovered nebula of PSR B1823-13, again finding pair injection rates greatly in excess
of the rates found for particle {\it outflows} from either polar cap/slot gap or outer gap
models\footnote{However, his conclusion requires extrapolation of the particle spectra inferred from the TeV emission to radio emitting energies, a big jump.} constructed using starvation electric fields shorted out by the pair creation\footnote{Some outer gap models applied to the Crab pulsar do find total particle production rates in the range $10^{39}-10^{40}$ s$^{-1}$, but these refer to particles flowing in toward the star, where they collide with and are absorbed by the surface and (over)heat it. See \cite{hiro06} for the most recent version of this kind of model.}. Only the early polar cap model by \cite{tademaru73}, in which the effect of pairs' ability to limit the voltage drop was completely neglected, comes even close to yielding the observed time average injection rates. Since incorporating the pairs' polarizability destroys Tademaru's model, his
empirical success has been ignored.

\begin{figure}[H]
\begin{center}
\unitlength = 0.0011\textwidth
\hspace{10\unitlength}
\begin{picture}(200,200)(0,15)
\put(-200,95){\makebox(100,100)[tl]{\includegraphics[width=3in]{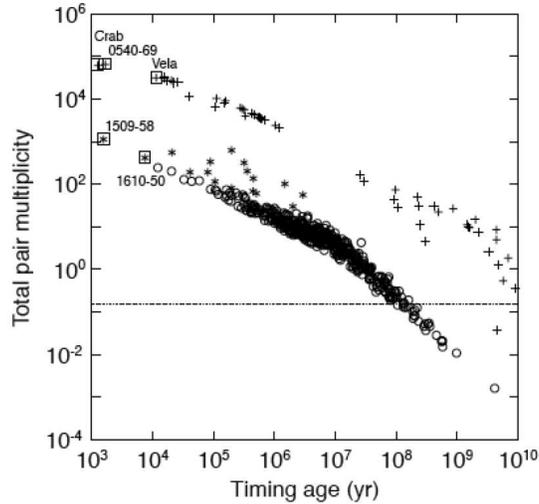}}}
\end{picture}
\end{center}
\vspace{4.5cm}
\caption{Multiplicities (number of pairs per particle in the Goldreich-Julian density) across the $P\dot{P}$ diagram,
from \cite{hibsch01b}. \cite{harding02} report similar results, using a more elaborate analysis.  Both used similar versions of space charge limited beam acceleration in the polar cap region, and both included the contribution of
synchrotron cascades to the total multiplicity.  Crosses refer to objects where curvature emission provides the gamma rays that convert to pairs, circles to objects where non-resonant inverse Compton upscatter of thermal X-rays (both from the polar cap heated by backflow bombardment and from the whole surface of he cooling neutron star) provides the gamma rays, while asterisks  show the more strongly magnetized stars where the cyclotron resonance in the scattering cross section makes a significant contribution to the gamma ray production rate.  All these calculations used a star centered dipole for the magnetic geometry, and can be substantially affected by surface magnetic anomalies, {\it e.g.} offsets of the dipole center from the stellar center, as in \cite{aro98}, or higher order multipoles, as in \cite{melik06}. The modern force-free model of the oblique rotator (\citealt{spit06})
offers the possibility of investigating the pair multiplicity within a self-conistent geometric setting, either with the traditional starvation electric fields or with improvements that take into account the full magnetospheric current system and charge densities, as outlined in \S \ref{sec:gapsubversion}. \label{fig:mult}}
\end{figure}

The starvation electric field models also have difficulty in coming  up with enough pair plasma to meet the desires of most (not all) models of radio emission over the whole $P\dot{P}$ diagram.  The results
of \cite{hibsch01b} appearing in Figure \ref{fig:mult} show clearly that for lower voltages ($\Phi < 10^{13.5}$ V), where most pulsars lie, the pair multiplicities drop well below unity, 

\vspace*{-1cm}

\noindent  far below the level
assumed in almost all models of the radio emission, and also required in models of propagation effects that have had some success in the interpretation of radio polarization and beaming structure (e.g. \citealt{mckinn97, barnard86}), or indeed needed to explain the death line in Figure \ref{fig:PPdot}.

It is possible that offset of the dipole from the stellar center, a central aspect of
the \cite{ruderman06} model for magnetic field evolution, might substantially
enhance a polar cap's pair yield, while still remaining consistent with the
apparently dipolar morphology of the low altitude magnetic field.  If the axis
of the offset dipole is tipped with respect to a radius vector, gravitational
bending of the gamma rays' orbits leads to much larger one photon pair creation
opacity in the magnetic field than is the case for the star centered dipole
(\citealt{aro98}). That opacity increase allows the more numerous low energy 
curvature  gamma rays to contribute to pair creation, thus enhancing the particle 
flux. This effect certainly does have a favorable impact on reconciling theoretical
with observational death lines; whether it seriously enhances the pair yield in
pulsars feeding plasma into young PWNe, which are far from death valley,
remains an unexplored topic.

The fact that pulsars (at least the young ones) must supply a plasma with
particle outflow rate  well above the Goldreich-Julian rate $c\Phi/e$ is
undeniable, based on the behavior of the PWNe - undeniable progress.  The
problems described here are quantitative, perhaps to be resolved by possible larger pair output if the polar flow is time dependent (``flicker'' flow), possibly to be resolved in a steady flow model if the charge density in the return current alters and increases the local electric field, perhaps by other effects not yet 
investigated.  These possibilities  have potential connections to time variability in the radio emission. Flickering of the polar current and pair creation might be connected to the radio microstructure ({\it e.g.} \citealt{jess05, rud75, benford77});
modification of the electric field by boundary layer space charge controlled by outer magnetsopheric unsteady reconnection (see above) may be connected to subpulse variability.  Improved sensitivity in infrared and shorter wavelength detection techniques that would allow probing for variability (in the corotating frame) of the higher frequency emission would be invaluable. Progress in this area in the next few years is to be expected. 

\subsection{Pulsar Wind Nebula Models}

\subsubsection{MHD Nebular Models}
Modeling the PWNe themselves has advanced greatly in the last decade. Driven by the
wealth of spatially and temporally resolved X-ray observations 
(\citealt{slane05}) of the ``torus-jet'' structures  shown in Figure \ref{fig:hfpwne} in 
the Crab and PSR B1509 nebulae and now known to be present in an 
increasing number of PWNe (\citealt{ng04, romani05}), modeling and simulation has advanced from the elementary ``spherical cow'' models of \cite{rees74} and \cite{ken84} to two dimensional, axisymmetric time dependent relativstic MHD simulations of the flow structure (\citealt{komiss03, delzanna04, bogo05}). 
\vspace*{-1.3cm}
\begin{figure}[H]
\begin{center}
\hspace{10\unitlength}
\begin{picture}(100,100)(0,15)
\put(-65,-5){\makebox(100,100)[tl]{\includegraphics[width=2.4in]{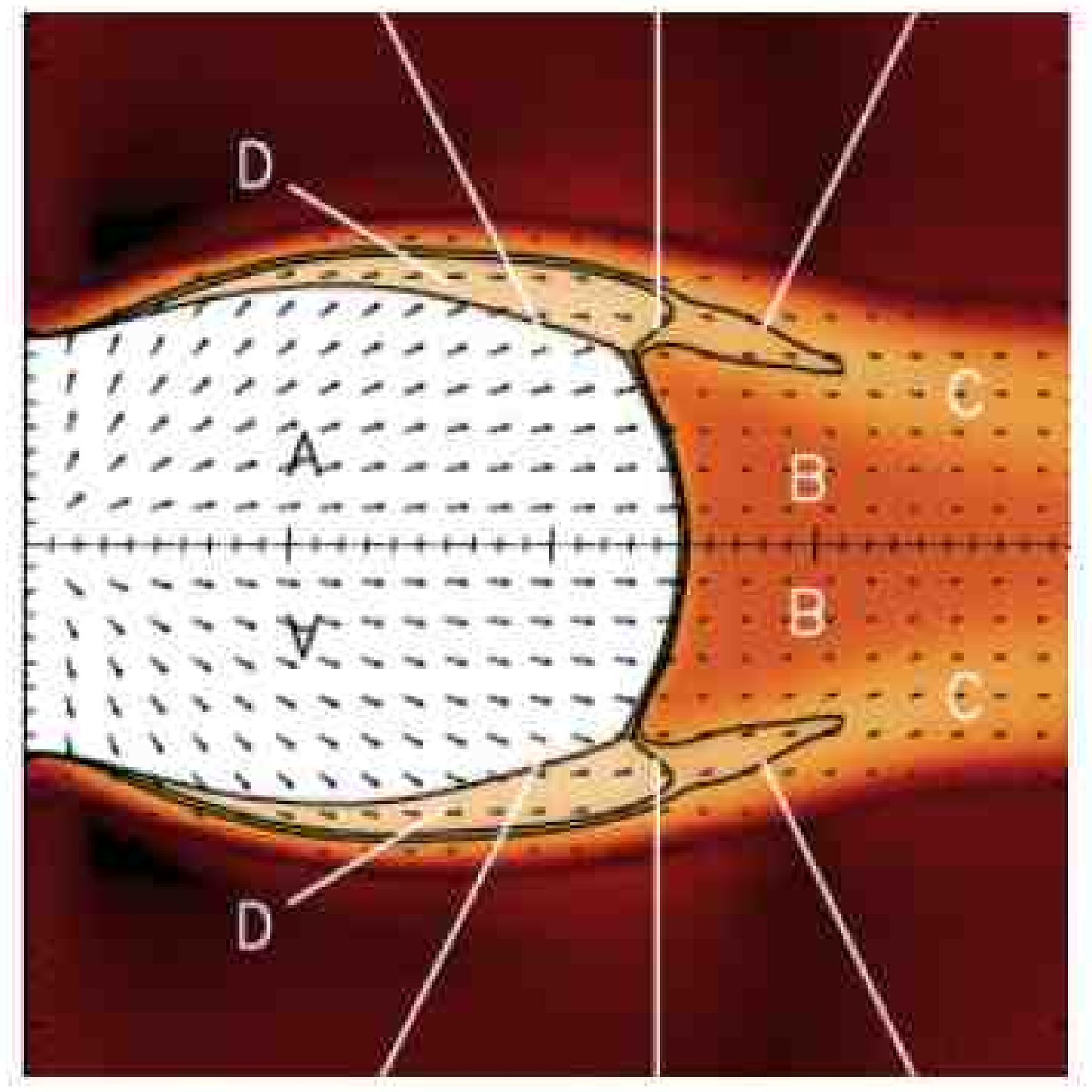}}}
\end{picture}
\hspace{10\unitlength}
\begin{picture}(100,100)(0,15)
\put(-20,-65){\makebox(150,150)[tl]{\includegraphics[width=2.4in]{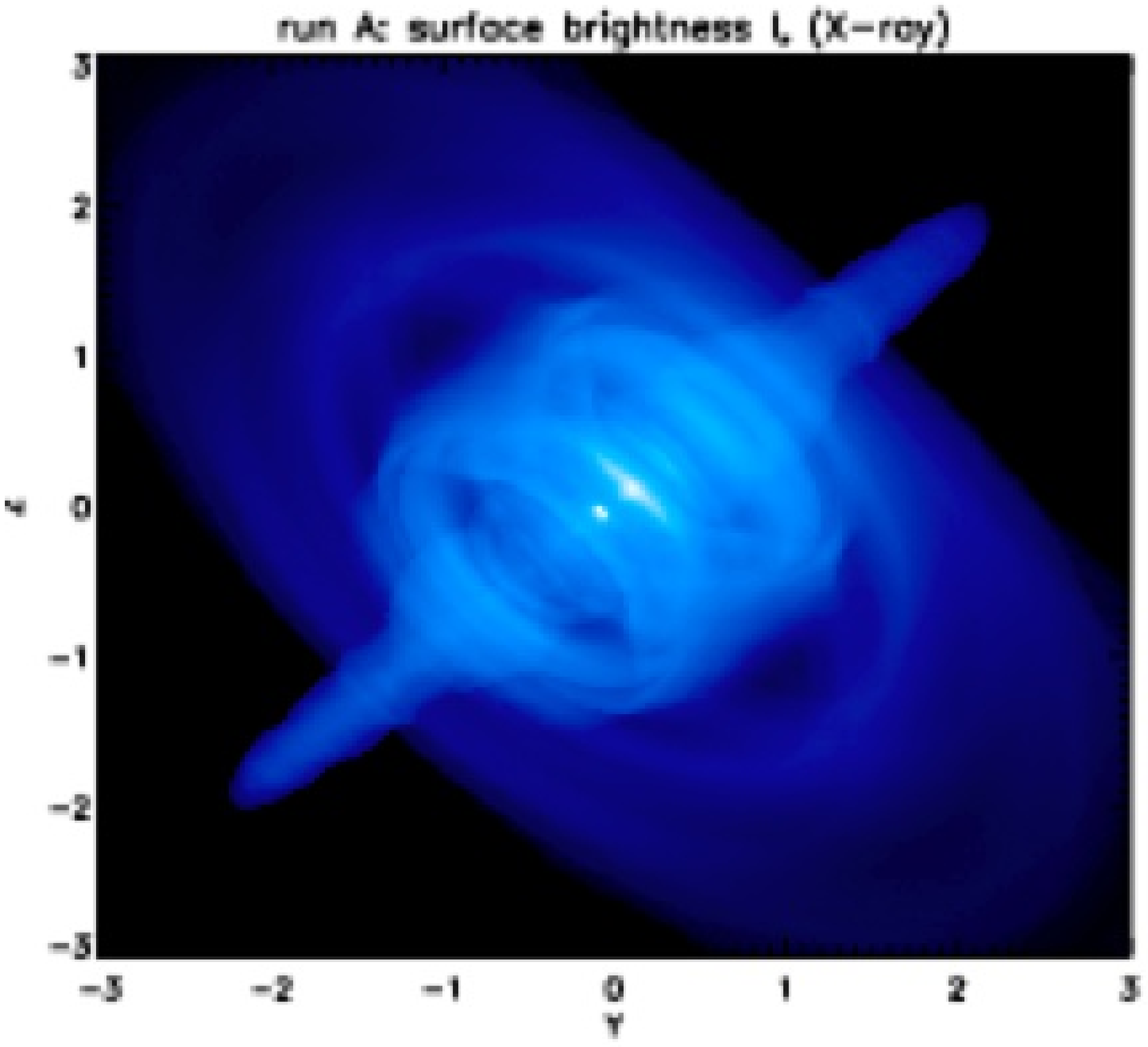}}}
\end{picture}
\end{center}
\vspace{3cm}
\caption{Left: Outflow structure in MHD models with energy injection concentrated in the equator. A: Upstream relativistic wind. B: Subsonic equatorial outflow, downstream of the equatorial termination shock. C: Fast downstream outflow emerging from the higher latitude oblique shock. D: Supersonic flow just downstream of the high latitude oblique shock. The backflow that focuses downstream plasma onto the axis is not shown. From \cite{delzanna04} Right: Synthetic torus-plume image, from \cite{delzanna06} \label{fig:mhd}}
\end{figure}

 These simulations
exploit the suggestions of \cite{bogo02} and \cite{lyub02}, that energy injected into these nebulae
follows the $\cos^2 \lambda$ profile ($\lambda$ = latitude with respect to a star's  rotational equator)   of the toroidal field
energy density exhibited  by the split monopole and oblique split monopole (\citealt{bogo99}) models of the Poynting fluxes from the neutron star.  They suggested that the consequence of such anisotropic energy injection into the surrounding nebulae would be greatly enhanced emission in a belt around the rotational equator - the ``torus'' appearing in the X-ray and optical imagery.  In addition, \cite{lyub02} suggested the outflow from the torus, since it is injected into the nonrelativistically expanding cavity formed by the supernova, would be deflected into a subsonic backflow at higher latitudes, where magnetic hoop stress could act to focus plasma into a magnetically compressed, outflowing, subsonic {\it plume} along the pulsar's rotation axis, thus creating the appearance of a jet.

\begin{figure}[H]
\begin{center}
\unitlength = 0.0011\textwidth
\hspace{10\unitlength}
\begin{picture}(200,200)(0,15)
\put(-375,95){\makebox(100,100)[tl]{\includegraphics[width=4.85in]{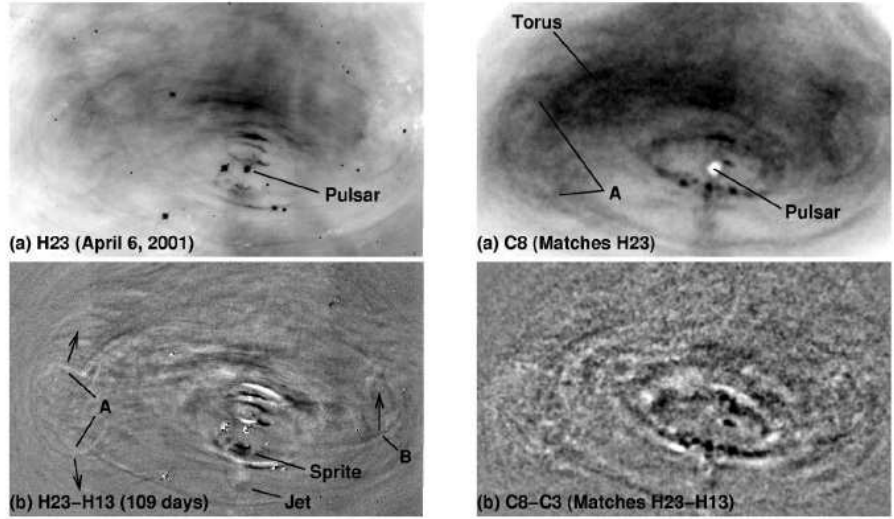}}}
\end{picture}
\end{center}
\vspace{4.5cm}
\caption{The torus and plume in the center of the Crab Nebula, as seen by HST and Chandra in 2001, from \cite{hester02}. Left column: HST; right column: Chandra. Upper row: total structure, showing the knotty inner ring in the Chandra image morphologically identified with the wind's termination shock in the rotational equator. The elliptical shape reflects the $60^\circ$ angle between the equatorial plane (which contains the torus) and the plane of the sky. Lower row: large scale structure subtracted by differencing between a a pair of early and later images in the series, producing a snapshot of the moving wisps emerging from the Chandra ring and expanding at speed $\sim 0.5c$\label{fig:crabtorus}}
\end{figure}

The simulations amply confirm the implications of the initial toy models, with flow velocities in the equator and the plume ($v \sim 0.5c$) comparable to those inferred from motions of features in the published time series of motions in the central regions of the Crab Nebula (\citealt{hester02}). Snapshots taken from that time series are shown in Figure \ref{fig:crabtorus}. 
\vspace*{0cm}

\subsubsection{The $\sigma$ Problem}

The MHD models do well at reproducing the torus-plume structure, as is shown in Figure
\ref{fig:mhd}, {\it if}  the wind upstream from the termination shock is weakly magnetized - the average of $\sigma = B^2/4\pi \rho \gamma c^2$ over the equatorial shock must
be $\sim 0.02$ in order to generate the good looking simulated image, a value $\sim 4 \times$ larger than what had been previously inferred from the 1D time stationary models of \cite{ken84}. But this value is still 
far below what the asymptotic $\sigma$ expected in ideal MHD outflow {\it of an unconfined wind}, exhibited in (\ref{eq:sigma})\footnote{Relativistic MHD jets accelerating within confining boundaries - ``walls'' - may have quite different behavior, as has been most recently exhibited by \cite{komiss07}. Here, the effect of a confining boundary, perhaps representing confining pressure in an outflow from a disk, forces poloidal field lines and stream lines to depart from monopolar form, which allows magnetic hoop stress to confine and accelerate a polar flow.  For a wind emerging from a star, essentially a point source, there is no analogue of confining walls to break
the balance between magnetic hoop stress and electric repulsion that lies 
behind the slow acceleration.},  and the asymptotic Lorentz factor and four velocity (in units of $c$) is 
\begin{eqnarray}
\sigma_\infty & \approx & \sigma_0^{2/3}, \; u_\infty \approx \sigma_0^{1/3}
  \label{eq:sigma}\\
\sigma_0 & \equiv & \frac{\Omega^2 \Psi_B^2}{\dot{M} c^3 \gamma_0 }
\nonumber \\
& = & \frac{ e\Phi}{2 m_{eff} c^2 \gamma_0}, \; \; m_{eff} = m_{ion}  + 2 m_\pm \kappa_{\pm}.
\label{eq:sigma0}
\end{eqnarray}
Here $\Psi_B = \mu / R_L = R_L \Phi, \; \kappa_\pm$ the pair multiplicity and
$\gamma_0$ the bulk four velocity of the plasma emerging from the plasma
source (pair creation at the polar cap, since outer gaps - if they exist as
particle production zones - send most of their plasma back toward the neutron star), itself an unknown function of magnetic latitude (possibly lower in the equator than at the poles, since pair creation should be weaker near the magnetospheric boundary layer).  For the Crab pulsar, pair creation theory suggests the multiplicity of the plasma feeding the nebular optical and X-ray source is  
$\kappa_{\pm OX} \sim 10^{4.5}$, based on spectral imaging modeling (\citealt{delzanna06}), a value consistent  with starvation gap modeling of the polar cap
(\citealt{hibsch01c}), who also find $\gamma_0 \sim 100$ for this star. If this piece of the mass loss budget corresponds to the total mass loss, $\sigma_0 \sim 1 \times 10^4$ and $u_\infty =22 \approx \gamma_\infty$. 

 If one includes the whole mass flow, $\dot{N}_\pm \sim 10^{40.5} $ s$^{-1}$ 
(\citealt{dejager07}), which includes the particles required to feed the Nebular radio emission, then $\kappa_{\pm} \sim 10^6$ and $\sigma_0$
is $\sim 10^{4.6} /\gamma_0$ - since the origin of the large mass flux is unknown, $\gamma_0$ is also unknown, although surely it is smaller
than the value $\sim 100$ found in existing gap models - then 
$u_\infty = 16 (10 /\gamma_0)^{1/3}$. 

The ideal MHD values of $\sigma$ and $\gamma_\infty$ are for a wind with 
monopolar poloidal field and flow geometry at large $r$.
Theory and simulation to date all support the poloidal field of the wind being
monopolar well outside the light  cylinder - see, for example, \cite{bucc06}, whose high $\sigma$ simulation of outflow from the aligned rotator extended to 
$r \approx 900 R_L$, well outside the fast magnetosonic surface, with the field
becoming closely monopolar with no polar focusing or hoop stress apparent. Then the asymptotic magnetization $\sigma_\infty$ and 4-velocity 
$c\gamma_\infty$
(outside the current sheet, which is infinitely thin in ideal MHD) 
are predicted to be as in (\ref{eq:sigma}).

The MHD models do answer the oft repeated question of just what is going on
at higher latitudes - if the torus structure is the manifestation of the shock termination of the wind, why don't we see evidence for the shock at high latitude (e.g., \citealt{blandford02})?  The MHD model asserts that the polar regions at distances we can resolve are occupied by the backflow that forms the plume. The shock shown in Figure
\ref{fig:mhd} curves down toward the star, reaching into radii too small to 
observationally resolve on the polar axis.  Furthermore, the shock is quite 
oblique at higher latitude, which weakens the efficacy of shock acceleration.
The MHD model and the curvature of the shock relies on the total energy flux being proportional to $\cos^2 \lambda$. In the Crab, the higher latitude parts of the curved shock do manifest themselves as the bright knots, which appear in projection as if they are right next to the pulsar (\citealt{komiss03}). Thus, qualitatively and semi-quantitatively, a satisfactory picture of PWNe plasma flow on nebular scales has appeared.

Ideal MHD models may also do well at reproducing the wisp structures shown 
in  Figure \ref{fig:crabtorus}, which are of interest for the diagnosis of the relativistic shock wave terminating the outflow. These are now known to be structures (probably waves, \citealt{scargle69}) appearing to be 
emitted from the Chandra ring with a periodicity $\sim$ 6 months, traveling out 
with  a speed $\sim 0.5c$, possibly with some deceleration with increasing 
radius  (\citealt{mori06}).  The
wisps occur on scales too small to be resolved by the published MHD simulations of the whole nebular flow. Of the various suggestions made over the years to interpret the wisps,
the most promising MHD model for these is that they are due to MHD Kelvin-Helmholtz instabilities occuring at the boundary between a fast equatorial and a slower high latitude flow (\citealt{begel99}). The global flow models have such a shear layer, as flow emerging from the equatorial shock in the nebular cavity
returns at higher latitudes toward the axis.  \cite{bucc06b} used high resolution MHD simulations of a shear layer in a box, repeated with periodic boundary 
conditions, to draw the conclusion that this hypothesis is quantitatively inadequate  to reproduce the observed variability. However, recent (summer
of 2007) high resolution MHD simulations of the whole nebular flow by
Bucciantini and by Komissarov, both still in progress, suggest that either
Kelvin-Helmholtz instability or, perhaps more likely, secondary instability of the vortices formed in the shear layer near the base of the plume, do show many
features similar to the observed moving wisp structures.  

If these models do
exhibit as much ability to reproduce the observations as has been found in
the preliminary work, the multi-dimensional MHD model of PWN structure will
have accumulated three major successes: 1) a model for thermal filament 
formation, through Rayleigh-Taylor instability of the boundary between the 
nonthermal bubble, as first pointed out by \cite{hester96} and developed
extensively by \cite{bucc04};  mildly relativistic plume (a.k.a. jet) formation,
as first suggested by \cite{lyub02} and modeled numerically by 
\cite{komiss03}, \cite{delzanna04} and \cite{bogo05}; and now the 
wisp variability near the termination shock.
Such models probably will trun out to be successful in interpreting the more slowly expanding outer structures of the torus - the current round of high resolution simulations will soon show whether or not these featurs of the 
nebular  ``weather'' can be captured numerically.

The MHD dynamics does have strong sensitivity to $\sigma_{wind}$ at
the termination shock.  The models are insensitive to the wind's
4-velocity (Lorentz factor) just upstream of the termination shock, and are
insensitive to the composition, 
other  than that  the particles must have small Larmor radii 
and that they be efficient radiators.  The last requirement leaves electron-
positron plasma as the only option, in the young systems with bright PWNe -
the particle injection rates greatly exceed the Goldreich-Julian value.

\subsection{Beyond MHD}

\subsubsection{Striped Winds}

\begin{figure}[H]
\begin{center}
\unitlength = 0.0011\textwidth
\hspace{10\unitlength}
\begin{picture}(200,200)(0,15)
\put(-400,120){\makebox(100,100)[tl]{\includegraphics[width=5in]{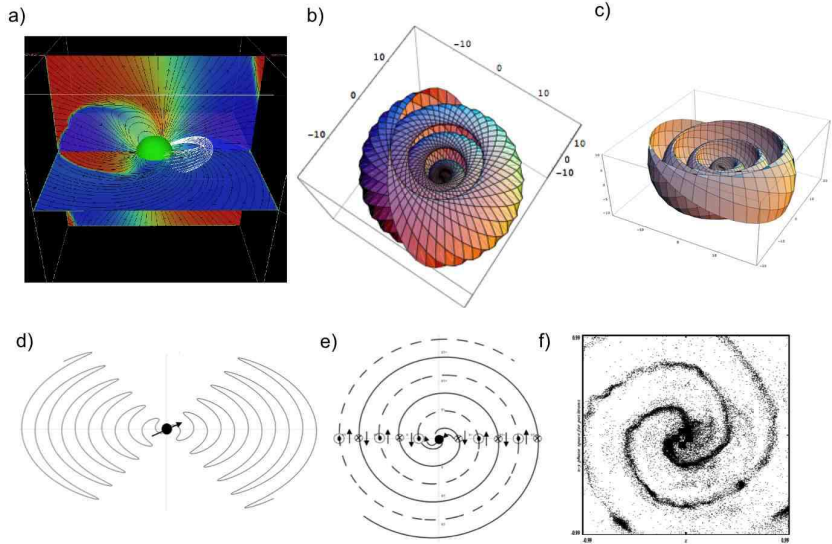}}}
\end{picture}
\end{center}
\vspace{7.25cm}
\caption{Magnetic geometry of the inner regions of a striped wind emerging from an oblique rotator with a large obliquity $i$. a) Magnetic structure of the force free rotator
for $i=60^\circ$, from \cite{spit06}. b) One of the two interleaved current sheets for the
$60^\circ$ rotator, derived from Bogovalov's oblique split monopole model (\citealt{bogo99}) c) The same as b) but for $i = 9^\circ$, shown for clarity. d) and e) Meridional and equatorial cross sections of the striped wind current sheet, for the $60^\circ$ rotator.
f) A snapshot of a 2D PIC simulation of the equatorial stripes, by \cite{spit02b}\label{fig:stripes}}
\end{figure}

However, the equator where the equatorial shock forms is a current sheet, a region notorious for breakdown of ideal MHD. Such breakdown has been assumed in the MHD models, which achieve their successful fits of nebular
appearance to observation only when there is a finite region around the
equator where the magnetic field at the shock is small compared to what
one would expect in the ideal MHD flow with an infinitely thin, flat current sheet.
\cite{coroniti90} suggested the apparent low value of $\sigma$ in the equator - inferred to be $\sim 0.005$ in the 1D, spherically symmetric \cite{ken84} model - is due to annilhilation of the equatorial magnetic field in the  current sheet. Close to the star but outside the light cylinder the magnetic field takes the form of the striped magnetic structure, with oppositely directed fields from the opposite poles of the dipole wound into a frozen in wave, shown in Figure \ref{fig:stripes}. 

\begin{figure}[H]
\begin{center}
\hspace{10\unitlength}
\begin{picture}(100,100)(0,15)
\put(-65,-45){\makebox(100,100)[tl]{\includegraphics[width=2.2in]{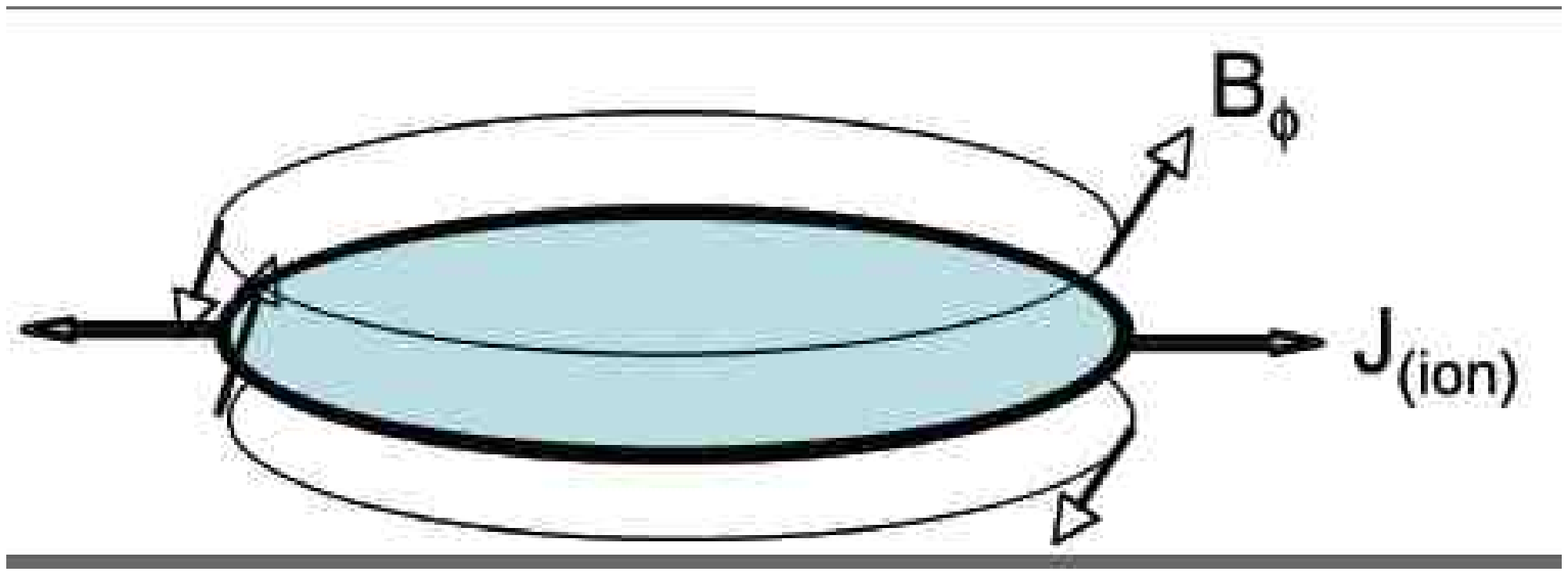}}}
\end{picture}
\hspace{10\unitlength}
\begin{picture}(100,100)(0,15)
\put(-20,-65){\makebox(150,150)[tl]{\includegraphics[width=2.2in]{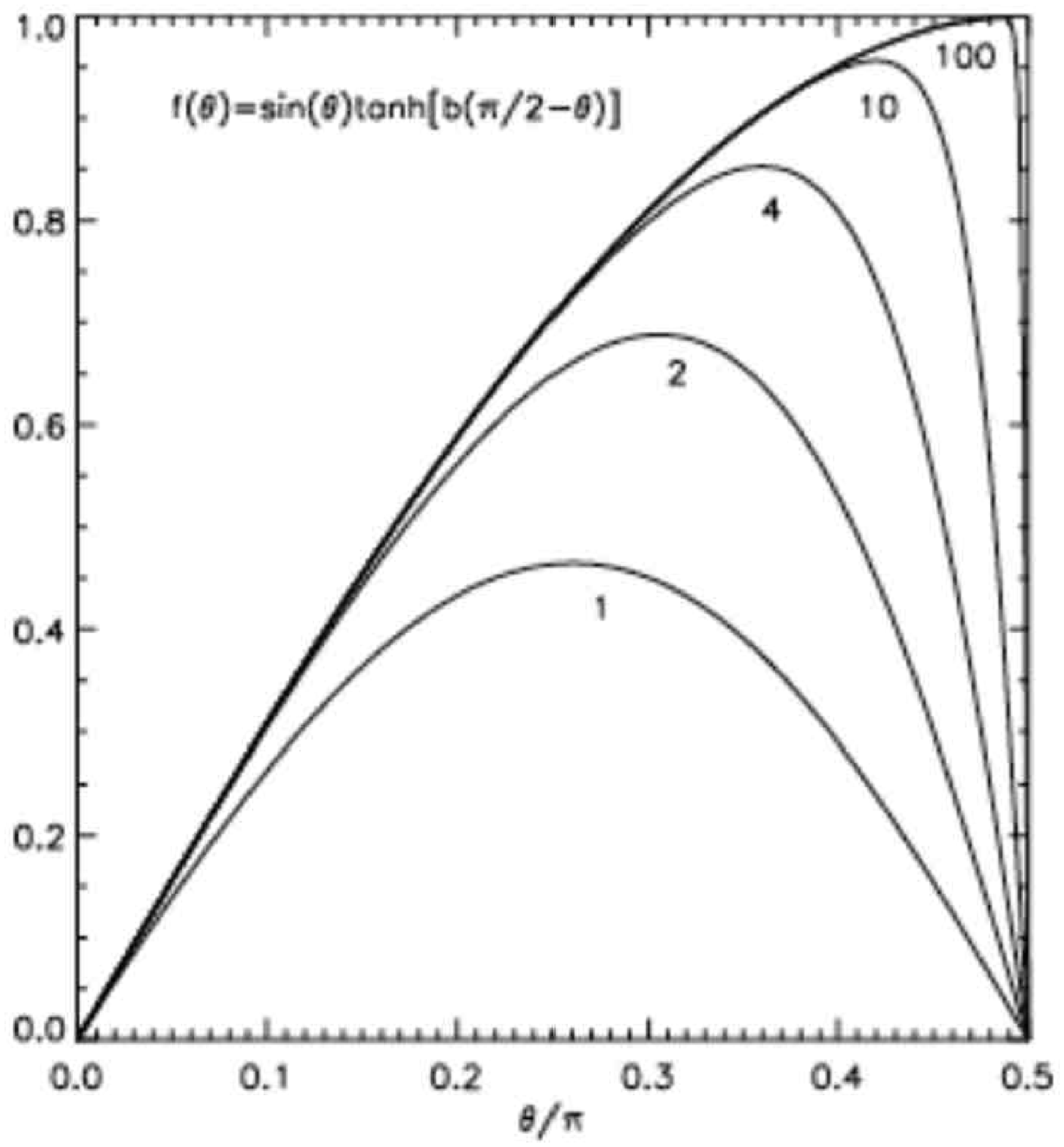}}}
\end{picture}
\end{center}
\vspace{3cm}
\caption{Left: Magnetic sandwich geometry of the equatorially concentrated outflow, with radial electric current flowing in a extremely weakly magnetized midplane between oppositely directed toroidal magnetic field at higher latitudes. Right: Typical magnetic profiles as a function of latitude, from \cite{delzanna04} \label{fig:sandwich}}
\end{figure}

Coroniti's idea was that some form of current sheet dissipation causes the magnetic energy to annihilate in the inner wind, causing conversion of magnetic energy to flow energy, and reducing the structure to something approximating that of an aligned
rotator's outflow with a magnetic field in the asymptotic wind $R_L \ll r \ll R_{shock}$ (= $10^9 \; R_L $ in the case of the Crab)  much weaker than that what one expects from ideal MHD transport of the light cylinder  field inferred 
from the star's spindown.
The resulting (dissipative) MHD model has an equatorial current sheet built in, since the $
\lambda > 0$ hemisphere has a toroidal magnetic field wound in the opposite direction to that 
found for $\lambda < 0$. Figure \ref{fig:sandwich} shows a cartoon of the resulting magnetic
``sandwich'' wind at large radius, along with magnetic field strength as a function of 
$\lambda$ considered in the MHD models of the nebulae beyond the shock.

One almost model independent constraint on this idea is that an acceptable theory of stripe dissipation in the wind zone necessarily leads to the wind's four 
velocity in the dissipation region being small compared to the value $\Gamma_{wind} \sim 10^{6.5} $ inferred from 1D dynamical models of the nebular high energy photon spectra (\citealt{ken84}). The reason is simple.

The magnetic field in the stripes, which have proper wavelength 
$\lambda^\prime =  \Gamma_{wind} R_L$, need proper dissipation time  
$T_d^\prime >  \lambda^\prime/c = \Gamma_{wind} R_L / c$, since the 
current sheets can't expand any faster than the speed of light; alternatively, in a reconnection model, the magnetic field flows into the sheets, to disappear in expanding islands of hot plasma around O-lines, with velocity 
$\epsilon_R v_A \approx \epsilon_R c$ (\citealt{lyubarsky05}), with $\epsilon_R$ expected\footnote{An expectation based on kinetic simulations and experiments on non-relativistic reconnection, as in 
\cite{gem01}; relativistic reconnection in a pair plasma, the case relevant here, has just started to receive
attention (\citealt{bessh05}).} to be on the order of 0.1 to 0.2.  In the pulsar's center of mass frame, the dissipation time then has the lower limit 
$T_d > \Gamma_{wind}^2 R_L /c$; in a reconnection model, 
$T_d \approx \Gamma_{wind}^2 R_L /\epsilon_R c$.  

\vspace*{0cm}
\noindent A successful model for 
the apparent low value of $\sigma$ at the termination shock in the Crab Nebula, 
where the equatorial shock occurs at $R_{shock} \approx 10^9 R_L$, requires 
that the 
dissipation go to completion in a region where 
$\Gamma_{wind}< 10^{4.5}$;  using the reconnection model reduces this upper limit 
to $\Gamma_{wind} < 10^4$.  In MSH15-52, where $R_{shock} \sim 0.4 - 0.5$ pc and $R_L = 7825$ km,  $\Gamma_{wind} < 10^{4.7}$ is a firm upper limit;
in a reconnection model, $\Gamma_{wind} < 10^{4.2}$. 

 \vspace*{-1.3cm}

It is interesting to note that the full average particle loss rate 
$\dot{N}_\pm = 10^{40.5} \dot{N}_{40.5}$
s$^{-1}$ inferred for the Crab Nebula requires, from energy conservation, that
the asymptotic value of the wind 4 velocity is 
$\Gamma_{wind\infty} \leq \dot{E}/\dot{N}_\pm m_\pm c^2 
= 10^{4.3}/\dot{N}_{40.5}$ in this system, much less than inferred in 1D models that neglect the plasma required for the radio emission such as that of
\cite{ken84} - the upper limit is achieved if the plasma is cold by the time it reaches the termination shock.   The acceleration from the inner wind,
launched by the pulsar's magnetic spring to $\Gamma_w = u_\infty $ 
given in (\ref{eq:sigma}), occurs if the current sheets dissipate and the wind
heats, accelerating from the resulting internal pressure gradient.  This is a
relatively slow process (\citealt{lyubarsky01}), reaching completion before
a fluid element collides with the termination shock in the Crab Nebula only
if $\Gamma_w < 10^{4.8}, \; \dot{N} > 10^{40}$ s$^{-1}$.  This requires
maximal dissipation of the current sheets - in a sheet broadening model,
as originally proposed by (\citealt{coroniti90}), the sheets must expand at a
substantial fraction of the speed of light in the proper frame of the
flow. Recently \cite{arons07} showed that the interaction of the
relativistic currents in neighboring sheets drives a Weibel-like instability in
each sheet, with a 
resulting anomalous resistivity that supports such maximal dissipation in the inner wind of the Crab pulsar, $r < 10^6 R_L$, a conclusion consistent with the maximal dissipation rate
model of  \cite{kirk03}. However, it is safe to say that 
the transition of the wind from high to low $\sigma$ and from low to high four velocity remains not fully understood, and not well constrained by direct observation of the winds. 

For a long time, winds have been modeled as having an asymptotic flow 
velocity $\Gamma_{w\infty} \sim 10^6$, with various arguments being
used, ranging from radiation modeling of the post-shock flow to the dynamics
of high energy particles injected by the wind at the shock, with Larmor radii
comparable to the termination shock's radius, invoked as a dynamical model
for the variable wisp dynamics near the shock.  Such particles, if they exist,
have 4 velocities (probably gained by acceleration in the wind) much larger 
than the maximum flow 4-velocity of the wind inferred from energy conservation.
They might occur due to acceleration in the current sheet, thus are confined to 
the equatorial sector.

Some of the energy dissipated in the wind might reappear as an unpulsed photon source superposed (because of relativistic beaming) on the pulsar.  The 
Crab pulsar
exhibits linearly polarized (33\%), unpulsed optical emission, with intensity $\sim 1.25$\% of the main pulse peak intensity with a fixed polarization position angle (\citealt{kanbach05}), polarization properties 
consistent with emission from the toroidal field in the wind zone (\citealt{barnard86a}).  If it proves possible to interpret such unpulsed flux as radiation from the wind, one might obtain significant observational constraints on this
difficult problem.  
 
 \subsubsection{Wisps as Ion Cyclotron/Magnetosonic Waves}

A non-MHD hypothesis based on kinetic structure in the current sheets' plasma (a ``beyond MHD'' model) does well at reproducing the observed variability near the wind termination shcok than MHD schemes. \cite{gallant94} proposed this current contains a high energy ion beam, accompanied by a flux of $e^{\pm}$ pairs, with ion energy/particle
approaching the total magnetospheric potential energy, and that this ion current carries
a large fraction of the spindown energy, while still being a minority population by number
density. 

It is important to note that as far as the dynamics is concerned, high energy ions
in the equatorial return current, expected to emerge from an ``acute'' pulsar
 - $ \angle ({\boldsymbol \Omega}, {\boldsymbol \mu})  < \pi/2 $ -  can be
 replaced by high energy electrons, expected to form the return current
 in an ``obtuse'' pulsar, 
 $\angle ({\boldsymbol \Omega}, {\boldsymbol \mu})  > \pi/2 $.  In both cases,
 the required acceleration to energy/particle comparable to $e\Phi$ must occur
 in the wind, perhaps as particle runaways in the current sheet's resistive 
 electric field (\citealt{arons07}). In the electron case, radiation reaction can limit
 the energy/particle that can be achieved, which makes the acute pulsar model slightly preferable, and for the rest of this discussion I confine discussion
 to the ion beam case.

Since such ions have Larmor radii  comparable to the radius, the compressions
induced in the pair plasma at the ion stream's turning points can appear as 
surface brightening spaced with separation comparable to the spacing of the 
wisps.  \cite{spit04}
implemented this idea in a time dependent simulation, showing that in a 1D 
model in 
a toroidal magnetic field winding in one direction in the equatorial flow ({\it i.e.},
ignoring the reversal of the field direction in latitude), the ions' deflection into circular motion in the abruptly increased magnetic field at the equatorial shock in the pairs is ion
cyclotron unstable, with gyrophase bunching forming a compressional limit cycle which launches finite amplitude magnetosonic waves in the pairs. This launching occurs approximately once per ion Larmor period, which is about six months for the parameters determined by comparing the model to the images. These waves travel out at speeds $\approx (0.3 - 0.5)c$, with the precise velocity value depending on the degree of isotropization of the pairs. 
 
\begin{figure}[H]
\hspace{10\unitlength}
\begin{picture}(100,100)(0,15)
\put(-55,85){\makebox(100,100)[tl]{\includegraphics[width=2.4in]{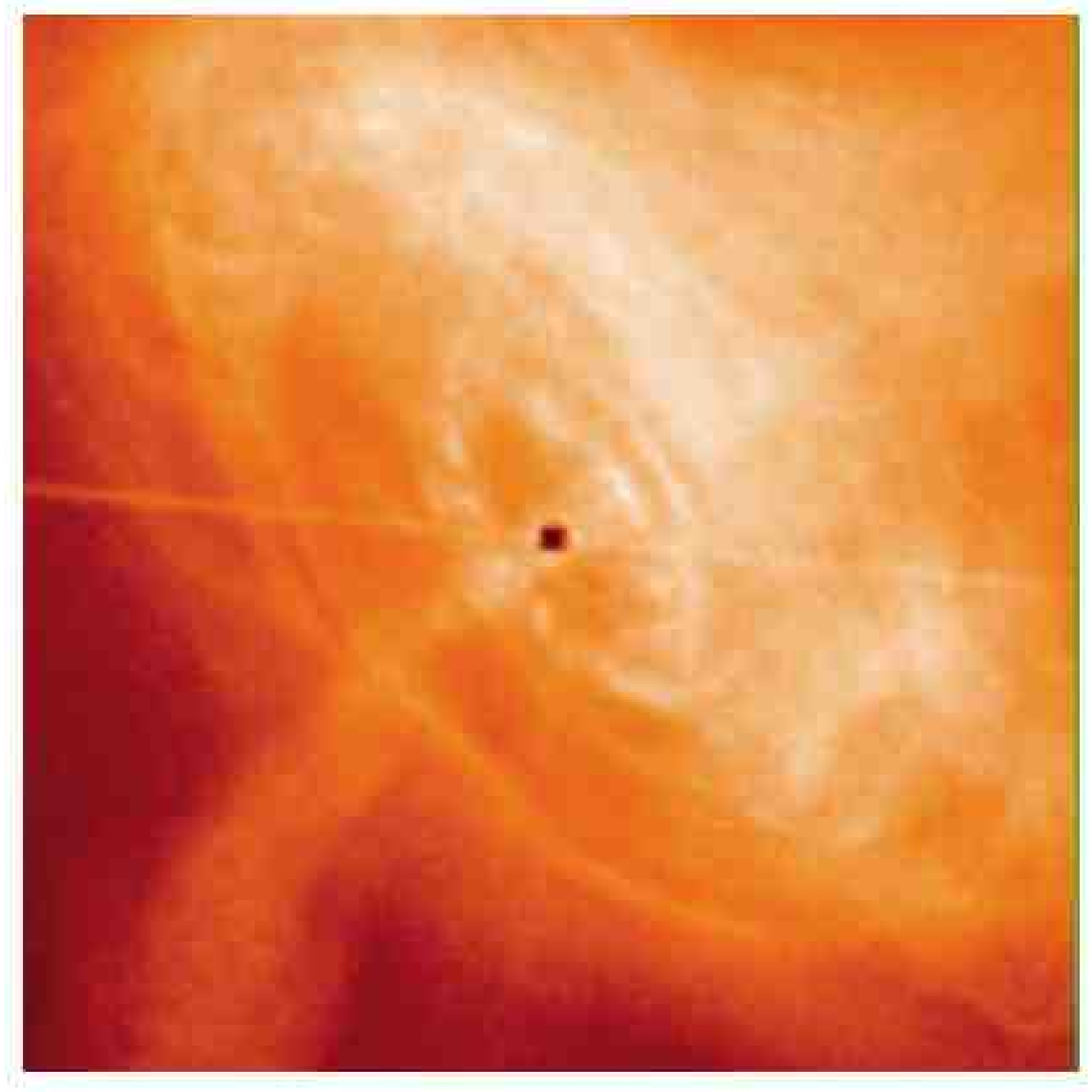}}}
\end{picture}
\hspace{10\unitlength}
\begin{picture}(100,100)(0,15)
\put(-60,83){\makebox(100,100)[tl]{\includegraphics[width=2.45in]{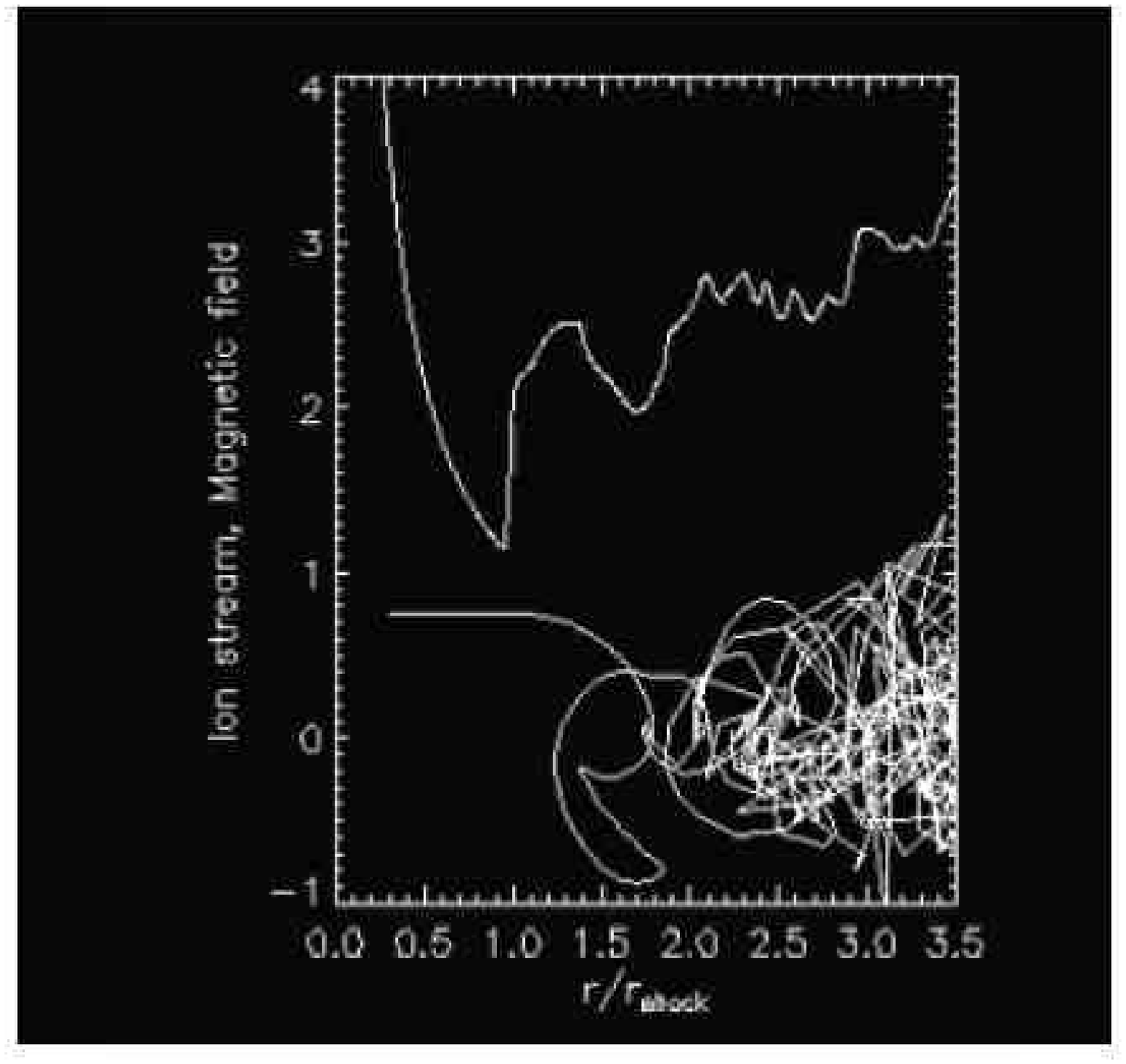}}}
\end{picture}
\hspace{10\unitlength}
\begin{picture}(100,100)(0,15)
\put(70,5){\makebox(100,100)[tl]{\includegraphics[width=2.5in]{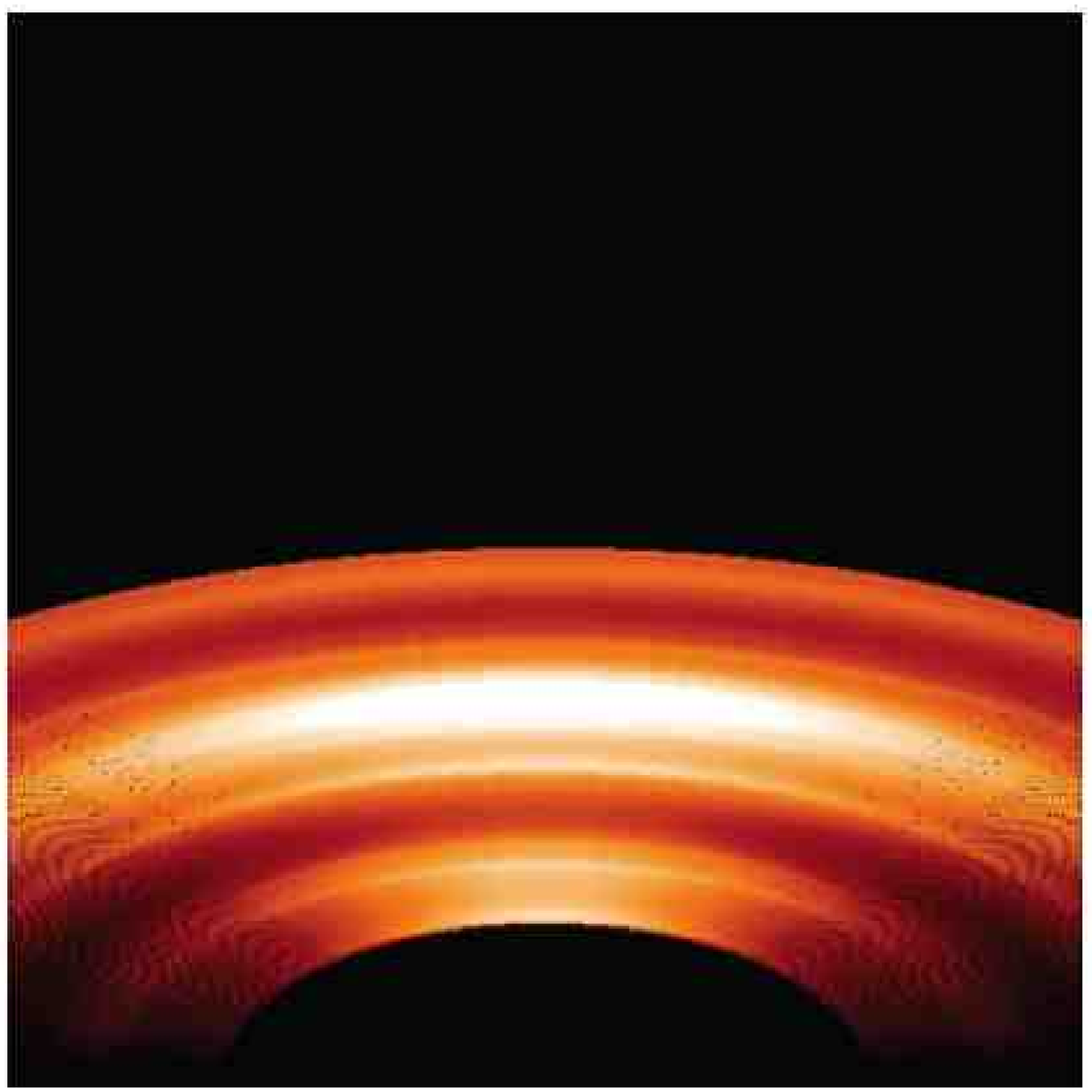}}}
\end{picture}
\vspace{3cm}
\caption{Ion cyclotron instability model of pulsar wind termination shock variability.
Top Left: Chandra snapshot of the X-ray ring and torus in the Crab Nebula. The ring is the location of the magnetosonic shock wave in the pairs, whose thickness is unresolved.
Top Right:  Snapshot of a 1D hybrid (PIC ions, MHD pairs) simulation of the equatorial
ion outflow as it encounters the enhanced magnetic field in the pairs.  The upper half shows the magnetic field with the ion induced compressions.The lower half shows the ion phase space ($p_r/\gamma_{upstream} m_{ion} c$ versus $r/R_{shock}$, $p_r$ = radial momentum of an ion), with the gyrophase bunch formed as the ions encounter enhanced $B$ and acquire a reduced Larmor radius. The rotating, reforming bunch (a limit cycle) emits a train of compressional magnetosonic waves that propagate to larger radius. Bottom: Snapshot of the synthetic surface brightness map, assuming the pairs emit synchrotron radiation in the compressed magnetic field with no nonthermal acceleration included. 
From \cite{spit04}. \label{fig:ions}}
\end{figure}

The resulting synthesized surface brightness map looks
more than a little like the observed waves emitted from the inner X-ray ring in the Crab Nebula, as shown in Figure \ref{fig:ions}. The observed wave emission period (\citealt{mori06}), announced at a conference {\it after} the model was developed and 
published, is in good accord with the model's predictions.  Application of the 
model to PSR B1509/G320 suggests that ``wisp'' variability on a time scale of 
years should be found. There is weak, but not very convincing,
evidence for such variability in the partial torus near this
pulsar (\citealt{delaney06}).

By far the most attractive theoretical feature of the model when it is fit to the HST and Chandra movies - fit by eye, there is little use in more elaborate fitting procedures, given the 1D model's departures from the observationally obvious requirement of at least 2D - is the inference that the ion flux required to produce the observed surface brightness enhancements is 
$\dot{N}_{ion} \approx c\Phi /e$, the equatorial return 
current of the force free rotator, in the case of the Crab pulsar.  Of course, since the magnetic field has 
largely dissipated in the wind, the pairs accompanying the ions must largely neutralize the electric current in the ions, but the result is an indication that back at the magnetosphere and the stellar surface, some piece of non-force-free electrodynamics does work to extract this ion flux so as to maintain the star's charge balance. In turn, that suggests $i < 90^\circ$, although $i$ certainly should be a large fraction of $\pi /2$. Another feature of the model is that the fact that the ions in the wind are inferred to have Larmor radii comparable to the wind's termination radius $R_s$ - this yields ion energy/particle  
$E_{ion} = e\Phi (m_p /m_{eff, eq}), $  and $\gamma_{ion} \approx 10^{6.5}$ with  the pair multiplicity evaluated  {\it in the equator}, a value close to the MHD wind 4-velocity inferred by \cite{ken84}. 

The model {\it assumes} the underlying acceleration from the neutron star is like MHD even in the current sheet, with all the particles - ions and pairs - traveling with a single (fluid) 4-velocity
until a fluid element encounters the shock, even though the flow in question is in the current sheet, where different plasma components may have different velocities. Thus Spitkovsky and Arons' inference
that $\Gamma_{wind} \approx 10^{6.5} $  is based on the {\it assumption} that the ions, which carry the
electric return current in this model, have the same 4-velocity as the underlying and surrounding MHD
wind.

\vspace*{2cm}

The upper limit  $\Gamma_{wind} < 10^{4.5}$ clearly is inconsistent with a 
single fluid velocity for ions and pairs. The 1D model averages over the whole equatorial sector that feels the equatorial belt shock, shown in Figure \ref{fig:mhd}, thus mixes the ion flux with the whole flux of pairs feeding the torus, $\sim 10^{38.5}$ pairs/sec for the optical, X- and $\gamma$-ray
emission from the Crab. It yields an average value of $\sigma$ within a factor of 3 of the average value of $\sigma$ inferred from the MHD nebular models. However, the assumption 
$\Gamma_{wind} = \gamma_{ion}$ clearly violates the upper limit on $\Gamma_{wind}$, a limit which comes from assuming the whole flux of pairs feeds the equatorial torus\footnote{\cite{aro98a} suggested 
the large flux of radio emitting particles might be in higher latitude flow, a possibility I now disfavor, 
both because continuity of the Crab Nebula's spectral energy distribution militates against the  accident that would be required, if the particle spectra were formed from such different flow components, and because the modern MHD models have no such distinction between low and high latitude particle fluxes.}, an assumption supported by the approximate correspondence of the radio wisps 
(\citealt{biet04}) with the optical features.  

The sandwich magnetic field model, a fundamentally 2D construct, suggests substantial alterations of
the \cite{spit04} scheme. The magnetic field inside the current sheet is weak 
compared to that in the MHD flow outside the sheet. That weakened $B$
alters the character of the 
ion orbits from simple magnetic reflections to partial deflections from radial flow, thus altering the 
momentum transfer to the pairs and therefore the compressions.  If most of the ions flow in an essentially unmagnetized region, $\gamma_{ion}$ is no longer coupled by the magnetic field to the flow 4-velocity
of the pairs, thus allowing $\gamma_{ion} \gg \Gamma_{wind}$
(\citealt{arons07}).   

Taking such improvements of the model into account is needed before one can realistically assess the model's consequences for observations, including  possible hadronic TeV gamma ray and neutrino emission from the nebulae, a
definite prediction of the model at some flux level.  At this writing,
models of the TeV SED based on the 1D dynamical model (\citealt{amato03, bednarek03}) show that the hadronic gamma rays at $\varepsilon < 20$ TeV
are masked by inverse Compton emission;  recent HESS observations
of the Crab (\citealt{aharonian06}) suggest that an interesting constraint 
on ions in the outflow may be attainable. Recently \cite{horns06} suggested that the TeV emission observed from the Vela-X PWN might be due to hadronic emission from the p-p interaction.  
Simple evolutionary models (Bucciantini and Arons, in preparation) 
suggest that such an interpretation is supported by the ion current sandwich model, when proper account is taken of the expansion and compression history 
of this older (age $\sim 10^{4.3}$ years) PWN.

Whatever the ultimate fate of models of this sort, they suggest the usefulness
of stepping beyond MHD (which does not by itself constrain the wind velocity or 
density) in modeling the observations with the goal of extracting the plasma content and constraining just what does come out of
pulsars, and other compact objects.  

See \cite{kirk07} for a parallel review of this subject.

\subsubsection{Shock Acceleration}

Starting with the seminal work of \cite{rees74}, the conversion of flow energy to
the nonthermal particles emitting nebular synchrotron and inverse Compton radiation has been attributed to some form of  ``shock acceleration''.  Most commonly, Diffusive Fermi Acceleration (DFA) has been invoked, even though in relativistic shocks it faces a number of substantial difficulties, especially when the magnetic field is transverse to the flow. Mechanisms have not been apparent to supply the very large amplitude turbulence required (\citealt{niemic06}), which must extend to large distances ($\sim R_{shock}$) both up and downstream of  the shock
so as to have a large ``optical depth'' for scattering of the largest Larmor radius particles both up- and down-stream.  The spectrum of test particles accelerated 
depends sensitively upon the isotropy (or lack thereof) of the scattering process.
However, in the case of isotropic scattering in the fluid frame, the test particle spectra are encouraging - Monte Carlo (\citealt{kirk00}) and analytic  (\citealt{keshet05}) calculations with assumed 
scattering rules and infinite optical depth for particle scattering up and downstream yield an accelerated particle spectrum in the downstream medium $N(E) \propto E^{-20/9} $, which is almost exactly that inferred by modeling the synchrotron emission in a 1D post-shock flow in the Crab Nebula (\citealt{ken84}) - this simplified flow model should be a not unreasonable approximation to flow right outside the equatorial belt shock shown in Figure \ref{fig:mhd}. The efficiency depends entirely on what is assumed for the particle injection rate into the process, and the acceleration rate depends entirely
on the assumed turbulence amplitude that goes into the scattering law adopted.

\cite{hoshino92} suggested an alternate process, especially well tuned to
the mixture of heavy ions and pairs injected in the equator with the magnetic field transverse to the flow\footnote{As in the high energy beam, cyclotron instability interpretation of the wisps, an electron current accelerated in the
equatorial current sheet of an ``obtuse'' pulsar can replace the ion beam accelerated in the current sheet of an ``acute'' pulsar without altering the
conclusions - at equal relativistic energy/particle, the only difference is the sense of gyration with respect to the unknown vector direction of the magnetic field.}. Using 1D PIC simulations, they, and, more recently, \cite{amato06} with higher mass ratio and resolution, showed that high harmonics of the ion  cyclotron waves generated by the ions as they pass through the shock in the pairs can be resonantly absorbed by the positrons 
{\it and} electrons, which are non-thermally heated, yielding power law downstream distributions with a spectral slope that depends on the ratio of the upstream ion energy density to that of the pairs. 

\begin{figure}[H]
\begin{center}
\unitlength = 0.0011\textwidth
\hspace{10\unitlength}
\begin{picture}(200,200)(0,15)
\put(-365,95){\makebox(100,100)[tl]{\includegraphics[width=4.8in]{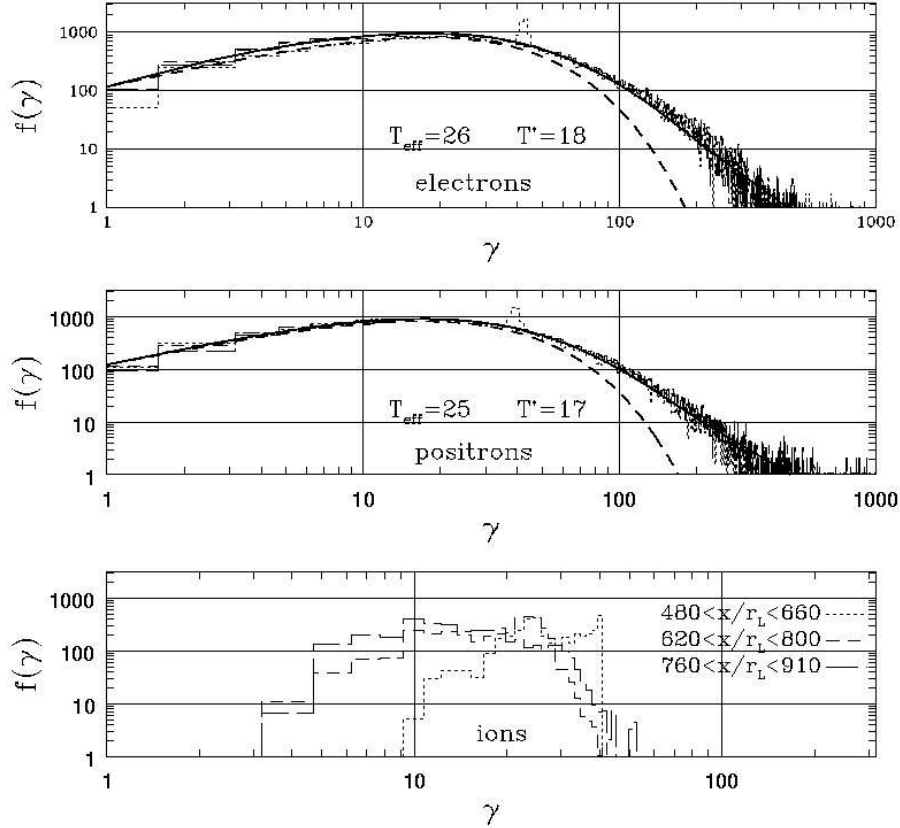}}}
\end{picture}
\end{center}
\vspace{9cm}
\caption{Downstream particle spectra of a PIC simulation of a shock with upstream magnetic field transverse to the flow in an electron-positron-proton plasma with mass ratio 
$m_p/m_\pm =100$ and upstream flow Lorentz factor of all species $\gamma_1= 40$.  The 
upstream Lorentz factor sets the scale; the results are otherwise independent of the specific value
of $\gamma_1$, so long as it is larger than 2 or 3. 
$f(\gamma)$ are the distribution functions, with $\int f(\gamma) \gamma d\gamma =$ species density.
The ratio of the proton number density to the number density of electrons plus positrons was 1/19, therefore the energy density in upstream protons was 2.1 times the energy density in pairs. The dashed curves in the panels showing the electron and positron spectra are the best fit Maxwellians, while the temperatures stated in the figure 
(which are in units of $m_\pm c^2$) rows are inferred from fitting a Maxwellian with a power law tail to 
the numerical data. The slope of the power suprathermal component is $\alpha = 3.2$, corresponding
to an energy space spectrum $dN/d\gamma \propto \gamma^{-(\alpha -1)} \propto \gamma^{-2.2}$, quite close to what is required 
in modeling the optical and X-ray synchrotron emission from the Crab Nebula and other young
pulsar wind nebulae.  The power law spectral index is a strong function of the upstream energy density
ratio, thus making the downstream nonthermality of the pairs a strong function of the upstream
composition. For the details, including the power law spectra index as a function of density ratio, see \cite{amato06}. \label{fig:cycaccel}}
\end{figure}

The nonthermal part of the
$e^\pm$ spectra shown in Figure \ref{fig:cycaccel} extends from the pairs' flow
energy/particle all the way up to the ions' energy/particle. 
 If all the species
have the same upstream flow velocity, the resulting spectra nicely span the range required for optical, X-ray and $\gamma$-ray emission from the Crab
(\citealt{hoshino92}), and for X-ray and $\gamma$-ray emission from G320 
\vspace*{0.0cm}
\noindent(\citealt{gaens02}). As demonstrated by \cite{amato06}, however, when applied to the 1D model of \cite{gallant94} of the Crab Pulsar wind's termination shock, this mechanism has trouble providing an accelerated spectrum of pairs in accord with the observation.
 
The ion flux by number is fixed at Goldreich-Julian value (it can hardly be 
anything else). If $\gamma_{ion} = \Gamma_{wind}$, the upstream energy density ratio 
$(U_\pm /U_{ion})_1 = (2n_\pm \Gamma_{wind} m_\pm /\gamma_{ion} n_{ion} m_i) \sim 10^3$ 
leads to the
pairs' particle energy distribution  hardly differing from a relativistic Maxwellian, 
the downstream distribution for a relativistic transverse shock in a pure pair plasma 
(\citealt{langdon88, gallant92, spit07}), not at all in accord
with the observations. 

\begin{figure}
\begin{center}
\unitlength = 0.0011\textwidth
\hspace{10\unitlength}
\begin{picture}(200,200)(0,15)
\put(-405,95){\makebox(100,150)[tl]{\includegraphics[width=4.9in]{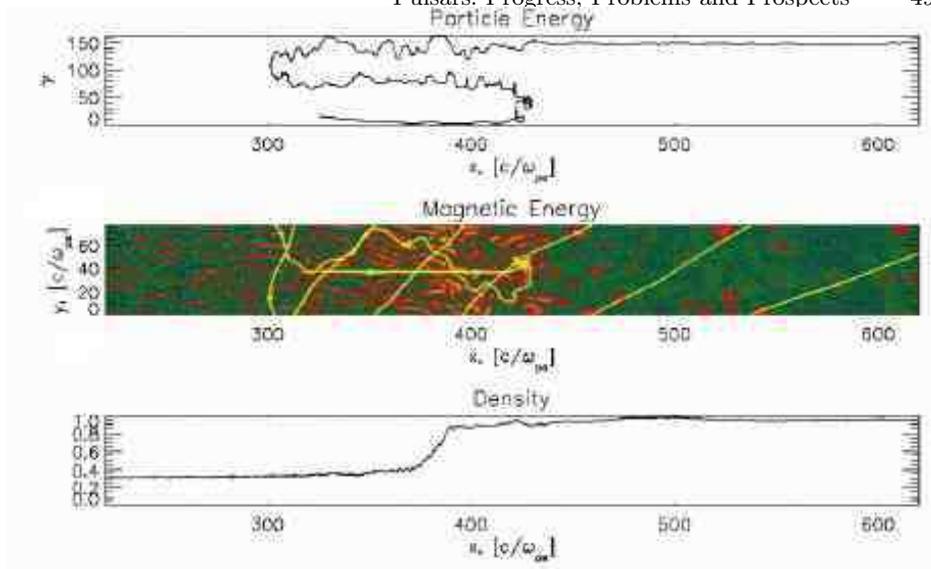}}}
\end{picture}
\end{center}
\vspace{4.5cm}
\caption{Shock structure and test particle gaining energy at the shock front, from 
a PIC simulation of an unmagnetized $e^\pm$ shock  by \cite{spit07a}. Lengths are measured in units of the upstream skin depth. Bottom panel:  Density as a function of position, exhibiting
the factor of 3 jump (properties of this shock in a 2D plasma are measured in the downstream frame).  Middle panel:  $(B^2)^{1/4}$, showing the upstream magnetic filaments characteristic of the
Weibel instability in the linear regime, the scrambled magnetic structures formed when the currents reach the Alfven critical current, magnetic trapping disrupts the flow
and the shock transition forms (\citealt{kato05}).  Note the transition of the magnetic structure to spatially intermittent (widely separated) islands in the downstream. The orbit of a test particle is superimposed. Because of the finite thickness of the strongly
turbulent scattering layer, particles escape downstream after a finite number of scatterings.  The sideways dimension is periodic, thus a particle leaving the
box at the top reappears at the lower bottom. Top panel:  Energy of the test particle, which started with the upstream flow
energy ($\gamma = 15$) and increased its energy by a factor of 10 before escaping.\label{fig:traceone}}
\end{figure}

If acceleration occurs near the sandwich midplane, the obstacles to acceleration by either mechanism may be reduced.  The cyclotron mechanism benefitsfrom $\gamma_{ion} \gg \Gamma_{wind}$.
With pair multiplicity $\kappa_\pm = 2n_\pm /n_{GJ} > 10^6$ 
(required to supply the radio emission of the Crab Nebula)
and therefore $\Gamma_w \leq 10^4$ (since most of the energy flux is carried 
by the pairs) while $\gamma_{ion} \sim 10^{6.5}$, now 
$(U_\pm /U_{ion} )_1 \sim 3$, which leads to a downstream
particle spectrum possibly as flat as the $E^{-1.5}$ radio emitting particles. DFA might benefit from a weaker transverse magnetic field -
for $\sigma_{local} < 10^{-3}$ within the current sheet, magnetized shocks in pair plasmas become indistinguishable from shocks formed in a flow with no magnetic field at all (\citealt{spit07}). Figure 
\ref{fig:traceone} shows a snapshot of a 2D PIC simulation of a shock in a 
$B=0, \; e^\pm$ plasma, exhibiting
a test particle gaining energy as it scatters in the magnetic turbulence in and 
behind the shock front, which forms due to the Weibel instability driven by the upstream flow penetrating into the heated downstream medium.

\vspace*{0cm}

\begin{figure}
\begin{center}
\unitlength = 0.0011\textwidth
\hspace{10\unitlength}
\begin{picture}(200,200)(0,15)
\put(-365,95){\makebox(100,100)[tl]{\includegraphics[width=4.65in]{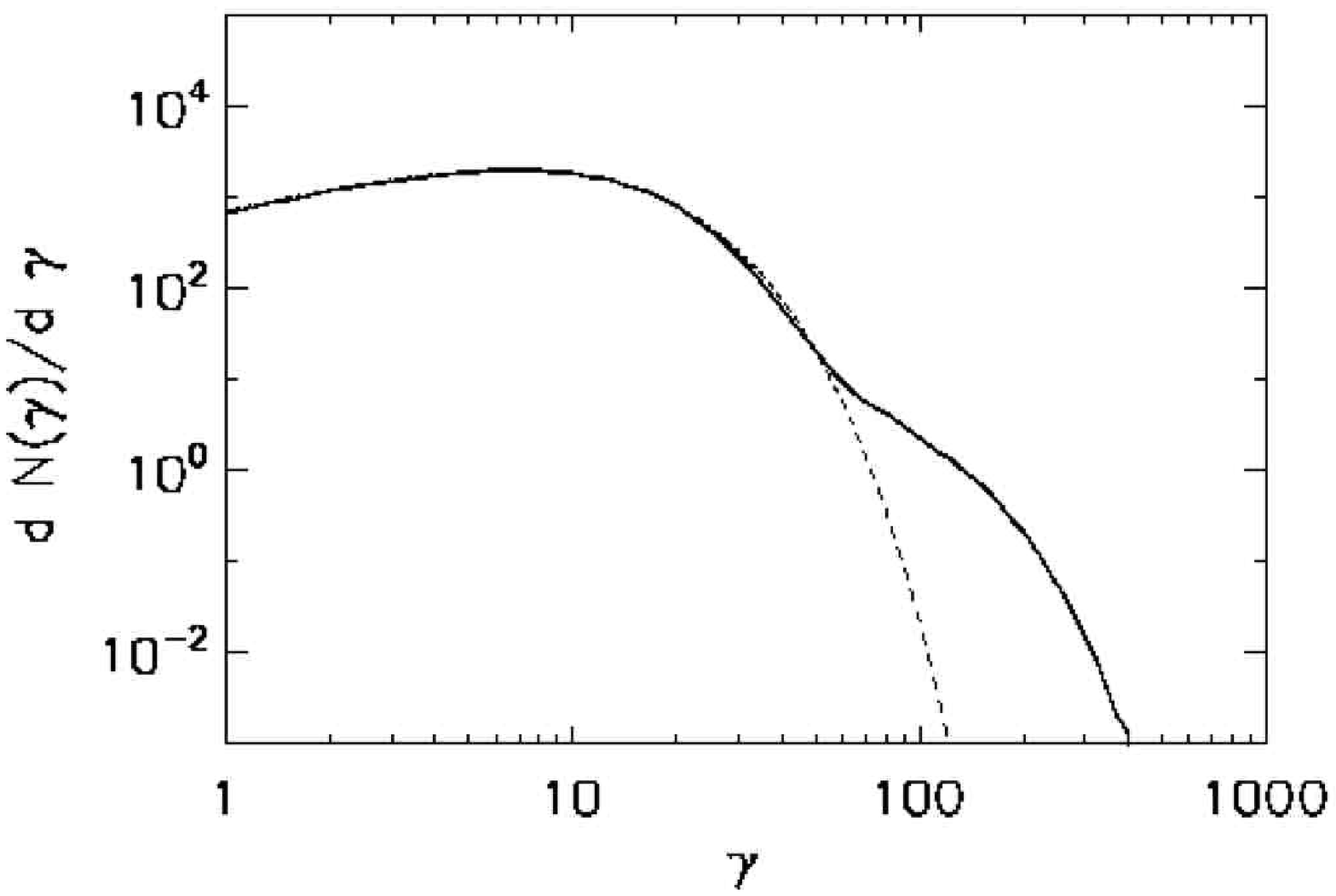}}}
\end{picture}
\end{center}
\vspace{5.5cm}
\caption{Downstream particle distribution function in the large PIC simulation of a pair
shock with no systematic magnetic field in the upstream medium used to create Figure \ref{fig:traceone}. 
The lowest dashed is a Maxwellian distribution fit. The solid curve is the actual particle distribution. Clearly, scattering in and around the shock front produces suprathermal particles (in this example, these are 5\% by number and 20\% by energy). \label{fig:spectmime}}
\end{figure}

The downstream pair spectra found in this simulation are shown in Figure \ref{fig:spectmime}, which exhibit the formation of a suprathermal particle spectrum through scattering in the turbulence in and near the shock front. To go to energies much higher requires turbulence that persists to much greater depths in the downstream than are studied in the simulations, with amplitudes that do not decay. Phase mixing between particles and fields may cause the decay of the downstream magnetic fields (\citealt{chang07}), 
 in completely unmagnetized shocks.
Nevertheless, these results, which show how nonthermal particles can be created out of the thermal 
pool in  a very weakly magnetized shock (upstream $\sigma < 10^{-3}$, as might be characteristic of the
central regions of the equatorial current sheet), there to act as seeds for DFA. Identifying the necessary scattering turbulence remains a challenge.

\vspace*{0cm}

These speculative possibilities suggest a solution to the vexing question of how the peculiar spectrum of
particles injected into the Crab Nebula  and other PWNe might be formed.  These systems all
have very flat power law distributions of particles $N(E) \propto E^{-p}, \; 1 \leq p \leq 1.5$ giving rise to
their radio synchrotron emission, while the inferred injection spectra of particles required for the
harder photon emissions (optical when seen, X-ray) have much steeper spectra, $p = 2.2 - 2.4$. 
The continuity observed between the radio and the harder photon spectra suggests the shock
injects a broken power law spectrum. Cyclotron resonant acceleration, driven by an ion flow in the current sheet with $\gamma_{ion} \gg \Gamma_w$, might be responsible for the very flat spectrum of radio and infrared emitting electrons, whose upper cutoff is determined by energy conservation to be not large compared to $\Gamma_{wind}$.  DFA might be responsible for continued acceleration to  optical, X-ray and gamma ray synchrotron emitting energies. This combination was first suggested by \cite{gallant02}. In their formulation, in which $\gamma_{ion} = \Gamma_{wind}$, the ions had negligible energy density compared to the much denser pairs and therefore could not act as the desired accelerator, while in the version suggested here, the ions having a 4 velocity much larger than that of the pairs 
allows them to act as an effective agent in creating the nonthermal radio emitting particles. A quantitative
expression of these thoughts is under investigation.

The scheme outlined above assumes the stripes in the wind do in fact disappear well upstream of the shock, as is suggested by the rapid dissipation results of \cite{arons07}.  \cite{lyubarsky03}, starting from the slow stripe dissipation model of \cite{lyubarsky01}, showed, using 1D and 2.5D kinetic 
simulations in $e^\pm$ plasma, that the striped field can annihilate in a broad ``shock'' region with strong plasma
heating - thus, the effect is as if $\sigma$ was low in the upstream medium, even when it was not in fact
small.  He also presented evidence that reconnection at individual sheets might create a flat particle
distribution $N \propto E^{-1.1}$ which might be of use in understanding the radio emission from PWNe.  However, a full 3D treatment is needed in order
to properly evaluate the nonthermal particle acceleration, since the periodicity
of the simulation in the electric field direction can lead to artificial results for
particle acceleration. 
 
A clear theoretical picture has yet to emerge, never mind models with readily checkable observational predictions. And all of these models and the observations underlying them tell us that a fully 
quantitative account of the pair plasma supply is still lacking. There will be
progress on the theoretical front in the next few years.  The much
anticipated launch of GLAST may add a new observational handle on these
problems, since the highest energy radiating particles have rapid radiation 
losses, therefore leading to interesting time series in the gamma ray emission
as the particles simultaneously accelerate and radiate in the time variable
termination shock region.

\section{Conclusion: Pulsar Problems and Prospects \label{sec:conclusion}}

Pulsar physics has made substantial progress in the last decade.  The emergence of the
MHD model of relativistic plasma flow in Pulsar Wind Nebulae has given a plausible account of the plume-torus  (a.k.a. jet-torus) structures observed in these systems, thus reconciling
the external response of the surrounding world to the well known difficulties in forming a
jet in the relativistic winds themselves. These models also explain the lack of shock
excitation of the nebulae at high latitude as a result of the shock's lack of spherical symmetry, itself a consequence of the anisotropy of the energy outflow found in the 
energy flux emerging from the magnetosphere.  

Application of force-free electrodynamics and relativistic MHD to the winds emerging from magnetospheres
with dipole magnetic fields has led to the first theory of the oblique rotator's energy loss that incorporates qualitative changes from the vacuum theory imposed by electric current flow, leading to the delightfully simple formula for the spindown energy loss  given in expression
(\ref{eq:spindown}).  For the model of the magnetopshere, the most important result is that the polar electric current distribution is close to that of the 
monopolar magnetopshere, reflecting the asymptotically monopolar poloidal 
magnetic field beyond the light cylinder.

The role of reconnection in the transfer of open to closed magnetic flux (and back again, since this is an unsteady process) has begun to be assessed, and is full of promise as a path to a physical theory of the boundary layer between the closed
and open magnetosphere, where existing gamma ray observations and gap models
suggest the most prominent photon emissions from pulsars occur. Prominent issues waiting assessment include reconciling the creation of pairs with the monopolar current distribution in expression (\ref{eq:polcapJ}) and Figure \ref{fig:polarcurrent}, with various solutions being on the table, awaiting surgery 
- these range
from rapid local current fluctuations (averaging to the force-free current) to manipulation of the polar cap electrostatics by return currents on the open flux tube boundaries.  These models all have consequences for long standing issues such as the origin of torque fluctuations, radio subpulse phase randomness and drifting, possibly for the origin of the departures of the braking index from its canonical value of 3, and for the origin of the large particle fluxes inferred from nebular radio emission. The modeling will be observationally illuminated 
by the results of the upcoming GLAST gamma ray mission.  If high sensitivity X-ray astronomy has a future, observations with the ability to inspect variability in the X-rays,
both nonthermal and  thermal from polar caps heated by magnetospheric currents
would be invaluable. 

The longstanding problem of the origin of the weak magnetic fields inferred downstream of pulsar winds' termination shocks in the young nebulae is still an outstanding question, with dissipation
of the magnetic stripes in the wind being the prime suspect. Whether this occurs in the wind far upstream form the shock, or in the shock itself, is an open question under active investigation.

Finally, the basic physics of relativistic shock waves is receiving significant attention, which opens the prospect of having a testable theory of the conversion of flow energy to non-thermal particle spectra in these relativistic systems within the next few years.
Significant issues that will receive attention include the mixture (if any) of Diffusive Fermi Acceleration and other shock related processes, the possible role of protons and other heavy ions as well as pairs in the acceleration physics (and testing for these ions' presence through VHE gamma ray and neutrino observations), and an understanding of how the nebular radio emitting electrons can be accelerated by the curved termination shock.

Exciting times lie ahead!

\vspace*{0.5cm}
\noindent {\bf Acknowledgments}

\vspace*{0.5cm}
\noindent I have benefitted from many discussions with A. Spitkovsky, P. Chang, N. Bucciantini, E. Amato, R. Blandford, F. Coroniti, D. Backer and E. Quataert. 
My research efforts on these topics have been supported by
NSF grant AST-0507813 and NASA grant NNG06G108G, both to the University of California, Berkeley; by the Department of Energy contract to the Stanford Linear Accelerator Center no. DE-AC3-76SF00515; and by the taxpayers of California.

\printindex

\end{document}